\documentclass[11pt, a4paper]{article}

\usepackage{jheppub}
\usepackage{tikz}
\usepackage{relsize}
\usepackage{subcaption}
\usepackage{youngtab}
\Yboxdim{8pt} 

\usetikzlibrary{calc, positioning}
\tikzset{
  baseline={([yshift=-0.75ex]current bounding box.center)},
  zerosep/.style={inner sep=0pt, outer sep=0pt, minimum size=0pt},
  node distance=8pt,
  align at top/.style={baseline=(current bounding box.north)},
}

\usetikzlibrary{decorations,decorations.markings}
\tikzset{
  strike out/.style={
    postaction=decorate,
    decoration={
      markings,
      mark=at position 0.5 with {
        \draw[-] (-2pt, -3pt) -- (2pt, 3pt);
      }
    }
  }
}

\tikzset{
vev/.style={strike out},
pdvev/.style={snake=snake},
}

\newcommand{\id}{\mathop{\mathrm{id}}\nolimits}

\newcommand{\vev}[1]{\langle #1 \rangle}

\newcommand{\Bigvev}[1]{\Bigl\langle #1 \Bigr\rangle}
\newcommand{\biggvev}[1]{\biggl\langle #1 \biggr\rangle}

\newcommand{\diag}{\mathop{\mathrm{diag}}\nolimits}

\newcommand{\Tr}{\mathop{\mathrm{Tr}}\nolimits}
\newcommand{\End}{\mathop{\mathrm{End}}\nolimits}
\newcommand{\Res}{\mathop{\mathrm{Res}}\nolimits}

\newcommand{\SU}{\mathrm{SU}}

\newcommand{\Spin}{\mathrm{Spin}}

\newcommand{\SL}{\mathrm{SL}}

\newcommand{\U}{\mathrm{U}}

\newcommand{\iso}{\cong}

\newcommand{\Z}{\mathbb{Z}}

\newcommand{\R}{\mathbb{R}}
\newcommand{\C}{\mathbb{C}}

\newcommand{\T}{\mathbb{T}}

\let\nc\newcommand
\let\renc\renewcommand
\nc{\wbar}{\overline}
\let\td\tilde
\let\wtd\widetilde
\let\wht\widehat
\let\mcl\mathcal

\nc{\ab}{{\bar{a}}} \nc{\at}{\tilde{a}} \nc{\ah}{\hat{a}}
\nc{\bb}{{\bar{b}}} \nc{\bt}{\tilde{b}} \nc{\bh}{\hat{b}}
\nc{\cb}{{\bar{c}}} \nc{\ct}{\tilde{c}} 
\nc{\db}{{\bar{d}}} \nc{\dt}{\tilde{d}} \renc{\dh}{\hat{d}}
\nc{\eb}{{\bar{e}}} \nc{\et}{\tilde{e}} \nc{\eh}{\hat{e}}
\nc{\fb}{{\bar{f}}} \nc{\ft}{\tilde{f}} \nc{\fh}{\hat{f}}
\nc{\gb}{{\bar{g}}} \nc{\gt}{\tilde{g}} \nc{\gh}{\hat{g}}
\nc{\hb}{{\bar{h}}} \nc{\hh}{\hat{h}} 
\nc{\ib}{{\bar{\imath}}} \nc{\ih}{\hat{\imath}} 
\nc{\jb}{{\bar{\jmath}}} \nc{\jt}{\tilde{\jmath}} \nc{\jh}{\hat{\jmath}}
\nc{\kb}{{\bar{k}}} \nc{\kt}{\tilde{k}} \nc{\kh}{\hat{k}}
\nc{\lb}{{\bar{l}}} \nc{\lt}{\tilde{l}} \nc{\lh}{\hat{l}}
\nc{\mb}{{\bar{m}}} \nc{\mt}{\tilde{m}} \nc{\mh}{\hat{m}}
\nc{\nb}{{\bar{n}}} \nc{\nt}{\tilde{n}} \nc{\nh}{\hat{n}}
\nc{\ob}{{\bar{o}}} \nc{\ot}{\tilde{o}} \nc{\oh}{\hat{o}}
\nc{\pb}{{\bar{p}}} \nc{\pt}{\tilde{p}} \nc{\ph}{\hat{p}}
\nc{\qb}{{\bar{q}}} \nc{\qt}{\tilde{q}} \nc{\qh}{\hat{q}}
\nc{\rb}{{\bar{r}}} \nc{\rt}{\tilde{r}} \nc{\rh}{\hat{r}}
\renc{\sb}{{\bar{s}}} \nc{\st}{\tilde{s}} \nc{\sh}{\hat{s}}
\nc{\tb}{{\bar{t}}} \renc{\th}{\hat{t}} 
\nc{\ub}{{\bar{u}}} \nc{\ut}{\tilde{u}} \nc{\uh}{\hat{u}}
\nc{\vb}{{\bar{v}}} \nc{\vt}{\tilde{v}} \nc{\vh}{\hat{v}}
\nc{\wb}{{\bar{w}}} \nc{\wt}{\tilde{w}} \nc{\wh}{\hat{w}}
\nc{\xb}{{\bar{x}}} \nc{\xt}{\tilde{x}} \nc{\xh}{\hat{x}}
\nc{\yb}{{\bar{y}}} \nc{\yt}{\tilde{y}} \nc{\yh}{\hat{y}}
\nc{\zb}{{\bar{z}}} \nc{\zt}{\tilde{z}} \nc{\zh}{\hat{z}}

\nc{\Ab}{{\wbar{A}}} \nc{\At}{{\wtd{A}}} \nc{\Ah}{{\wht{A}}}
\nc{\Bb}{{\wbar{B}}} \nc{\Bt}{{\wtd{B}}} \nc{\Bh}{{\wht{B}}}
\nc{\Cb}{{\wbar{C}}} \nc{\Ct}{{\wtd{C}}} \nc{\Ch}{{\wht{C}}}
\nc{\Db}{{\wbar{D}}} \nc{\Dt}{{\wtd{D}}} \nc{\Dh}{{\wht{D}}}
\nc{\Eb}{{\wbar{E}}} \nc{\Et}{{\wtd{E}}} \nc{\Eh}{{\wht{E}}}
\nc{\Fb}{{\wbar{F}}} \nc{\Ft}{{\wtd{F}}} \nc{\Fh}{{\wht{F}}}
\nc{\Gb}{{\wbar{G}}} \nc{\Gt}{{\wtd{G}}} \nc{\Gh}{{\wht{G}}}
\nc{\Hb}{{\wbar{H}}} \nc{\Ht}{{\wtd{H}}} \nc{\Hh}{{\wht{H}}}
\nc{\Ib}{{\bar{I}}} \nc{\It}{{\wtd{I}}} \nc{\Ih}{{\wht{I}}}
\nc{\Jb}{{\bar{J}}} \nc{\Jt}{{\wtd{J}}} \nc{\Jh}{{\wht{J}}}
\nc{\Kb}{{\wbar{K}}} \nc{\Kt}{{\wtd{K}}} \nc{\Kh}{{\wht{K}}}
\nc{\Lb}{{\wbar{L}}} \nc{\Lt}{{\wtd{L}}} \nc{\Lh}{{\wht{L}}}
\nc{\Mb}{{\wbar{M}}} \nc{\Mt}{{\wtd{M}}} \nc{\Mh}{{\wht{M}}}
\nc{\Nb}{{\wbar{N}}} \nc{\Nt}{{\wtd{N}}} \nc{\Nh}{{\wht{N}}}
\nc{\Ob}{{\wbar{O}}} \nc{\Ot}{{\wtd{O}}} \nc{\Oh}{{\wht{O}}}
\nc{\Pb}{{\wbar{P}}} \nc{\Pt}{{\wtd{P}}} \nc{\Ph}{{\wht{P}}}
\nc{\Qb}{{\wbar{Q}}} \nc{\Qt}{{\wtd{Q}}} \nc{\Qh}{{\wht{Q}}}
\nc{\Rb}{{\wbar{R}}} \nc{\Rt}{{\wtd{R}}} \nc{\Rh}{{\wht{R}}}
\nc{\Sb}{{\wbar{S}}} \nc{\St}{{\wtd{S}}} \nc{\Sh}{{\wht{S}}}
\nc{\Tb}{{\wbar{T}}} \nc{\Tt}{{\wtd{T}}} \nc{\Th}{{\wht{T}}}
\nc{\Ub}{{\wbar{U}}} \nc{\Ut}{{\wtd{U}}} \nc{\Uh}{{\wht{U}}}
\nc{\Vb}{{\wbar{V}}} \nc{\Vt}{{\wtd{V}}} \nc{\Vh}{{\wht{V}}}
\nc{\Wb}{{\wbar{W}}} \nc{\Wt}{{\wtd{W}}} \nc{\Wh}{{\wht{W}}}
\nc{\Xb}{{\wbar{X}}} \nc{\Xt}{{\wtd{X}}} \nc{\Xh}{{\wht{X}}}
\nc{\Yb}{{\wbar{Y}}} \nc{\Yt}{{\wtd{Y}}} \nc{\Yh}{{\wht{Y}}}
\nc{\Zb}{{\wbar{Z}}} \nc{\Zt}{{\wtd{Z}}} \nc{\Zh}{{\wht{Z}}}

\nc{\CA}{{\mcl{A}}} \nc{\CAb}{{\wbar{\CA}}} \nc{\CAt}{{\wtd{\CA}}} \nc{\CAh}{{\wht{\CA}}}
\nc{\CB}{{\mcl{B}}} \nc{\CBb}{{\wbar{\CB}}} \nc{\CBt}{{\wtd{\CB}}} \nc{\CBh}{{\wht{\CB}}}
\nc{\CC}{{\mcl{C}}} \nc{\CCb}{{\wbar{\CC}}} \nc{\CCt}{{\wtd{\CC}}} \nc{\CCh}{{\wht{\CC}}}
\nc{\cD}{{\mcl{D}}} \nc{\cDb}{{\wbar{\cD}}} \nc{\cDt}{{\wtd{\cC}}} \nc{\cDh}{{\wht{\cD}}}
\nc{\CE}{{\mcl{E}}} \nc{\CEb}{{\wbar{\CE}}} \nc{\CEt}{{\wtd{\CE}}} \nc{\CEh}{{\wht{\CE}}}
\nc{\CF}{{\mcl{F}}} \nc{\CFb}{{\wbar{\CF}}} \nc{\CFt}{{\wtd{\CF}}} \nc{\CFh}{{\wht{\CF}}}
\nc{\CG}{{\mcl{G}}} \nc{\CGb}{{\wbar{\CG}}} \nc{\CGt}{{\wtd{\CG}}} \nc{\CGh}{{\wht{\CG}}}
\nc{\CH}{{\mcl{H}}} \nc{\CHb}{{\wbar{\CH}}} \nc{\CHt}{{\wtd{\CH}}} \nc{\CHh}{{\wht{\CH}}}
\nc{\CI}{{\mcl{I}}} \nc{\CIb}{{\wbar{\CI}}} \nc{\CIt}{{\wtd{\CI}}} \nc{\CIh}{{\wht{\CI}}}
\nc{\CJ}{{\mcl{J}}} \nc{\CJb}{{\wbar{\CJ}}} \nc{\CJt}{{\wtd{\CJ}}} \nc{\CJh}{{\wht{\CJ}}}
\nc{\CK}{{\mcl{K}}} \nc{\CKb}{{\wbar{\CK}}} \nc{\CKt}{{\wtd{\CK}}} \nc{\CKh}{{\wht{\CK}}}
\nc{\CL}{{\mcl{L}}} \nc{\CLb}{{\wbar{\CL}}} \nc{\CLt}{{\wtd{\CL}}} \nc{\CLh}{{\wht{\CL}}}
\nc{\CM}{{\mcl{M}}} \nc{\CMb}{{\wbar{\CM}}} \nc{\CMt}{{\wtd{\CM}}} \nc{\CMh}{{\wht{\CM}}}
\nc{\CN}{{\mcl{N}}} \nc{\CNb}{{\wbar{\CN}}} \nc{\CNt}{{\wtd{\CN}}} \nc{\CNh}{{\wht{\CN}}}
\nc{\CO}{{\mcl{O}}} \nc{\COb}{{\wbar{\CO}}} \nc{\COt}{{\wtd{\CO}}} \nc{\COh}{{\wht{\CO}}}
\nc{\CP}{{\mcl{P}}} \nc{\CPb}{{\wbar{\CP}}} \nc{\CPt}{{\wtd{\CP}}} \nc{\CPh}{{\wht{\CP}}}
\nc{\CQ}{{\mcl{Q}}} \nc{\CQb}{{\wbar{\CQ}}} \nc{\CQt}{{\wtd{\CQ}}} \nc{\CQh}{{\wht{\CQ}}}
\nc{\CR}{{\mcl{R}}} \nc{\CRb}{{\wbar{\CR}}} \nc{\CRt}{{\wtd{\CR}}} \nc{\CRh}{{\wht{\CR}}}
\nc{\CS}{{\mcl{S}}} \nc{\CSb}{{\wbar{\CS}}} \nc{\CSt}{{\wtd{\CS}}} \nc{\CSh}{{\wht{\CS}}}
\nc{\CT}{{\mcl{T}}} \nc{\CTb}{{\wbar{\CT}}} \nc{\CTt}{{\wtd{\CT}}} \nc{\CTh}{{\wht{\CT}}}
\nc{\CU}{{\mcl{U}}} \nc{\CUb}{{\wbar{\CU}}} \nc{\CUt}{{\wtd{\CU}}} \nc{\CUh}{{\wht{\CU}}}
\nc{\CV}{{\mcl{V}}} \nc{\CVb}{{\wbar{\CV}}} \nc{\CVt}{{\wtd{\CV}}} \nc{\CVh}{{\wht{\CV}}}
\nc{\CW}{{\mcl{W}}} \nc{\CWb}{{\wbar{\CW}}} \nc{\CWt}{{\wtd{\CW}}} \nc{\CWh}{{\wht{\CW}}}
\nc{\CX}{{\mcl{X}}} \nc{\CXb}{{\wbar{\CX}}} \nc{\CXt}{{\wtd{\CX}}} \nc{\CXh}{{\wht{\CX}}}
\nc{\CY}{{\mcl{Y}}} \nc{\CYb}{{\wbar{\CY}}} \nc{\CYt}{{\wtd{\CY}}} \nc{\CYh}{{\wht{\CY}}}
\nc{\CZ}{{\mcl{Z}}} \nc{\CZb}{{\wbar{\CZ}}} \nc{\CZt}{{\wtd{\CZ}}} \nc{\CZh}{{\wht{\CZ}}}

\let\eps\epsilon
\let\ups\upsilon
\let\veps\varepsilon
\let\vtht\vartheta

\let\vsgm\varsigma
\let\vphi\varphi
\let\vrho\varrho

\nc{\alphab}{{\bar{\alpha}}} \nc{\alphat}{{\td{\alpha}}} \nc{\alphah}{{\hat{\alpha}}}
\nc{\betab}{{\bar{\beta}}}   \nc{\betat}{{\td{\beta}}}   \nc{\betah}{{\hat{\beta}}} 
\nc{\gammab}{{\bar{\gamma}}} \nc{\gammat}{{\td{\gamma}}} \nc{\gammah}{{\hat{\gamma}}} 
\nc{\deltab}{{\bar{\delta}}} \nc{\deltat}{{\td{\delta}}} \nc{\deltah}{{\hat{\delta}}} 
\nc{\epsilonb}{{\bar{\eps}}} \nc{\epsilont}{{\td{\eps}}} \nc{\epsilonh}{{\hat{\eps}}} 
\nc{\vepsb}{{\bar{\veps}}}   \nc{\vepst}{{\td{\veps}}}   \nc{\vepsh}{{\hat{\veps}}} 
\nc{\zetab}{{\bar{\zeta}}}   \nc{\zetat}{{\td{\zeta}}}   \nc{\zetah}{{\hat{\zeta}}} 
\nc{\etab}{{\bar{\eta}}}     \nc{\etat}{{\td{\eta}}}     \nc{\etah}{{\hat{\eta}}} 
\nc{\thetab}{{\bar{\theta}}} \nc{\thetat}{{\td{\theta}}} \nc{\thetah}{{\hat{\theta}}} 
\nc{\vthetab}{{\bar{\vtht}}} \nc{\vthetat}{{\td{\vtht}}} \nc{\vthetah}{{\hat{\vtht}}} 
\nc{\lambdab}{{\bar{\lambda}}} \nc{\lambdat}{{\td{\lambda}}} \nc{\lambdah}{{\hat{\lambda}}} 
\nc{\iotab}{{\bar{\iota}}}   \nc{\iotat}{{\td{\iota}}}   \nc{\iotah}{{\hat{\iota}}} 
\nc{\kappab}{{\bar{\kappa}}} \nc{\kappat}{{\td{\kappa}}} \nc{\kappah}{{\hat{\kappa}}} 
\nc{\lmdb}{{\bar{\lmd}}}     \nc{\lmdt}{{\td{\lmd}}}     \nc{\lmdh}{{\hat{\lmd}}} 
\nc{\mub}{{\bar{\mu}}}       \nc{\mut}{{\td{\mu}}}       \nc{\muh}{{\hat{\mu}}} 
\nc{\nub}{{\bar{\nu}}}       \nc{\nut}{{\td{\nu}}}       \nc{\nuh}{{\hat{\nu}}} 
\nc{\xib}{{\bar{\xi}}}       \nc{\xit}{{\td{\xi}}}       \nc{\xih}{{\hat{\xi}}} 
\nc{\pib}{{\bar{\pi}}}       \nc{\pit}{{\td{\pi}}}       \nc{\pih}{{\hat{\pi}}} 
\nc{\vpib}{{\bar{\vpi}}}     \nc{\vpit}{{\td{\vpi}}}     \nc{\vpih}{{\hat{\vpi}}} 
\nc{\rhob}{{\bar{\rho}}}     \nc{\rhot}{{\td{\rho}}}     \nc{\rhoh}{{\hat{\rho}}} 
\nc{\vrhob}{{\bar{\vrho}}}   \nc{\vrhot}{{\td{\vrho}}}   \nc{\vrhoh}{{\hat{\vrho}}} 
\nc{\sigmab}{{\bar{\sigma}}} \nc{\sigmat}{{\td{\sigma}}} \nc{\sigmah}{{\hat{\sigma}}} 
\nc{\vsigmab}{{\bar{\vsgm}}} \nc{\vsigmat}{{\td{\vsgm}}} \nc{\vsigmah}{{\hat{\vsgm}}} 
\nc{\taub}{{\bar{\tau}}}     \nc{\taut}{{\td{\tau}}}     \nc{\tauh}{{\hat{\tau}}} 
\nc{\upsb}{{\bar{\ups}}} \nc{\upst}{{\td{\ups}}} \nc{\upsh}{{\hat{\ups}}} 
\nc{\phib}{{\bar{\phi}}}     \nc{\phit}{{\td{\phi}}}     \nc{\phih}{{\hat{\phi}}} 
\nc{\varphib}{{\bar{\vphi}}}   \nc{\varphit}{{\td{\vphi}}}   \nc{\varphih}{{\hat{\vphi}}} 
\nc{\chib}{{\bar{\chi}}}     \nc{\chit}{{\td{\chi}}}     \nc{\chih}{{\hat{\chi}}} 
\nc{\psib}{{\bar{\psi}}}     \nc{\psit}{{\td{\psi}}}     \nc{\psih}{{\hat{\psi}}} 
\nc{\omegab}{{\bar{\omega}}} \nc{\omegat}{{\td{\omega}}} \nc{\omegah}{{\hat{\omega}}} 

\nc{\Gammab}{{\wbar{\Gamma}}}     \nc{\Gammat}{{\wtd{\Gamma}}}     \nc{\Gammah}{{\wht{\Gamma}}}
\nc{\Deltab}{{\wbar{\Delta}}}     \nc{\Deltat}{{\wtd{\Delta}}}     \nc{\Deltah}{{\wht{\Delta}}}
\nc{\Thetab}{{\wbar{\Theta}}}     \nc{\Thetat}{{\wtd{\Theta}}}     \nc{\Thetah}{{\wht{\Theta}}}
\nc{\Lambdab}{{\wbar{\Lambda}}}   \nc{\Lambdat}{{\wtd{\Lambda}}}   \nc{\Lambdah}{{\wht{\Lambda}}}
\nc{\Xib}{{\wbar{\Xi}}}           \nc{\Xit}{{\wtd{\Xi}}}           \nc{\Xih}{{\wht{\Xi}}}
\nc{\Pib}{{\wbar{\Pi}}}           \nc{\Pit}{{\wtd{\Pi}}}           \nc{\Pih}{{\wht{\Pi}}}
\nc{\Sigmab}{{\wbar{\Sigma}}}     \nc{\Sigmat}{{\wtd{\Sigma}}}     \nc{\Sigmah}{{\wht{\Sigma}}}
\nc{\Upsilonb}{{\wbar{\Upsilon}}} \nc{\Upsilont}{{\wtd{\Upsilon}}} \nc{\Upsilonh}{{\wht{\Upsilon}}}
\nc{\Phib}{{\wbar{\Phi}}}         \nc{\Phit}{{\wtd{\Phi}}}         \nc{\Phih}{{\wht{\Phi}}}
\nc{\Psib}{{\wbar{\Psi}}}         \nc{\Psit}{{\wtd{\Psi}}}         \nc{\Psih}{{\wht{\Psi}}}
\nc{\Omegab}{{\wbar{\Omega}}}     \nc{\Omegat}{{\wtd{\Omega}}}     \nc{\Omegah}{{\wht{\Omega}}}

\newcommand{\rmd}{\mathrm{d}}

\newlength{\qsep}
\usetikzlibrary{decorations.markings, arrows}
\tikzset{
  x=\qsep, y=\qsep,  font=\smaller,
  >=latex,
  ->-/.style={decoration={
      markings, mark=at position #1 with
      {\arrow{>}}},postaction={decorate}},
  -<-/.style={decoration={
      markings, mark=at position #1 with
      {\arrow{<}}},postaction={decorate}},
  node/.style={draw, fill=white, shape=circle, minimum size=12pt, inner
    sep=0pt},
  gnode/.style={node},
  fnode/.style={node, shape=rectangle},
  tnode/.style={fnode, double, minimum size=12pt},
  q-/.style={-},
  q->/.style={->, shorten >=1pt, font=\smaller[2]},
  q<-/.style={q->, <-, shorten >=0pt, shorten <=1pt},
  eq-/.style={double, double distance=2pt},
}

\newcommand{\fnode}[1]{
  \mathbin{
    \mathchoice
    {\tikz[baseline=(x.base)]{\node[fnode] (x) at (0,0) {$#1$};}}
    {\tikz[baseline=(x.base)]{\node[fnode] (x) at (0,0) {$#1$};}}
    {\tikz[baseline=(x.base)]{\node[fnode] (x) at (0,0) {$#1$};}}
    {\tikz[baseline=(x.base)]{\node[fnode] (x) at (0,0) {$#1$};}}
}}

\newcommand{\gnode}[1]{
  \mathbin{
    \mathchoice
    {\tikz[baseline=(x.base)]{\node[gnode] (x) at (0,0) {$#1$};}}
    {\tikz[baseline=(x.base)]{\node[gnode] (x) at (0,0) {$#1$};}}
    {\tikz[baseline=(x.base)]{\node[gnode, minimum size=7pt, font=\scriptsize] (x) at (0,0) {$#1$};}}
    {\tikz[baseline=(x.base)]{\node[gnode] (x) at (0,0) {$#1$};}}
}}

\newcommand{\RM}{\check{R}}
\newcommand{\LM}{\check{L}}
\newcommand{\CRM}{\check{\CR}}

\tikzset{
  r-/.style={-, thick},
  r->/.style={r-, ->},
  r<-/.style={r-, <-},
  r->-/.style={->-=#1, thick},
}

\tikzset{
  z-/.style={-},
  z->/.style={z-, ->},
  z<-/.style={z->, <-},
  dz-/.style={z-, densely dashed},
  dz->/.style={dz-, ->},
  dz<-/.style={dz-, <-},
  dz<->/.style={dz-, <->},
  shaded/.style={fill=black!20},
  lshaded/.style={fill=black!10},
  dshaded/.style={fill=black!30},
}

\tikzset{
  minp/.style={draw, shape=circle, minimum size=3pt, inner 
    sep=0pt, font=\tiny},
  maxp/.style={minp, double, double distance=1pt, fill=white, minimum
    size=5pt},
}
\newcommand{\IV}{\mathcal{I}_{\mathrm{V}}}
\newcommand{\IB}{\mathcal{I}_{\mathrm{B}}}
\newcommand{\IBt}{\widetilde{\mathcal{I}}_{\mathrm{B}}}
\newcommand{\V}{\mathbb{V}}
\newcommand{\W}{\mathbb{W}}
\newcommand{\dia}{{\tikz[baseline=(current bounding box.base)]{\draw[scale=0.06, semithick] (0,0) -- (1,1) -- (0,2) -- (-1,1) -- cycle;}}}
\newcommand{\tri}{{\tikz[baseline=(current bounding box.base)]{\draw[scale=0.06, semithick] (0,0) -- (2,0) -- (1,2) -- cycle;}}}
\newcommand{\Vdia}{\V^\dia}
\newcommand{\Vtri}{\V^\tri}
\newcommand{\Rdia}{\RM^\dia}
\newcommand{\Rtri}{\RM^\tri}
\newcommand{\CRdia}{\CRM^\dia}

\newcommand{\Ldia}{\LM^\dia}

\newcommand{\LSkl}{\LM^{\text{S}}}
\newcommand{\LDS}{\LM^{\text{DS}}}
\newcommand{\RBax}{\CRM^{\text{B}}}
\newcommand{\RDS}{\RM^{\text{DS}}}
\newcommand{\UV}{\mathrm{UV}}
\newcommand{\IR}{\mathrm{IR}}

\newcommand{\pc}{\chi}
\newcommand{\po}{\mathrm{o}}

\title{Surface defects as transfer matrices}

\author[a,b]{Kazunobu Maruyoshi}
\author[c]{and Junya Yagi}

\emailAdd{maruyoshi@st.seikei.ac.jp, junya.yagi@fuw.edu.pl}

\affiliation[a]{Department of Physics, Imperial College London\\
  Blackett Laboratory, Prince Concert Road, South Kensington, London,
  SW7 2AZ, UK}

\affiliation[b]{Faculty of Science and Technology, Seikei University\\
  3--3--1 Kichijoji-Kitamachi, Musashino-shi, Tokyo 180--8633 Japan}

\affiliation[c]{Faculty of Physics, University of Warsaw\\
  ul.\ Pasteura 5, 02--093 Warsaw, Poland}

\abstract{The supersymmetric index of the 4d $\CN = 1$ theory realized
  by a brane tiling coincides with the partition function of an
  integrable 2d lattice model.  We argue that a class of half-BPS
  surface defects in brane tiling models are represented on the
  lattice model side by transfer matrices constructed from
  L-operators.  For the simplest surface defects in theories with
  $\SU(2)$ flavor groups, we identify the relevant L-operator as that
  discovered by Sklyanin in the context of the eight-vertex model.  We
  verify this identification by computing the indices of class-$\CS$
  and -$\CS_k$ theories in the presence of the surface defects.}
 

\arxivnumber{}

\preprint{IMPERIAL-TP-16-KM-01}

\begin{document}
\maketitle

\section{Introduction}

Remarkable connections have been uncovered in the past several years
between supersymmetric field theories and integrable lattice
models~\cite{Spiridonov:2010em, Yamazaki:2012cp, Terashima:2012cx,
  Costello:2013zra, Yamazaki:2013nra, Costello:2013sla, Yagi:2015lha,
  Yamazaki:2015voa, Gahramanov:2015cva}.  A prominent example is the
correspondence~\cite{Spiridonov:2010em, Yamazaki:2012cp,
  Terashima:2012cx, Yagi:2015lha} between 4d $\CN = 1$ quiver gauge
theories realized by certain brane configurations in string theory,
known as brane tilings~\cite{Hanany:2005ve, Franco:2005rj,
  Hanany:2005ss}, and the 2d lattice models that Bazhanov and
Sergeev~\cite{Bazhanov:2010kz, Bazhanov:2011mz} constructed using
elliptic hypergeometric integrals discovered by
Spiridonov~\cite{MR1846786, MR2044635, MR2076912, MR2325931,
  MR2479997}.  In this correspondence, the supersymmetric index of a
brane tiling model is equated with the partition function of a lattice
model, and the integrability of the latter follows from Seiberg
duality.

In this paper we study a class of half-BPS surface defects in 4d
$\CN=1$ theories from the perspective of the above correspondence.  We
argue that a surface defect in this class is represented on the
lattice model side by a transfer matrix, an object which we depict as
\begin{equation}
  \label{eq:TM-L}
  \begin{tikzpicture}[scale=0.5, baseline=(x.base)]
    \node (x) at (0,0) {\vphantom{x}};

    \draw[dz->, right hook->] (-0.1,0) -- (1.1,0);
    \draw[dz->, >=left hook] (1,0) -- (4.1,0);
    \node[fill=white, inner sep=1pt] at (2.5,0) {$\,\dots$};

    \begin{scope}[shift={(0.5,-0.5)}]
      \draw[r->] (0,0) -- (0,1.2);
      \draw[r->] (1,0) -- (1,1.2);
      \draw[r->] (3,0) -- (3,1.2);
    \end{scope}
  \end{tikzpicture}
  \ .
\end{equation}
Each crossing of a solid line with a dashed one,
\begin{equation}
  \label{eq:L}
  \begin{tikzpicture}[scale=0.5, baseline=(x.base)]
    \node (x) at (0,1) {\vphantom{x}};

    \draw[dz->] (0,1) -- (2,1);
    \draw[r->] (1,0) -- (1,2);
  \end{tikzpicture}
  \ ,
\end{equation}
represents an object we call an L-operator.  In the simplest case, we
identify the concrete form of the relevant L-operator and present a
formula for the corresponding transfer matrices.  We compare our
formula with computations based on a different approach developed in
connection with class-$\CS$ and -$\CS_k$ theories, and find that they
agree.

Since the results obtained in the present work bridge two areas of
physics in a way that may be unfamiliar to many readers, in this
introduction we provide a somewhat detailed overview.

To begin with, let us briefly review where the correspondence between
brane tilings and integrable lattice models comes from, following the
discussion in~\cite{Yagi:2015lha}.  We will give a more thorough
explanation in section~\ref{sec:BT}.

In fact, the correspondence in question is a combination of two
correspondences: one between brane tilings and 2d topological quantum
field theories (TQFTs) equipped with line operators, and another one
between 2d TQFTs with line operators localized in extra dimensions and
integrable lattice models~\cite{Costello:2013zra, Costello:2013sla}.

The first correspondence has its origin in six dimensions.  A brane
tiling model is constructed from a stack of D5-branes on
$\R^{3,1} \times \Sigma$, intersected by NS5-branes that occupy
$\R^{3,1}$ and, roughly speaking, are supported on curves in the
Riemann surface $\Sigma$.  From the viewpoint of the 6d theory living
on the D5-branes, the NS5-branes create codimension-$1$ defects or
domain walls.  At low energies, the 6d theory compactified on $\Sigma$
in the presence of these codimension-$1$ defects is described by a 4d
$\CN = 1$ theory.  Under nice circumstances, it is a gauge theory
characterized by a quiver diagram with $\SU(N)$ nodes drawn on
$\Sigma$, where $N$ is the number of the D5-branes.

The situation is therefore similar to the construction of 4d $\CN = 2$
theories of class~$\CS$~\cite{Witten:1997sc, Gaiotto:2009we,
  Gaiotto:2009hg}.  A class-$\CS$ theory is defined by
compactification of a 6d $\CN = (2,0)$ superconformal field theory on
a Riemann surface in the presence of codimension-$2$ defects.  The
locations of these defects (``punctures'') in the surface are
parameters of the theory, and generally physical quantities depend on
them.  If, however, we place the theory on $S^3 \times S^1$ and
compute its partition function, the result is the supersymmetric
index~\cite{Romelsberger:2005eg, Kinney:2005ej} which is a protected
quantity independent of continuous parameters and hence determined by
the topological data of the punctured surface.  It follows that the
index of a class-$\CS$ theory is captured by a correlation function of
local operators in a TQFT on the associated
surface~\cite{Gadde:2009kb}.

By the same logic, the index of a brane tiling model coincides with a
topological correlator on~$\Sigma$, albeit of line operators in
another TQFT.  We get a different TQFT because of the different 6d
origin, namely 6d $\CN = (1,1)$ super Yang--Mills theory instead of a
6d $\CN = (2,0)$ theory, and line operators rather than local ones
since our defects are of codimension-$1$ and not of codimension-$2$.

The second correspondence relates TQFTs and integrable lattice models
defined on the same surface $\Sigma$.  Nevertheless, a
higher-dimensional point of view is also crucial here.

Given a configuration of line operators in a TQFT on $\Sigma$, we can
express its correlation function in the form of the partition function
of a lattice model.  Generically, these operators form a lattice with
no three lines meeting at a point.  Therefore, we can cut $\Sigma$
into square pieces in such a way that each of them contains a crossing
of two lines,
\begin{equation}
  \label{eq:R}
  \begin{tikzpicture}[scale=0.5, baseline=(x.base)]
    \node (x) at (0,1) {\vphantom{x}};

    \draw[r->] (0,1) -- (2,1);
    \draw[r->] (1,0) -- (1,2);
  \end{tikzpicture}
  \ ,
\end{equation}
and perform the path integral separately on these pieces first.  The
path integral on a single piece defines the R-operator, or the
Boltzmann weight, assigned to the vertex of the lattice contained in
that piece.  To reconstruct the original surface, we glue these pieces
back together, which amounts to multiplying the R-operators from all
vertices and summing over all states on the boundaries of the pieces,
or in other words, computing the partition function.  Hence, the
structure of a 2d TQFT equipped with line operators gives rise to some
lattice model.

This fact may not be noteworthy---it is merely rewriting of the path
integral---were it not for the following observation by
Costello~\cite{Costello:2013zra, Costello:2013sla}: the lattice model
thus obtained would be integrable if there exist extra dimensions in
the TQFT.

For a lattice model to be integrable, the lattice lines must carry
continuous parameters, called spectral parameters, and the R-operator
should satisfy the Yang--Baxter equation
\begin{equation}
  \label{eq:YBE}
  \begin{tikzpicture}[scale=0.5, baseline=(x.base)]
    \node (x) at (30:2) {\vphantom{x}};

    \draw[r->] (0,0) -- ++(30:3);
    \draw[r->] (0,2) -- ++(-30:3);
    \draw[r->] (-30:1) -- ++(0,3);
  \end{tikzpicture}
  \ = \
  \begin{tikzpicture}[scale=0.5, baseline=(x.base)]
    \node (x) at (30:1) {\vphantom{x}};

    \draw[r->] (0,0) -- ++(30:3);
    \draw[r->] (0,1) -- ++(-30:3);
    \draw[r->] (-30:2) -- ++(0,3);
  \end{tikzpicture}
\end{equation}
(or more generally, transfer matrices must commute).  Imagine that
there are extra dimensions hidden in this picture and the lines sit at
different points there.  Then, the topological invariance on~$\Sigma$
implies the Yang--Baxter equation, since it allows us to move any one
of the three lines past the intersection of the other two without
possibly causing a phase transition.  Moreover, the lines naturally
come equipped with continuous parameters, as their locations can vary
in the extra dimensions.

Thus, a correlation function of line operators in a 2d TQFT with extra
dimensions is equal to the partition function of an integrable lattice
model.  The model is defined on the lattice formed by the line
operators, whose coordinates in the extra dimensions provide the
spectral parameters.

We have to ask whether the 2d TQFT arising from brane tiling models
has desired hidden extra dimensions.  It actually has one: the 11th
dimension that emerges when the brane system is embedded into M-theory
via string dualities.  Consequently, the index of a brane tiling model
is the partition function of an integrable lattice model.

The main theme of this paper is to incorporate surface defects into
the above story of connections between brane tilings, TQFTs and
integrable lattice models.  We address this question in
section~\ref{sec:SD}.

In order to create a surface defect in our 4d theory, we add to the
brane system a D3-brane supported on a plane in $\R^{3,1}$ and ending
on the D5-branes along a curve in~$\Sigma$.  The total brane
configuration preserves $\CN = (0,2)$ supersymmetry on the plane.
Inside the 4d theory, the D3-brane appears as a half-BPS surface
defect.  This is the most basic example of a class of half-BPS surface
defects, all of which admit a similar, if slightly more elaborate,
brane construction.  Surface defects in this class are specified by a
representation of $\SU(N)$.  The one just described corresponds to the
fundamental representation.

Since the D3-brane ends on the D5-branes along a curve in $\Sigma$, it
creates a line operator in the 2d TQFT.  In the lattice model, the
introduction of the surface defect is therefore translated to
insertion of an extra line, which we represent by a dashed line:
\begin{equation}
  \begin{tikzpicture}[scale=0.5, baseline=(x.base)]
    \node (x) at (0,0.5) {\vphantom{x}};

    \begin{scope}[shift={(0,0)}]
      \draw[r->, right hook->] (-0.1,0) -- (1.1,0);
      \draw[r->, >=left hook] (1,0) -- (4.1,0);
      \node[fill=white, inner sep=1pt] at (2.5,0) {$\,\dots$};
    \end{scope}

    \begin{scope}[shift={(0,1)}]
      \draw[r->, right hook->] (-0.1,0) -- (1.1,0);
      \draw[r->, >=left hook] (1,0) -- (4.1,0);
      \node[fill=white, inner sep=1pt] at (2.5,0) {$\,\dots$};
    \end{scope}

    \begin{scope}[shift={(0.5,-0.5)}]
      \draw[r->] (0,0) -- (0,2.2);
      \draw[r->] (1,0) -- (1,2.2);
      \draw[r->] (3,0) -- (3,2.2);
    \end{scope}
  \end{tikzpicture}
  \
  \longrightarrow
  \
  \begin{tikzpicture}[scale=0.5, baseline=(x.base)]
    \node (x) at (0,1) {\vphantom{x}};

    \begin{scope}[shift={(0,0)}]
      \draw[r->, right hook->] (-0.1,0) -- (1.1,0);
      \draw[r->, >=left hook] (1,0) -- (4.1,0);
      \node[fill=white, inner sep=1pt] at (2.5,0) {$\,\dots$};
    \end{scope}

    \begin{scope}[shift={(0,1)}]
      \draw[dz->, right hook->] (-0.1,0) -- (1.1,0);
      \draw[dz->, >=left hook] (1,0) -- (4.1,0);
      \node[fill=white, inner sep=1pt] at (2.5,0) {$\,\dots$};
    \end{scope}

    \begin{scope}[shift={(0,2)}]
      \draw[r->, right hook->] (-0.1,0) -- (1.1,0);
      \draw[r->, >=left hook] (1,0) -- (4.1,0);
      \node[fill=white, inner sep=1pt] at (2.5,0) {$\,\dots$};
    \end{scope}

    \begin{scope}[shift={(0.5,-0.5)}]
      \draw[r->] (0,0) -- (0,3.2);
      \draw[r->] (1,0) -- (1,3.2);
      \draw[r->] (3,0) -- (3,3.2);
    \end{scope}
  \end{tikzpicture}
  \ .
\end{equation}
This picture tells us that a surface defect acts on the Hilbert space
of the lattice model as the transfer matrix~\eqref{eq:TM-L}.

With lines of a different type in hand, we can write down different
versions of the Yang--Baxter equation~\eqref{eq:YBE}.  In particular,
we have the relation
\begin{equation}
  \begin{tikzpicture}[scale=0.5, baseline=(x.base)]
    \node (x) at (30:2) {\vphantom{x}};

    \draw[z->] (0,0) -- ++(30:3);
    \draw[dz->] (0,2) -- ++(-30:3);
    \draw[r->] (-30:1) -- ++(0,3);
  \end{tikzpicture}
  \ = \
  \begin{tikzpicture}[scale=0.5, baseline=(x.base)]
    \node (x) at (30:1) {\vphantom{x}};

    \draw[z->] (0,0) -- ++(30:3);
    \draw[dz->] (0,1) -- ++(-30:3);
    \draw[r->] (-30:2) -- ++(0,3);
  \end{tikzpicture}
  \ ,
\end{equation}
which involves both the L-operator~\eqref{eq:L} and the
R-operator~\eqref{eq:R}.  It is called an RLL relation.

It turns out that when the dashed line is labeled with the fundamental
representation of $\SU(2)$, the above RLL relation was studied by
Derkachov and Spiridonov in~\cite{Derkachov:2012iv}.  According to
their work, an L-operator that solves the RLL relation is essentially
Sklyanin's L-operator~\cite{Sklyanin:1983ig}, a $2 \times 2$ matrix
whose entries are difference operators acting on meromorphic
functions.  This observation allows us to infer that in this simplest
case, the L-operator~\eqref{eq:L} for our theory is Sklyanin's
L-operator.  If this is true, one consequence is that the integrable
model realized on a lattice consisting solely of dashed lines should
be the eight-vertex model~\cite{Baxter:1971cr, Baxter:1972hz}.

As the Yang--Baxter equations do not determine the L-operator
uniquely, it is important to check our proposal by comparing it with
independent computations from the gauge theory side.  Fortunately,
such checks can be performed, as we do in sections~\ref{sec:class-S}
and~\ref{sec:class-Sk}.

The brane tiling models we mainly consider in this paper are also
examples of $\CN = 1$ theories of class $\CS_k$~\cite{Gaiotto:2015usa,
  Franco:2015jna, Hanany:2015pfa, Coman:2015bqq}, which are
generalizations of class-$\CS$ theories.  As such, their indices in
the presence of surface defects can be computed by the method
developed in~\cite{Gaiotto:2012xa, Gaiotto:2015usa}.

Briefly, the procedure goes as follows.  A class-$\CS_k$ theory of
type $A_{N-1}$ arises from $N$ M5-branes placed on a $\C^2/\Z_k$
orbifold singularity and further compactified on a punctured surface.
To this surface we introduce an extra puncture carrying a $\U(1)$
flavor symmetry (known as a ``minimal'' puncture).  The addition of
the puncture modifies the 4d theory.  The index of the new theory has
a series of poles in the fugacity parameter associated with the flavor
symmetry of the puncture.  These poles are labeled with a pair of
nonnegative integers, and the residue at the pole $(r,s)$ encodes the
index of the original theory in the presence of two surface defects,
supported on different tori inside $S^3 \times S^1$ and labeled with the
$r$th and $s$th symmetric representations, respectively.

String dualities map the class-$\CS_k$ setup to the brane tiling
setup.  Under this map, the addition of a minimal puncture corresponds
to the introduction of an NS5-brane.  In turn, the latter operation
inserts a lattice line in the lattice model.  Taking the residue
converts this line to a dashed one representing the surface defect.
Incidentally, the integrability of the lattice model is nothing but
the statement that the index is invariant under interchange of the
positions of minimal punctures.

We carry out the residue computation for $N = 2$ and $(r,s) = (0,1)$,
first for $k = 1$ (i.e., for class-$\CS$ theories) in
section~\ref{sec:class-S}, then for general $k$ in
section~\ref{sec:class-Sk}.  In each case, the index in the presence
of the surface defect is obtained by letting a difference operator act
on the bare index, in the absence of the surface defect.  We find that
this operator precisely matches the corresponding transfer matrix
calculated based on our proposal.

A reader familiar with the class-$\CS_k$ story may ask the following
question: The residue method described above produces $2k$ distinct
difference operators for a given pair $(r,s)$~\cite{Gaiotto:2015usa}.
How can they all be accommodated in a single transfer matrix?  The
answer is that although there is only one transfer matrix, it carries
a continuous parameter, namely the spectral parameter of the dashed
line.  The $2k$ difference operators are unified into this
one-parameter family as $2k$ values of the spectral parameter.

While we focus on a particular class of surface defects in this paper,
4d $\CN = 1$ theories have many other aspects that should be equally
illuminated by the integrability structure.  In this sense, the
present work may be regarded as a first step in the broader program of
studying 4d $\CN = 1$ theories through integrability.  In
section~\ref{sec:con} we suggest a couple of possible directions to be
taken for next steps.  Clearly, though, they are only a tiny fraction
of the long list of interesting topics for future research in this
ambitious program.

\section{Brane tilings and integrable lattice models}
\label{sec:BT}

In this section we review the correspondence between 4d $\CN = 1$
theories realized by brane tilings and integrable 2d lattice models.
A central role is played by a 2d TQFT equipped with line operators
that are localized in a hidden extra dimension emerging from M-theory.
Our presentation follows~\cite{Yagi:2015lha}, to which we refer the
reader for more details.

\subsection{Quiver gauge theories and their supersymmetric indices}

Throughout our discussion we will encounter gauge and flavor groups
that are either $\U(1)$ or $\SU(N)$.%
\footnote{In this paper the term ``flavor symmetry'' refers to any
  global symmetry that commutes with supersymmetry and is not 
  a spacetime symmetry.  It does not necessarily mix matter fields of
  different flavors.}
To each such group, we assign fugacities parameterizing the maximal
torus.  For instance, an element in the maximal torus of an $\SU(N)$
group takes the form $\diag(z_1, \dotsc, z_N)$, hence
$z = \{z_1, \dotsc, z_N\}$ is a set of fugacities for this group,
obeying the constraint $\prod_{I=1}^N z_I = 1$.  Fugacities are also
used to label the groups themselves; thus $\SU(N)_z$ is an $\SU(N)$
gauge or flavor group whose associated set of fugacities is $z$.  The
quiver diagrams we will deal with involve $\SU(N)$ gauge and flavor
nodes.  Since $N$ is fixed in each quiver, we label the nodes with the
fugacities of the corresponding $\SU(N)$ groups, rather than the rank:
$\gnode{z}$ is a gauge group $\SU(N)_z$, while $\fnode{z}$ is a flavor
group $\SU(N)_z$.

Building blocks of 4d $\CN = 1$ supersymmetric quiver gauge theories
are vector multiplets and bifundamental chiral multiplets.  A vector
multiplet is present at a gauge node.  A bifundamental chiral
multiplet has two flavor groups, say $\SU(N)_z$ and $\SU(N)_w$, and
transforms in the fundamental representation under $\SU(N)_z$ and in
the antifundamental representation under $\SU(N)_w$.  We represent it
by an arrow going from $\gnode{w}$ to $\gnode{z}$ (if $\SU(N)_z$ and
$\SU(N)_w$ are gauged).  In general, a bifundamental chiral multiplet
is also charged under an R-symmetry group $\U(1)_R$, which we assume
to exist and be anomaly-free, and under additional flavor groups
$\U(1)_{u_\alpha}$.  When we need to indicate the charges under these
$\U(1)$ symmetries, we mark the arrow with
\begin{equation}
  (pq)^{R/2} \prod_\alpha u_\alpha^{F_\alpha}
  \,,
\end{equation}
where $R$ is the R-charge (and not the fugacity for $\U(1)_R$),
$F_\alpha$ is the charge of $\U(1)_{u_\alpha}$, and $p$, $q$ are
complex parameters to be introduced momentarily.  We refer to this
value as the fugacity of the multiplet.

Given a theory $\CT$ with flavor group $\SU(N)_w$ and another theory
$\CT'$ with flavor group $\SU(N)_{w'}$, we can couple them to obtain a
new theory $(\CT \times \CT')/\SU(N)_z$ by gauging the diagonal
subgroup $\SU(N)_z$ of $\SU(N)_w \times \SU(N)_{w'}$.  To construct a
quiver gauge theory, we take a number of bifundamental chiral
multiplets and couple them by gauging all or part of the flavor nodes.

In what follows, we will mainly study the supersymmetric indices of
$\CN = 1$ quiver gauge theories formulated on the Euclidean spacetime
$S^3 \times S^1$~\cite{Romelsberger:2005eg, Kinney:2005ej,
  Festuccia:2011ws}.  The index of an $\CN = 1$ theory is defined by
the trace
\begin{equation}
  \label{eq:I}
  \CI(p,q, \{u_f\})
  =
  \Tr_{\CH_{S^3}}\biggl((-1)^F p^{j_1 + j_2 + R/2} q^{j_1 - j_2 + R/2}
  \prod_f u_f^{F_f}\biggr)
  \,,
\end{equation}
taken over the space $\CH_{S^3}$ of states on $S^3$.  Here $(-1)^F$ is
the fermion parity, and $j_1$, $j_2$ are generators of the maximal
torus $\U(1)_1 \times \U(1)_2$ of (the double cover of) the isometry
group $\Spin(4) \iso \SU(2)_1 \times \SU(2)_2$ of $S^3$.  If $S^3$ is
described by the equation $|\zeta_1|^2 + |\zeta_2|^2 = 1$ with
$(\zeta_1, \zeta_2) \in \C^2$, then $\U(1)_p$ and $\U(1)_q$, generated
by $j_p = j_1 + j_2$ and $j_q = j_1 - j_2$, rotate $\zeta_1$ and
$\zeta_2$ by phase, respectively.  Also, the index $f$ runs over all
$\U(1)$ flavor symmetries with charges $F_f$, including the maximal
tori of nonabelian flavor groups.

Thanks to supersymmetry, the index~\eqref{eq:I} receives contributions
only from those states whose energies belong to a certain discrete
spectrum determined by the R-charge assignment.  As a result, it
remains invariant under continuous changes of the parameters of the
theory.  This protected nature of the index will be important for our
argument.

The index $\CI_\CT$ of a 4d $\CN = 1$ theory $\CT$ with flavor group
$\SU(N)_z$ is a symmetric meromorphic function of the fugacities
$z_1$, $\dotsc$, $z_N$.  The symmetricity property reflects the gauge
invariance of the index.  At the level of the index, gauging of a
flavor group is realized by introduction of the corresponding vector
multiplet and integration over its fugacities.  In particular,
\begin{equation}
  \label{eq:gauging-I}
  \CI_{(\CT \times \CT')/\SU(N)_z}
  =
  \int_{\T^{N-1}} \prod_{I=1}^{N-1} \frac{\rmd z_I}{2\pi i z_I}
  \IV(z) \CI_{\CT}(z) \CI_{\CT'}(z)
  \,,
\end{equation}
with the integration performed over the unit circle $\T$ for each
variable $z_I$.  The index $\IV(z)$ of the vector multiplet is given
by elliptic gamma functions:%
\footnote{As in this equation, we will often use a quiver to mean the
  index of the corresponding theory.  It should be clear from the
  context whether a given quiver represents a theory or its index.}
\begin{equation}
  \gnode{z}
  =
  \IV(z; p,q)
  =
  \frac{(p;p)_\infty^{N-1} (q;q)_\infty^{N-1}}{N!}
  \prod_{\substack{I, J = 1 \\ I \neq J}}^N \frac{1}{\Gamma(z_I/z_J;
    p, q)}
  \,.
\end{equation}
See the appendix for the definition of the elliptic gamma function and
various identities it satisfies.  From now on we fix $p$, $q$ and
omit them from the notation unless needed.

The index of a bifundamental chiral multiplet with fugacity $a$ is
given by
\begin{equation}
  \label{eq:I_B}
  \begin{tikzpicture}[baseline=(z.base)]
    \node[fnode] (z) at (0,0) {$z$};
    \node[fnode] (w) at (1,0) {$w$};
    \draw[q->] (z) -- node[above] {$a$} (w);
  \end{tikzpicture}
  \ =
  \IB(z, w; a)
  =
  \prod_{I,J=1}^N
  \Gamma\biggl(a\frac{w_I}{z_J}\biggr)
  \,.
\end{equation}
This function satisfies
\begin{equation}
  \IB(z, w; a) \IB\Bigl(w, z; \frac{pq}{a}\Bigr) 
  =
  1
  \,.
\end{equation}
This identity says that as far as the index is concerned, we can
cancel a pair of arrows making a loop if their R-charges add up to $2$
and flavor charges add up to $0$:
\begin{equation}
  \label{eq:II=0}
  \begin{tikzpicture}[baseline=(z.base)]
    \node[fnode] (z) at (0,0) {$z$};
    \node[fnode] (w) at (1,0) {$w$};
    \draw[q->] (z) to[bend left] node[above] {$a$} (w);
    \draw[q->] (w) to[bend left] node[below] {$pq/a$} (z);
  \end{tikzpicture}
  \ = \
  \begin{tikzpicture}[baseline=(z.base)]
    \node[fnode] (z) at (0,0) {$z$};
    \node[fnode] (w) at (1,0) {$w$};
  \end{tikzpicture}
  \ .
\end{equation}
Physically, the reason is that we can turn on a mass term for such a
pair.  The index is invariant under this deformation, and the
bifundamental chiral multiplets decouple from the theory if we send
the mass to infinity, leaving a trivial contribution to the index.  We
will make use of this identity frequently.

Another useful fact is that if we define the ``delta function''
\begin{equation}
  \begin{tikzpicture}[baseline=(z.base)]
    \node[fnode] (z) at (0,0) {$z$};
    \node[fnode] (w) at (1,0) {$w$};
    \draw[eq-] (z) -- (w);
  \end{tikzpicture}
\end{equation}
by the relation
\begin{equation}
  \label{eq:z=w-def}
  \CI_{\CT}(w)
  =
  \int_{\T^{N-1}} \prod_{I=1}^{N-1} \frac{\rmd z_I}{2\pi i z_I}
  \IV(z)  \CI_{\CT}(z)
  \
  \begin{tikzpicture}[baseline=(z.base)]
    \node[fnode] (z) at (0,0) {$z$};
    \node[fnode] (w) at (1,0) {$w$};
    \draw[eq-] (z) -- (w);
  \end{tikzpicture}
  \ ,
\end{equation}
then we have
\begin{equation}
  \label{eq:I-inv}
  \begin{split}
  \begin{tikzpicture}[baseline=(z.base)]
    \node[fnode] (z) at (0,0) {$z$};
    \node[gnode] (x) at (1,0) {$x$};
    \node[fnode] (w) at (2,0) {$w$};
    \draw[q->] (z) -- node[above] {$a$} (x);
    \draw[q->] (x) -- node[above] {$a^{-1}$} (w);
  \end{tikzpicture}
  \ &=
  \int_{\T^{N-1}} \prod_{I=1}^{N-1} \frac{\rmd x_I}{2\pi i x_I}
  \IV(x) \IB(z,x; a) \IB\bigl(x,w; a^{-1}\bigr)
  \\
  &=
  \Gamma\bigl(a^{\pm N}\bigr)
  \
  \begin{tikzpicture}[baseline=(z.base)]
    \node[fnode] (z) at (0,0) {$z$};
    \node[fnode] (w) at (1,0) {$w$};
    \draw[eq-] (z) -- (w);
  \end{tikzpicture}
  \ .
  \end{split}
\end{equation}
This is a consequence of confinement and chiral symmetry
breaking~\cite{Seiberg:1994bz, Spiridonov:2014cxa}.  At low energies
the theory on the left-hand side is described by the mesons and the
baryons.  It has a vacuum in which the mesons take nonzero expectation
values and the flavor symmetry $\SU(N)_w \times \SU(N)_z$ is broken to
the diagonal subgroup.  In this vacuum the fugacities $w$ and $z$ are
identified, so we get the quiver on the second line.  The factor
$\Gamma(a^{\pm N}) = \Gamma(a^N) \Gamma(a^{-N})$ is the contribution
from the baryons.

We can readily write down the formula for the index of a general
quiver gauge theory.  For simplicity, suppose that the theory is
described by a quiver that contains no flavor node.  Then, the index
is computed by
\begin{equation}
  \label{eq:I-quiver}
  \prod_{\gnode{z}}
  \int_{\T^{N-1}}
  \prod_{I=1}^{N-1} \frac{\rmd z_I}{2\pi i z_I}
  \IV(z)
  \prod_{
    \tikz{
      \node[gnode, font=\scriptsize, minimum size=8pt] (x) at (0,0) {$x$};
      \node[gnode, font=\scriptsize, minimum size=8pt] (y) at (1.8em,0) {$y$};
      \draw[q->] (x) -- (y);        
    }}
  \IB(x, y)
  \,,
\end{equation}
where the two products are taken over all nodes and all arrows,
respectively.  The index is a function of the parameters $p$, $q$ and
the flavor fugacities $u_\alpha$, which are suppressed in the above
expression.  If the quiver contains flavor nodes, the index is also a
function of their fugacities.

\subsection{Supersymmetric index and integrable lattice models}

The supersymmetric index \eqref{eq:I-quiver} of a quiver gauge theory
may be interpreted as the partition function of a statistical
mechanics model with continuous spins.  Indeed, this formula precisely
computes the partition function of a spin model in which spins are
placed at the gauge nodes.  The spin variables at $\gnode{z}$ are the
fugacities $z_1$, $\dotsc$, $z_N$, and they interact among themselves
as well as with spins at nearest-neighbor nodes, namely those
connected by arrows.  The Boltzmann weights for the self-interaction
and the nearest-neighbor interaction are $\IV$ and $\IB$,
respectively.

This is not particularly surprising in view of the fact that the index
is a protected quantity and can be computed in the free theory limit.
In this limit, vector and bifundamental chiral multiplets decouple, so
their contributions factorize.  What is remarkable is that for a
certain class of $\CN = 1$ theories, the index is equal to the
partition function of an \emph{integrable} model defined on a 2d
lattice.

The connection between the supersymmetric index and the lattice model
comes from higher dimensions.  Consider a 6d supersymmetric theory
$\CT_{\text{6d}}$ equipped with codimension-$1$ defects.  (Here we
have in mind the 6d theory living on a stack of D5-branes, though our
argument is more general.)  We compactify this theory on a
two-manifold $\Sigma$ and place codimension-$1$ defects $\CW_i$ along
various curves $C_i$ in $\Sigma$.  Suppose that this kind of
configuration preserves four supercharges for any choice of $\Sigma$
and $\{C_i\}$.  Then, at low energies, the system is described
effectively by a 4d $\CN = 1$ theory
$\CT_{\text{4d}}[\Sigma, \{C_i\}; \{\CW_i\}]$.  We can place it on
$S^3 \times S^1$ and perform the path integral to compute its index.
The index is invariant under continuous changes of the parameters of
the theory, and the geometric data of the curves $C_i$ are such
parameters.  As a consequence, this procedure defines a map from the
set of \emph{topological} configurations of curves to the set of
supersymmetric indices, given a choice of codimension-$1$ defects.

Now start again from the same 6d theory, placed on the spacetime
$S^3 \times S^1 \times \Sigma$ with the same configuration of defects.
In the previous paragraph it was implicitly assumed that the size of
$\Sigma$ is much smaller than $S^3$ and $S^1$ so that the description
by the 4d theory is sensible.  This time, let us make $\Sigma$ much
larger than $S^3 \times S^1$; the index remains invariant under the
rescaling of the metric of $\Sigma$.  In this case, the low-energy
physics is described instead by a 2d theory
$\CT_{\text{2d}}[S^3 \times S^1]$ on $\Sigma$, and the codimension-$1$
defects $\CW_i$ inserted on $S^3 \times S^1 \times C_i$ become line
operators $\CL_i$ supported along $C_i$ in this theory.  This
consideration leads to the relation
\begin{equation}
  \label{eq:I=L}
  \CI_{\CT_{\text{4d}}[\Sigma, \{C_i\}; \{\CW_i\}]}
  =
  \Bigvev{\prod_i \CL_i(C_i)}_{\CT_{\text{2d}}[S^3 \times S^1],
    \Sigma}
  \,.
\end{equation}
Moreover, the right-hand side depends only on the topology of the
configuration of the line operators, as we just explained.  Thus,
$\CT_{\text{2d}}[S^3 \times S^1]$ is a 2d TQFT.

We can compute the above correlation function by dividing $\Sigma$ into
square pieces, each containing segments of two line operators crossing
in the middle:%
\footnote{This decomposition is always possible by inserting
  ``identity'' or ``invisible'' line operators if necessary, which is
  the same as doing nothing at all.}
\begin{equation}
  \label{eq:RM-PI}
  \begin{tikzpicture}[scale=0.5, baseline=(x.base)]
    \node (x) at (0,1) {\vphantom{x}};

    \draw (0,0) rectangle (2,2);
    \draw[r->] (0,1) node[left] {$i$} -- (2,1);
    \draw[r->] (1,0) node[below] {$j$} -- (1,2);
  \end{tikzpicture}
  \ .
\end{equation}
We use arrows to represent line operators together with their
orientation, and label them with numbers.  This picture is
topologically equivalent to the following one:
\begin{equation}
  \label{open-string-pic}
  \begin{tikzpicture}[scale=0.5, baseline=(x.base)]
    \node (x) at (0,1) {\vphantom{x}};

    \draw (0,0) -- (1,0);
    \draw (1.5,0) -- (2.5,0);
    \draw (0,2) -- (1,2);
    \draw (1.5,2) -- (2.5,2);
    
    \draw (1,0) .. controls (1.25, 0.75) .. (1.5,0);
    \draw (1,2) .. controls (1.25, 1.25) .. (1.5,2);
    \draw (0,0) .. controls (0.5, 1) .. (0,2);
    \draw (2.5,0) .. controls (2, 1) .. (2.5,2);

    \draw[r->] (0.5,0) node[below] {$i$}
    .. controls (0.5,1) and (2,1) .. (2, 2);
    \draw[r->] (2,0) node[below] {$j$}
    .. controls (2,1) and (0.5,1) .. (0.5, 2);
  \end{tikzpicture}
  \ .
\end{equation}
Intuitively, we can view it as the worldsheet of a scattering process
involving two ``open strings,'' each of which carries a ``particle''
whose worldline is either of the line operators.  Let $\V_i$ be the
space of states on an interval intersected by $\CL_i$; this is the
Hilbert space for the ``open string'' with the $i$th ``particle''
attached.  Then, the path integral on the above piece produces the
S-matrix%
\footnote{Note that the two factors in the tensor product are swapped
  in the target space.  The h\'a\v cek~$\check{}$ is used to stress
  this fact.}
\begin{equation}
  \RM_{ij}\colon \V_i \otimes \V_j \to \V_j \otimes \V_i
  \,.
\end{equation}
To reconstruct the correlation function for the original configuration
of line operators on~$\Sigma$, we simply glue these pieces back
together.

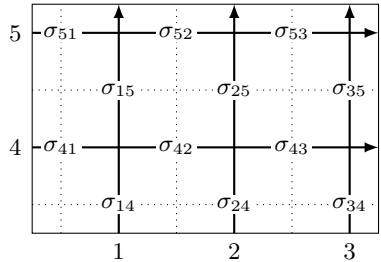
\begin{figure}
  \centering
  \begin{tikzpicture}[scale=1.2]
    \draw (0,0) rectangle (3,2);
      
    \begin{scope}[shift={(0.75,0)}]
      \draw[r->] (0,0) node[below] {$1$} -- (0,2);
      \draw[r->] (1,0) node[below] {$2$} -- (1,2);
      \draw[r->] (2,0) node[below] {$3$} -- (2,2);
    \end{scope}
    
    \begin{scope}[shift={(0,0.75)}]
      \draw[r->] (0,0) node[left] {$4$} -- (3,0);
      \draw[r->] (0,1) node[left] {$5$}  -- (3,1);
    \end{scope}

    \begin{scope}[shift={(0,0.25)}]
      \draw[dotted] (0,0) -- (3,0);
      \draw[dotted] (0,1) -- (3,1);
    \end{scope}
    
    \begin{scope}[shift={(0.25,0)}]
      \draw[dotted] (0,0) -- (0,2);
      \draw[dotted] (1,0) -- (1,2);
      \draw[dotted] (2,0) -- (2,2);
    \end{scope}

    \begin{scope}[shift={(0.25,0.75)}]
      \node[fill=white, inner sep=1pt] at (0,0) {$\sigma_{41}$};
      \node[fill=white, inner sep=1pt] at (1,0) {$\sigma_{42}$};
      \node[fill=white, inner sep=1pt] at (2,0) {$\sigma_{43}$};
      \node[fill=white, inner sep=1pt] at (0,1) {$\sigma_{51}$};
      \node[fill=white, inner sep=1pt] at (1,1) {$\sigma_{52}$};
      \node[fill=white, inner sep=1pt] at (2,1) {$\sigma_{53}$};
    \end{scope}
  
    \begin{scope}[shift={(0.75,0.25)}]
      \node[fill=white, inner sep=1pt] at (0,0) {$\sigma_{14}$};
      \node[fill=white, inner sep=1pt] at (0,1) {$\sigma_{15}$};
      \node[fill=white, inner sep=1pt] at (1,0) {$\sigma_{24}$};
      \node[fill=white, inner sep=1pt] at (1,1) {$\sigma_{25}$};
      \node[fill=white, inner sep=1pt] at (2,0) {$\sigma_{34}$};
      \node[fill=white, inner sep=1pt] at (2,1) {$\sigma_{35}$};
    \end{scope}
  \end{tikzpicture}
  
  \caption{Decomposition of a $2 \times 3$ lattice formed by line
    operators on a torus.  Spin variables $\sigma_{ij}$ are placed on
    the sides of the squares, or the edges of the lattice.}
  \label{fig:2x3}
\end{figure}

We will mainly study the case where $\Sigma$ is either a cylinder or
torus and line operators form a square lattice.  To be specific, let
us take $\Sigma = T^2$ and wrap line operators around $1$-cycles
$C_i$, $i = 1$, $\dotsc$, $m+n$ making up an $m \times n$ lattice.  We
divide the torus into $m \times n$ square pieces as above, and to each
side of these squares assign a variable $\sigma_{ij}$ that labels
basis vectors for the state space on that side.  The situation for
$(m,n) = (2,3)$ is illustrated in Fig.~\ref{fig:2x3}.  For the
computation of the correlation function, first we take the product of
the matrix elements of $\RM$ from all squares for each configuration
of state variables, then sum over all configurations:
\begin{equation}
  \label{Z}
  \biggvev{\prod_{i=1}^{m+n} \CL_i(C_i)}_{\CT_{\text{2d}}[S^3 \times S^1], T^2}
  =
  \sum_{\{\sigma_{ij}\}}
  \prod_{l=1}^n \prod_{k=n+1}^{n+m}
  (\RM_{kl})_{\sigma_{kl}, \sigma_{lk}}^{\sigma_{l,k+1},
    \sigma_{k,l+1}}
  \,,
\end{equation}
where $\sigma_{k,n+1} = \sigma_{k1}$ and
$\sigma_{l,n+m+1} = \sigma_{l,n+1}$.

Let us shift our perspective slightly and look at the operator $\RM$
as assigned to the vertices of the lattice, not the square pieces
containing them.  Also, we think of the state variables $\sigma_{ij}$
as living on the edges of the lattice, not the sides of the squares.
If we view the system in this way, the above formula is precisely the
partition function of a \emph{vertex model} in statistical mechanics:
spins $\sigma_{ij}$ live on the edges of a lattice, and interact at
the vertices with the Boltzmann weight $\RM$.  In the context of
vertex models, the S-matrix $\RM$ is known as the \emph{R-matrix} or
\emph{R-operator}.

Thus, we find that the correlation function in question is equal to
the partition function of a vertex model
$\CV[S^3 \times S^1; \{\CL_i\}]$ defined on the lattice formed by the
curves $\{C_i\}$:
\begin{equation}
  \label{CFLO}
  \biggvev{\prod_i \CL_i(C_i)}_{\CT_{\text{2d}}[S^3 \times S^1], \Sigma}
  =
  Z_{\CV[S^3 \times S^1; \{\CL_i\}], \{C_i\}}
  \,.
\end{equation}
Combined with the equality~\eqref{eq:I=L}, this relation implies that
the index of the 4d theory is equal to the partition function of the
lattice model.

So far we have discussed the connection between the supersymmetric
index and vertex models.  We now explain how integrability comes into
the picture.

To talk about the integrability of a vertex model, we should consider
the situation that each lattice line carries a continuous parameter,
called the \emph{spectral parameter} assigned to that line.
Correspondingly, the R-operator depends on two spectral parameters in
general:
\begin{equation}
  \RM_{ij}(r_i, r_j)
  =
  \begin{tikzpicture}[scale=0.5, baseline=(x.base)]
    \node (x) at (0,1) {\vphantom{x}};

    \draw[r->] (0,1) node[left] {$r_i$} -- (2,1);
    \draw[r->] (1,0) node[below] {$r_j$} -- (1,2);
  \end{tikzpicture}
  \ .
\end{equation}
By concatenating the R-operators horizontally, we obtain the
\emph{transfer matrix}
\begin{equation}
  \label{eq:TM}
  T(r; r_1, \dotsc, r_n) 
  =
  \begin{tikzpicture}[scale=0.5, baseline=(x.base)]
    \node (x) at (0,0) {\vphantom{x}};

    \draw[r->, right hook->] (-0.1,0)
    node[left] {$r$} -- (1.1,0);
    \draw[r->, >=left hook] (1,0) -- (4.1,0);
    \node[fill=white, inner sep=1pt] at (2.5,0) {$\,\dots$};

    \begin{scope}[shift={(0,-0.5)}]
      \draw[r->] (0.5,0)
      node[below] {$r_1$} -- (0.5,1.2);
      \draw[r->] (1.5,0)
      node[below] {$r_2$} -- (1.5,1.2);
      \draw[r->] (3.5,0)
      node[below] {$r_n$} -- (3.5,1.2);
    \end{scope}
  \end{tikzpicture}
  \ .
\end{equation}
More precisely,%
\footnote{
  In matrix elements,
  $
    T(r; r_1, \dotsc, r_n)_{\sigma_1 \dotso \sigma_n}
    ^{\sigma_1' \dotso \sigma_n'}
    =
    \RM_n(r, r_n)_{\sigma''_n \sigma_n}^{\sigma_n' \sigma''_1}
    \dotsm
    \RM_2(r, r_2)_{\sigma''_2 \sigma_2}^{\sigma_2' \sigma''_3}
    \RM_1(r, r_1)_{\sigma''_1 \sigma_1}^{\sigma_1' \sigma''_2}
    $.
}
\begin{equation}
  T(r; r_1, \dotsc, r_n)
  =
  \Tr_\V\bigl(\RM_n(r, r_n) \circ_\V \dotsb
  \circ_\V \RM_2(r, r_2) \circ_\V \RM_1(r, r_1)\bigr)
  \,,
\end{equation}
where $\V$ is the vector space assigned to the unnumbered horizontal
line and $\RM_i\colon \V \otimes \V_i \to \V_i \otimes \V$ is the
R-operator at the crossing of the horizontal and $i$th vertical lines.
The hooks on the ends of the horizontal line indicate the periodic
boundary condition, which leads to the trace in the above formula.
The transfer matrix is an endomorphism of $\bigotimes_{i=1}^n \V_i$
that maps a state just below the horizontal line to another state just
above it.

A vertex model is said to be \emph{integrable} if transfer matrices at
different values of the spectral parameter for the horizontal line
commute:
\begin{equation}
  \label{eq:[T,T]=0}
  [T(r; r_1, \dotsc, r_n), T(r'; r_1, \dotsc, r_n)] = 0
  \,.
\end{equation}
By expanding $T(r; r_1, \dotsc, r_n)$ in powers of $r$ (or of a
related parameter in which $T$ is analytic), we get a series of
endomorphisms of $\bigotimes_{i=1}^n \V_i$ as expansion coefficients.
This equation ensures that these infinitely many linear operators
mutually commute.

Therefore, for the vertex model constructed from line operators in a
2d TQFT to be integrable, the line operators must carry spectral
parameters, and transfer matrices must commute.  These two features
arise naturally if there are extra dimensions in the theory.

Suppose that our 2d TQFT is really a higher-dimensional theory
compactified on some manifold $M$, and the line operators $\CL_i$ in
the correlation function \eqref{CFLO} are placed at some points $p_i$
in $M$, which we assume for simplicity to be all different.  Suppose
also that the correlation function is topological on $\Sigma$, but
varies nontrivially along $M$.  Then, the lattice lines carry
continuous parameters, namely their locations in $M$.  Furthermore,
the commutativity of transfer matrices holds:
\begin{equation}
  \label{eq:[T,T]=0-G}
  \begin{tikzpicture}[scale=0.5, baseline=(x.base)]
    \node (x) at (0,0.5) {\vphantom{x}};

    \begin{scope}[shift={(0,0)}]
      \draw[r->, right hook->] (-0.1,0)
      node[left] {$p$} -- (1.1,0);
      \draw[r->, >=left hook] (1,0) -- (4.1,0);
      \node[fill=white, inner sep=1pt] at (2.5,0) {$\,\dots$};
    \end{scope}

    \begin{scope}[shift={(0,1)}]
      \draw[r->, right hook->] (-0.1,0)
      node[left] {$p'$} -- (1.1,0);
      \draw[r->, >=left hook] (1,0) -- (4.1,0);
      \node[fill=white, inner sep=1pt] at (2.5,0) {$\,\dots$};
    \end{scope}

    \begin{scope}[shift={(0.5,-0.5)}]
      \draw[r->] (0,0)
      node[below] {$p_1$} -- (0,2.2);
      \draw[r->] (1,0)
      node[below] {$p_2$} -- (1,2.2);
      \draw[r->] (3,0)
      node[below] {$p_n$} -- (3,2.2);
    \end{scope}
  \end{tikzpicture}
  \ = \
  \begin{tikzpicture}[scale=0.5, baseline=(x.base)]
    \node (x) at (0,0.5) {\vphantom{x}};

    \begin{scope}[shift={(0,0)}]
      \draw[r->, right hook->] (-0.1,0)
      node[left] {$p'$} -- (1.1,0);
      \draw[r->, >=left hook] (1,0) -- (4.1,0);
      \node[fill=white, inner sep=1pt] at (2.5,0) {$\,\dots$};
    \end{scope}

    \begin{scope}[shift={(0,1)}]
      \draw[r->, right hook->] (-0.1,0)
      node[left] {$p$} -- (1.1,0);
      \draw[r->, >=left hook] (1,0) -- (4.1,0);
      \node[fill=white, inner sep=1pt] at (2.5,0) {$\,\dots$};
    \end{scope}

    \begin{scope}[shift={(0.5,-0.5)}]
      \draw[r->] (0,0)
      node[below] {$p_1$} -- (0,2.2);
      \draw[r->] (1,0)
      node[below] {$p_2$} -- (1,2.2);
      \draw[r->] (3,0)
      node[below] {$p_n$} -- (3,2.2);
    \end{scope}
  \end{tikzpicture}
  \ .
\end{equation}
The TQFT structure itself is not strong enough to imply the
commutativity, simply because the two configurations of line operators
are topologically distinct; even though we can slide the horizontal
lines freely in a generic situation, a phase transition may occur when
the two lines meet and pass each other.  In the presence of extra
dimensions, such a singular situation is avoided as the lines do not
actually meet.

Even better, in this setup the R-operator satisfies the unitarity
relation
\begin{equation}
  \label{eq:RR=id}
  \RM_{ji}(p_j, p_i) \RM_{ij}(p_i, p_j)   = \id_{\V_i \otimes \V_j}
\end{equation}
and the Yang--Baxter equation%
\footnote{The is a relation between linear maps from
  $\V_1 \otimes \V_2 \otimes \V_3$ to
  $\V_3 \otimes \V_2 \otimes \V_1$.  Each $\RM_{ij}$ in the equation
  acts as the R-operator as described above on the factor
  $\V_i \otimes \V_j$ in the triple product
  $\V_i \otimes \V_j \otimes \V_k$ or
  $\V_k \otimes \V_i \otimes \V_j$, and trivially on the remaining
  space $\V_k$.}
\begin{equation}
  \label{eq:YBE-RRR}
  \RM_{12}(p_1, p_2) \RM_{13}(p_1, p_3) \RM_{23}(p_2, p_3)
  =
  \RM_{23}(p_2, p_3) \RM_{13}(p_1, p_3) \RM_{12}(p_1, p_2)
  \,.
\end{equation}
Graphically, the unitarity relation~\eqref{eq:RR=id} can be expressed as
\begin{equation}
  \label{eq:invert}
  \begin{tikzpicture}[baseline=(x.base)]
    \node (x) at (0,0.25) {\vphantom{x}};

    \draw[r->] (0,0) node[left] {$p_j$}
    to[out=0, in=180] (1,0.5) to[out=0, in=180] (2,0);

    \draw[r->] (0,0.5) node[left] {$p_i$}
    to[out=0, in=180] (1,0) to[out=0, in=180] (2,0.5);
  \end{tikzpicture}
  \ = \
  \begin{tikzpicture}[baseline=(x.base)]
    \node (x) at (0,0.25) {\vphantom{x}};

    \draw[r->] (0,0) node[left] {$p_j$} -- (2,0);
    \draw[r->] (0,0.5) node[left] {$p_i$} -- (2,0.5);
  \end{tikzpicture}
  \ ,
\end{equation}
while the Yang--Baxter equation~\eqref{eq:YBE-RRR} takes the form
\begin{equation}
  \label{eq:YBE-G}
  \begin{tikzpicture}[scale=0.5, baseline=(x.base)]
    \node (x) at (30:2) {\vphantom{x}};

    \draw[r->] (0,0) node[left] {$p_2$} -- ++(30:3);
    \draw[r->] (0,2) node[left] {$p_1$} -- ++(-30:3);
    \draw[r->] (-30:1) node[below] {$p_3$} -- ++(0,3);
  \end{tikzpicture}
  \ = \
  \begin{tikzpicture}[scale=0.5, baseline=(x.base)]
    \node (x) at (30:1) {\vphantom{x}};

    \draw[r->] (0,0) node[left] {$p_2$} -- ++(30:3);
    \draw[r->] (0,1) node[left] {$p_1$} -- ++(-30:3);
    \draw[r->] (-30:2) node[below] {$p_3$} -- ++(0,3);
  \end{tikzpicture}
  \ .
\end{equation}
Again, these relations hold in our theory since the line operators sit
at different points in $M$ and therefore no phase transition occurs.
The commutativity of transfer matrices \eqref{eq:[T,T]=0-G} follows
from the above two relations.  There is a nice diagrammatic proof of
this property:
\begin{equation}
  \begin{tikzpicture}[scale=0.5, baseline=(x.base)]
    \node (x) at (0,0.5) {\vphantom{x}};

    \draw[r->, right hook->] (-0.1,0) node[left] {$p$} -- (0,0)
    to[out=0, in=180] (1,1) -- (1.1,1);
    \draw[r->, >=left hook] (1,1) to[out=0, in=180] (2,0)
    -- (4.6,0);

    \draw[r->, right hook->] (-0.1,1) node[left] {$p'$} -- (0,1)
    to[out=0, in=180] (1,0) -- (1.1,0);
    \draw[r->, >=left hook] (1,0) to[out=0, in=180] (2,1)
    -- (4.6,1);

    \begin{scope}[shift={(2,-0.5)}, xscale=2/3]
      \draw[r->] (0,0) node[below] {$p_1$} -- (0,2.2);
      \draw[r->] (1,0) node[below] {$p_2$} -- (1,2.2);
      \draw[r->] (3,0) node[below] {$p_n$} -- (3,2.2);
      \node[fill=white, inner sep=1pt] at (2,0.5) {$\,\dots$};
      \node[fill=white, inner sep=1pt] at (2,1.5) {$\,\dots$};
    \end{scope}
  \end{tikzpicture}
  =
  \begin{tikzpicture}[scale=0.5, baseline=(x.base)]
    \node (x) at (0,0.5) {\vphantom{x}};

    \draw[r->, right hook->] (-0.1,0) node[left] {$p$} -- (0,0)
    to[out=0, in=180] (1,1) -- (3.6,1);
   \draw[r->, >=left hook] (3.5,1) to[out=0, in=180] (4.5,0) -- (4.6,0);

    \draw[r->, right hook->] (-0.1,1) node[left] {$p'$} -- (0,1)
    to[out=0, in=180] (1,0) -- (3.6,0);
   \draw[r->, >=left hook] (3.5,0) to[out=0, in=180] (4.5,1) -- (4.6,1);

    \begin{scope}[shift={(1,-0.5)}, xscale=2/3]
      \draw[r->] (0,0) node[below] {$p_1$} -- (0,2.2);
      \draw[r->] (1,0) node[below] {$p_2$} -- (1,2.2);
      \draw[r->] (3,0) node[below] {$p_n$} -- (3,2.2);
      \node[fill=white, inner sep=1pt] at (2,0.5) {$\,\dots$};
      \node[fill=white, inner sep=1pt] at (2,1.5) {$\,\dots$};
    \end{scope}
  \end{tikzpicture}
  =
  \begin{tikzpicture}[scale=0.5, baseline=(x.base)]
    \node (x) at (0,0.5) {\vphantom{x}};

    \draw[r->, right hook->] (-0.1,0) node[left] {$p'$} -- (2.5,0)
    to[out=0, in=180] (3.5,1) -- (3.6,1);
    \draw[r->, >=left hook] (3.5,1) to[out=0, in=180] (4.5,0) -- (4.6,0);

    \draw[r->, right hook->] (-0.1,1) node[left] {$p$} -- (2.5,1)
    to[out=0, in=180] (3.5,0) -- (3.6,0);
    \draw[r->, >=left hook] (3.5,0) to[out=0, in=180] (4.5,1) -- (4.6,1);

    \begin{scope}[shift={(0.5,-0.5)}, xscale=2/3]
      \draw[r->] (0,0) node[below] {$p_1$} -- (0,2.2);
      \draw[r->] (1,0) node[below] {$p_2$} -- (1,2.2);
      \draw[r->] (3,0) node[below] {$p_n$} -- (3,2.2);
      \node[fill=white, inner sep=1pt] at (2,0.5) {$\,\dots$};
      \node[fill=white, inner sep=1pt] at (2,1.5) {$\,\dots$};
    \end{scope}
  \end{tikzpicture}
  \ .
\end{equation}
In the second equality we used the cyclic property of trace; we
``rotated'' the cylinder until the crossing on the left comes to the
right.

Let us recapitulate the logic of our argument.  We consider a 4d $\CN
= 1$ theory that is constructed from a 6d theory by compactification
on a two-manifold $\Sigma$, in the presence of codimension-$1$ defects
supported on curves in $\Sigma$.  Due to its protected nature, the
supersymmetric index of the theory is captured by the correlation
function of a lattice of line operators in a 2d TQFT on $\Sigma$.  In
turn, by dividing $\Sigma$ into square pieces, the correlation
function can be mapped to the partition function of a vertex model
defined on the same lattice.  This vertex model is furthermore
integrable if the 2d TQFT has hidden extra dimensions along which the
correlation function varies nontrivially.

It is clear that the above argument applies to any protected
quantities, not just the supersymmetric index on $S^3 \times S^1$.
For example, we can use the index on $M \times S^1$ with the
$3$-manifold $M$ different from $S^3$.  For each protected quantity,
there is a corresponding TQFT and hence an integrable lattice model.
The case when $M$ is a lens space was investigated
in~\cite{Yamazaki:2013nra}.  In this paper we focus on the $S^3$ index
since this is a well-understood quantity and, accordingly, there are
nice mathematical results available.

\subsection{Brane tilings}
\label{sec:BT-BT}

Now we turn to a specific class of 4d $\CN = 1$ theories that have the
desired properties described above.  These theories are constructed
using branes in string theory.

Consider a stack of $N$ D5-branes extending along the $012346$
directions.  We introduce a number of NS5-branes intersecting these
D5-branes and occupying either the $012345$ or $012367$ directions:
\begin{equation}
  \label{eq:BB}
  \begin{tabular}{|l|cccccccccc|}
    \hline
    & 0 & 1 & 2 & 3 & 4 & 5 & 6 & 7 & 8 & 9
    \\ \hline
    D5 & $\times$ & $\times$ & $\times$ & $\times$ & $\times$ & & $\times$
    &&&
    \\
    NS5 & $\times$ & $\times$ & $\times$ & $\times$ & $\times$ & $\times$
    &&&&
    \\
    NS5 & $\times$ & $\times$ & $\times$ & $\times$ & & & $\times$ & $\times$
    &&
    \\ \hline
  \end{tabular}
\end{equation}
All branes are located at the same point on the $89$-plane.  On the
$46$-plane, an NS5-brane intersects the D5-branes along the $X^4$- or
$X^6$-direction.  Each of the three types of branes breaks half of the
$32$ supercharges.  Altogether, the system preserves $4$
supercharges.  They are acted on by the $\U(1)$ R-symmetry group
originating from the rotational symmetry on the $89$-plane.  If
NS5-branes of either type are absent, this brane setup is T-dual to
the Hanany--Witten brane configuration~\cite{Hanany:1996ie}.

Let us replace the $46$-plane with an arbitrary Riemann surface
$\Sigma$.  To preserve supersymmetry, we take the background spacetime
to be $\R^{3,1} \times T^*\Sigma \times \R^2$ and place the D5-branes
on $\R^{3,1} \times \Sigma \times \{0\}$, where $T^*\Sigma$ is the
$4567$-space and $\Sigma$ is embedded in it as the zero section.
NS5-branes intersect the D5-branes along curves $C_i \subset \Sigma$.
More precisely, they are placed on
$\R^{3,1} \times \Sigma_i \times \{0\}$, where $\Sigma_i$ are surfaces
in $T^*\Sigma$ such that they restrict to $C_i$ on $\Sigma$.  Provided
that $\Sigma_i$ are chosen appropriately, this system preserves $4$
supercharges.%
\footnote{In a neighborhood of the zero section
  $\Sigma \subset T^*\Sigma$, there exists a hyperk\"ahler structure
  compatible with the canonical holomorphic symplectic structure:
  there are complex structures $J_1$, $J_2$, $J_3$ satisfying the
  quaternion relations, with $J_3$ being the canonical complex
  structure of $T^*\Sigma$.  If we decompose the canonical holomorphic
  symplectic form $\Omega_3$ into the sum of real two-forms as
  $\Omega_3 = \omega_1 + i\omega_2$, then $\omega_1$ and $\omega_2$
  are the K\"ahler forms associated with the complex structures $J_1$
  and $J_2$, respectively.  By construction, $\Sigma$ is a complex
  Lagrangian submanifold, i.e., $\Omega_3$ vanishes on $\Sigma$.  It
  follows that $\Sigma$ is a special Lagrangian submanifold in the
  complex structure $J_2$ since $\omega_2$ and the imaginary part of
  the holomorphic symplectic form $\Omega_2 = \omega_3 + i\omega_1$
  associated with $J_2$ vanish on $\Sigma$.  As such, $\Sigma$ is a
  supersymmetric cycle.  Similarly, for any given analytic curve
  $C_i \subset \Sigma$, we can find a supersymmetric cycle
  $\Sigma_i \subset T^*\Sigma$ such that $\Sigma_i \cap \Sigma = C_i$.
  To see this, pick a complex structure~$J$ different from $J_3$, and
  take complex coordinates $(z, w)$ on $T^*\Sigma$ such that their
  real parts $(x, y)$ define coordinates on $\Sigma$.  If $C_i$ is
  given by $y = y(x)$, then its analytic continuation $z = z(w)$
  defines a complex Lagrangian submanifold in the complex
  structure~$J$.}

On the D5-branes lives 6d $\CN=(1,1)$ super-Yang--Mills theory with
gauge group $\SU(N)$.  The theory is placed on
$\R^{3,1} \times \Sigma$ and topologically twisted along $\Sigma$, as
can be seen by noting that two of its four scalar fields describing
fluctuations of the D5-branes are not really scalars, but rather
sections of $T^*\Sigma$.  Thanks to the twisting, $8$ of the $16$
supercharges are left unbroken by the curvature of $\Sigma$.  In this
6d theory the NS5-branes create half-BPS codimension-$1$ defects, or
domain walls, supported on $\R^{3,1} \times C_i$.  (The four
supercharges preserved by the two types of NS5-branes are compatible
with the twisting~\cite{Yagi:2015lha}.)  If $\Sigma$ is compact, the
6d theory is effectively described by a 4d $\CN = 1$ theory.  We call
4d theories constructed in this way \emph{brane box
  models}~\cite{Hanany:1997tb}.

We are now in the situation considered before: we have a 6d theory
that produces 4d $\CN = 1$ theories by compactification in the
presence of codimension-$1$ defects.  By following the same logic, we
conclude that the index of a brane box model is given by a correlation
function of line operators in a 2d TQFT, and coincides with the
partition function of a lattice model.

Furthermore, an extra dimension emerges if the brane system is
embedded into M-theory.  For the computation of the index, we take the
spacetime of the 4d theory to be $S^3 \times S^1$.  Thus, we are
considering type IIB string theory on
$S^3 \times S^1 \times T^*\Sigma \times \R^2$.  We can apply T-duality
along the $S^1$ and lift the resulting type IIA system to M-theory.
In this process, the D5-branes are transformed to M5-branes wrapping
the 11th dimension, the M-theory circle.  On the other hand, the
NS5-branes become M5-branes supported at points on the circle.  Hence,
the M-theory circle provides the extra dimension along which
NS5-branes, or line operators in the 2d TQFT, can avoid one another.
The existence of the extra dimension implies that the lattice model is
integrable.

In fact, theories we will consider in this paper are not really brane
box models.  Rather, we will study \emph{brane tiling
  models}~\cite{Hanany:2005ve, Franco:2005rj} whose brane construction
is slightly more complicated.

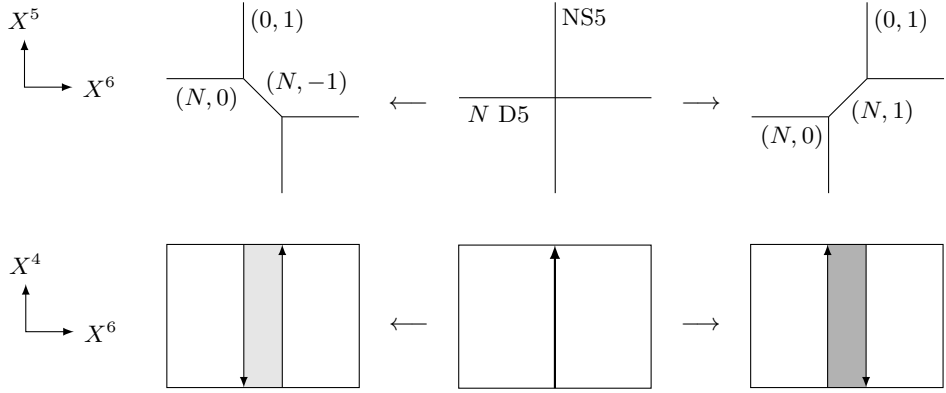
\begin{figure}
  \centering

  \begin{tikzpicture}[scale=0.5, align at top]
    \draw[<->] (1,0) node[right] {$X^6$}
    -- (0,0) -- (0,1) node[above] {$X^5$};
  \end{tikzpicture}
  \quad
  \begin{tikzpicture}[scale=2/2.5, align at top]
      \draw (0,1) -- (-1,1);
      \draw (-1,0) -- (-1,1);
      \draw (-1.5,1.5) -- (-1.5, 2.5) node[below right=-2pt] {$(0,1)$};
      \draw (-1.5,1.5) -- (-2.5, 1.5) node[below right=-1pt] {$(N,0)$};
      \draw (-1,1) -- node[above right=-4pt] {$(N,-1)$} (-1.5,1.5);          
  \end{tikzpicture}
  \
  \begin{tikzpicture}[align at top]
    \node at (0,0) {};
    \node at (0,2) {};
    \node at (0,1) {$\longleftarrow$};
  \end{tikzpicture}
  \
  \begin{tikzpicture}[align at top]
    \draw (0,1) node[below right=-1pt] {$N$ D5} -- (2,1);
    \draw (1,0) -- (1,2) node[below right=-2pt] {NS5};
  \end{tikzpicture}
  \
  \begin{tikzpicture}[align at top]
    \node at (0,0) {};
    \node at (0,2) {};
    \node at (0,1) {$\longrightarrow$};
  \end{tikzpicture}
  \
  \begin{tikzpicture}[scale=2/2.5, align at top]
    \draw (0,1) node[below right=-1pt] {$(N,0)$} -- (1,1);
    \draw (1,0) -- (1,1);
    \draw (1.5,1.5) -- (1.5, 2.5) node[below right=-2pt] {$(0,1)$};
    \draw (1.5,1.5) -- (2.5, 1.5);
    \draw (1,1) -- node[below right=-4pt] {$(N,1)$} (1.5,1.5);
  \end{tikzpicture}

  \bigskip

  \begin{tikzpicture}[scale=0.5, align at top]
    \draw[<->] (1,0) node[right] {$X^6$} -- (0,0) --
    (0,1) node[above] {$X^4$};
  \end{tikzpicture}
  \quad
  \begin{tikzpicture}[align at top]
    \fill[lshaded] (-0.8,0) rectangle (-1.2,1.5);
    \draw (0,0) rectangle (-2,1.5);
    \draw[z->] (-0.8,0) -- (-0.8,1.5);
    \draw[z<-] (-1.2,0) -- (-1.2,1.5);
  \end{tikzpicture}
  \
  \begin{tikzpicture}[align at top]
    \node at (0,0) {};
    \node at (0,1.5) {};
    \node at (0,0.75) {$\longleftarrow$};
  \end{tikzpicture}
  \
  \begin{tikzpicture}[align at top]
    \draw (0,0) rectangle (2,1.5);
    \draw[r->] (1,0) -- (1,1.5);
  \end{tikzpicture}
  \
  \begin{tikzpicture}[align at top]
    \node at (0,0) {};
    \node at (0,1.5) {};
    \node at (0,0.75) {$\longrightarrow$};
  \end{tikzpicture}
  \
  \begin{tikzpicture}[align at top]
    \fill[dshaded] (0.8,0) rectangle (1.2,1.5);
    \draw (0,0) rectangle (2,1.5);
    \draw[z->] (0.8,0) -- (0.8,1.5);
    \draw[z<-] (1.2,0) -- (1.2,1.5);
  \end{tikzpicture}

  \caption{An intersection of $N$ D5-branes and an NS5-brane may be
    resolved to trivalent junctions involving either an $(N,1)$
    5-brane (right) or an $(N,-1)$ 5-brane (left).  Illustrated here
    is the case when $\Sigma$ is flat and an NS5-brane extends along
    the $012345$ directions.  The $(N,\pm1)$ 5-branes are supported on
    the shaded regions.}
  \label{fig:5BW}
\end{figure}

In a brane box model, the NS5-branes intersect the D5-branes along
curves $C_i \subset \Sigma$.  We can resolve the intersections to
trivalent junctions.  Each junction connects $N$ D5-branes, a single
NS5-brane, and their bound state; in the terminology of $(p,q)$
5-branes, they are $(N,0)$, $(0,1)$ and $(N, \pm 1)$ 5-branes,
respectively.  Upon this resolution, the intersection curves $C_i$
spilt into pairs of curves representing the 5-brane junctions.  See
Fig.~\ref{fig:5BW} for illustration.  These curves, which separate
$\Sigma$ into regions supporting different values of the 5-brane
charge~$q$, are called \emph{zigzag paths}.  We orient them in such a
way that $q$ increases by $1$ as we cross a zigzag path from left to
right.

We can consider more general configurations of zigzag paths, not
necessarily those obtained by resolving D5--NS5 intersections.  Each
configuration encodes a 5-brane system: the NS5-branes approach the
D5-branes from transverse directions, meet them along the zigzag
paths, and together make bound states over some regions.  Such a brane
configuration is known as a \emph{brane tiling}.  Provided that the
NS5-branes wrap appropriate surfaces in $T^*\Sigma$, a brane tiling
configuration preserves $4$ supercharges.

If the string coupling is strong enough, the tension of D5-branes is
much larger than that of NS5-branes.  In that situation, the shape of
the D5-branes is unaffected by the NS5-branes, which simply make 90
degree turns when they hit the D5-branes.  (Away from the D5-branes,
the NS5-branes wrap the same kinds of supersymmetric cycles as in the
case of brane box models.)  Hence, the 6d theory on the D5-branes may
be regarded as formulated on the fixed spacetime
$\R^{3,1} \times \Sigma$, irrespective of the precise configuration of
the NS5-branes.  From the point of view of the 6d theory, the latter
branes create codimension-$1$ defects supported on the zigzag paths.

For compact $\Sigma$, the 6d theory in the presence of these defects
is described at low energies by a 4d $\CN = 1$ theory.  This
construction therefore defines a map from brane tilings on compact
Riemann surfaces to 4d $\CN = 1$ theories.  Composing it with the
supersymmetric index, we get a 2d TQFT equipped with line operators
and the associated lattice model.

As before, an extra dimension emerges via embedding into M-theory,
implying integrability of the lattice model.  Accordingly, each zigzag
path naturally carries a circle-valued spectral parameter which is the
$X^{10}$ coordinate of the corresponding M5-brane.  If $X^{10}$ has
period $2\pi$, then $\exp(iX^{10})$ is identified with a $\U(1)$ flavor
fugacity in the index of the 4d theory.  (Flavor fugacities are often
analytically continued to complex parameters.)  The relevant flavor
symmetry comes from the $\U(1)$ gauge symmetry on the NS5-brane.  Via
the boundary condition on the brane junction, this symmetry is related
to a $\U(1)$ gauge symmetry on the D5-branes, which gets frozen at low
energies and becomes a global symmetry in the field theory.

\subsection{Integrable lattice models from quiver gauge theories}

In order to actually write down the R-operator of the integrable
lattice model arising from brane tilings, we need to know more
precisely what 4d theory results from a given configuration of zigzag
paths.  The answer is known when there is no region supporting $(N,q)$
5-brane with $|q| > 1$ on $\Sigma$.%
\footnote{The case $N=2$ seems to be somewhat special, as we will
  explain in section~\ref{sec:BS}.}
In this case, the 4d theory is a quiver gauge theory.

The rule for reading off the quiver is as follows~\cite{Franco:2005rj,
  Hanany:2005ss, Franco:2012mm}.  We indicate regions with $q = +1$ by
dark shading and those with $q = -1$ by light shading.  Regions with
$q = 0$ are left unshaded.  On an unshaded region there lies an
$\SU(N)$ node, produced by open strings attached on this region.  This
is a flavor node if the region contains part of the boundary of
$\Sigma$, and otherwise a gauge node.  (We allow $\Sigma$ to have
boundary components where the $N$ D5-branes end separately on $N$
D7-branes.)  A crossing of two zigzag paths gives rise to a
bifundamental chiral multiplet, produced by open strings that start
from one unshaded region and end on another:
\begin{equation}
  \begin{tikzpicture}[baseline=(x.base), scale=0.8*2/3]
    \node (x) at (0,1) {\vphantom{x}};

    \fill[lshaded] (0,1) rectangle (1,2);
    \fill[dshaded] (1,0) rectangle (2,1);
    \draw[z->] (0,1) node[left] {$a$} -- (2,1);
    \draw[z->] (1,0) node[below] {$b$} -- (1,2);
    \node at (0.5,0.5) {$z$};
    \node at (1.5,1.5) {$w$};
  \end{tikzpicture}
  \ =
  \begin{tikzpicture}[baseline=(x.base), scale=0.8*2/3]
    \node (x) at (0,1) {\vphantom{x}};

    \node[fnode, above right] (z) at (0,0) {$z$};
    \node[fnode, below left] (w) at (2,2) {$w$};
    \draw[q->] (z) -- node[above left=-4pt] {$(pq)^{R/2} b/a$} (w);
  \end{tikzpicture}
  \ .
\end{equation}
For clarity, we have labeled the unshaded regions with the fugacities for
the corresponding nodes.  As explained at the end of
section~\ref{sec:BT-BT}, there is a $\U(1)$ flavor symmetry for each
zigzag path.  We choose the convention that the above arrow has charge
$-1$ and $+1$ under the flavor symmetries $\U(1)_a$ and $\U(1)_b$
associated with the horizontal and vertical zigzag paths,
respectively.

Since every arrow is oppositely charged under two different $\U(1)$
flavor symmetries coming from zigzag paths, the diagonal combination
of all $\U(1)$ flavor symmetries associated with zigzag paths acts on
the theory trivially.  Therefore, the zigzag paths provide as many
$\U(1)$ flavor symmetries as their number minus $1$, and these
generate the nonanomalous $\U(1)$ flavor symmetries of the
theory~\cite{Imamura:2006ie}.

\begin{figure}
\centering 
\subcaptionbox{\label{fig:zigzag-loop-a}}{
  \begin{tikzpicture}[scale=0.5]
    \draw[z<-] (0.2,0) .. controls (-1,0.7) .. (-2,2);
    \draw[z<-] (-1.5,0.5) .. controls (-1.5,1.8) and (-1,2.5) .. (-1.4,4);
    \draw[z<-] (-2,2.8) .. controls (-0.5,3.8) .. (0.5,4);
    \draw[z<-] (-0.5,4.5) .. controls (0.8,3) .. (1.5,2);
    \draw[z<-] (1,3.5) .. controls (1.2,2.5) and (0.6,1.5) .. (0.8,0.8);
    \draw[z<-] (1.2,1.5) .. controls (0.5,1.4) and (0.2,1) .. (-0.6,-0.2);
    \node[node, minimum size=5pt, fill=black] at (-0.2,2) {$$};
    \node[node, minimum size=5pt] at (-0.2,-0.1) {$$};
    \node[node, minimum size=5pt] at (1,1.15) {$$};
    \node[node, minimum size=5pt] at (1.3,2.8) {$$};
    \node[node, minimum size=5pt] at (0.1,4.2) {$$};
    \node[node, minimum size=5pt] at (-1.6,3.4) {$$};
    \node[node, minimum size=5pt] at (-1.8,1.1) {$$};
 \end{tikzpicture}
 \ = \
  \begin{tikzpicture}[scale=0.5]
    \node[fnode] (a) at (0.4,0.6) {$$};
    \node[fnode] (b) at (1.15,1.85) {$$};
    \node[fnode] (c) at (0.7,3.6) {$$};
    \node[fnode] (d) at (-0.8,4) {$$};
    \node[fnode] (e) at (-1.7,2.3) {$$};
    \node[fnode] (f) at (-1,0.3) {$$};
    \draw[q->] (a) -- (b);
    \draw[q->] (b) -- (c);
    \draw[q->] (c) -- (d);
    \draw[q->] (d) -- (e);
    \draw[q->] (e) -- (f);
    \draw[q->] (f) -- (a);
 \end{tikzpicture}
}
\qquad
\subcaptionbox{\label{fig:zigzag-loop-b}}{
  \begin{tikzpicture}[scale=0.5]
    \draw[z<-] (0.2,0) .. controls (-1,0.7) .. (-2,2);
    \draw[z->] (-1.5,0.5) .. controls (-1.5,1.8) and (-1,2.5) .. (-1.4,4);
    \draw[z<-] (-2,2.8) .. controls (-0.5,3.8) .. (0.5,4);
    \draw[z->] (-0.5,4.5) .. controls (0.8,3) .. (1.5,2);
    \draw[z<-] (1,3.5) .. controls (1.2,2.5) and (0.6,1.5) .. (0.8,0.8);
    \draw[z->] (1.2,1.5) .. controls (0.5,1.4) and (0.2,1) .. (-0.6,-0.2);
    \node[node, minimum size=5pt, fill=black] (a) at (0.4,0.6) {$$};
    \node[node, minimum size=5pt] (b) at (1.15,1.85) {$$};
    \node[node, minimum size=5pt, fill=black] (c) at (0.7,3.6) {$$};
    \node[node, minimum size=5pt] (d) at (-0.8,4) {$$};
    \node[node, minimum size=5pt, fill=black] (e) at (-1.7,2.3) {$$};
    \node[node, minimum size=5pt] (f) at (-1,0.3) {$$};
 \end{tikzpicture}
 \ = \
  \begin{tikzpicture}[scale=0.5]
    \node[gnode] (g) at (-0.2,2) {$$};
    \node[fnode] (a) at (-0.2,-0.1) {$$};
    \node[fnode] (b) at (1,1.15) {$$};
    \node[fnode] (c) at (1.3,2.8) {$$};
    \node[fnode] (d) at (0.1,4.2) {$$};
    \node[fnode] (e) at (-1.6,3.4) {$$};
    \node[fnode] (f) at (-1.8,1.1) {$$};

    \draw[q->] (g) -- (a);
    \draw[q<-] (g) -- (b);
    \draw[q->] (g) -- (c);
    \draw[q<-] (g) -- (d);
    \draw[q->] (g) -- (e);
    \draw[q<-] (g) -- (f);
 \end{tikzpicture}
}
\caption{Zigzag paths bounding (a) a shaded region and (b) an unshaded
  region.  Here we indicate dark and light shading by placing white
  and black dots instead.}
 \label{fig:zigzag-loop}
\end{figure}
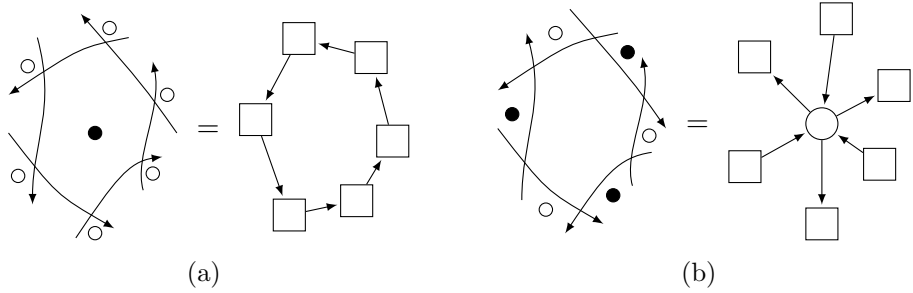

The R-charge $R$ is not uniquely determined since it can be shifted by
$\U(1)$ flavor charges.  From the point of view of the index, the
shift amounts to a redefinition of flavor fugacities by some factors.
That said, the R-charge assignment is constrained by two conditions.%
\footnote{For flat $\Sigma$, one way to satisfy these conditions is to
  make the zigzag paths straight and set the R-charge of a
  bifundamental chiral multiplet to $R = \theta/\pi$, where $\theta$
  is the angle between two zigzag paths through which the arrow
  goes~\cite{Hanany:2005ss}.  Then the two conditions are satisfied
  since the sum of the interior angles of an $n$-gon is equal to
  $(n - 2)\pi$, while the exterior angles add up to $2\pi$.  This
  prescription is not desirable for our purposes, however.  The
  supersymmetric index depends on the R-charge assignment.  We want
  the index to be a topological invariant of the brane tiling, so the
  R-charges should not change as zigzag paths are deformed.  We will
  describe our R-charge assignment for specific classes of brane
  tilings that we study.}
From zigzag paths bounding a shaded region, we get a sequence of
arrows making a loop.  For example, in our brane tiling we may have a
configuration of zigzag paths shown in Fig.~\ref{fig:zigzag-loop-a}.
For each such loop, worldsheet instantons induce a superpotential term
given by the product of the bifundamental chiral multiplets, with sign
determined by the orientation of the arrows.  Thus, the R-charges of
the arrows must add up to $2$.  Likewise, from zigzag paths bounding
an unshaded region, we get arrows starting from or ending at a gauge
node, as in Fig.~\ref{fig:zigzag-loop-b}.  For $\U(1)_R$ to be free of
anomaly, the sum of the R-charges of the arrows must equal the number
of the arrows minus $2$.

As already said, we would not be able to compute the supersymmetric
index without detailed knowledge of the theory.  For the purpose of
identifying the integrable lattice model, we should therefore study
brane tilings in which no regions with $|q| > 1$ appear, at least as a
first step of more general analysis.  However, even if we do restrict
to that case and identify the lattice model, it is not possible to
check the integrability directly using the index
formula~\eqref{eq:I-quiver}.  Unfortunately, the Yang--Baxter equation
for three zigzag paths always involves regions with $|q| > 1$.

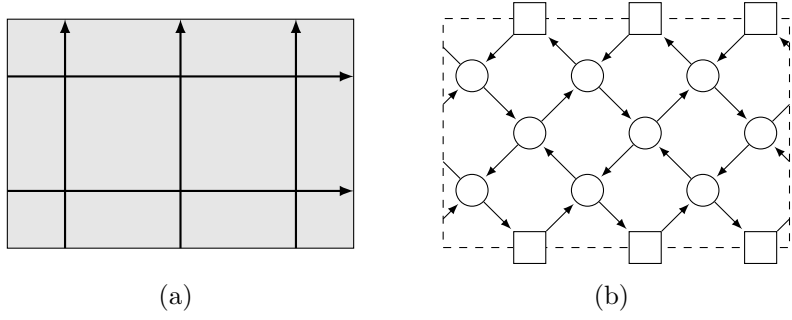
\begin{figure}
  \centering
  \subcaptionbox{\label{fig:dia-quiver-a}}{
    \begin{tikzpicture}[scale=1.2]
      \node[fnode, draw=none, fill=none] at (1,0) {};

      \draw[lshaded] (0,0) rectangle (3,2);

      \begin{scope}[shift={(0.5,0)}]
        \draw[r->] (0,0) -- (0,2);
        \draw[r->] (1,0) -- (1,2);
        \draw[r->] (2,0) -- (2,2);
      \end{scope}

      \begin{scope}[shift={(0,0.5)}]
        \draw[r->] (0,0) -- (3,0);
        \draw[r->] (0,1) -- (3,1);
      \end{scope}

      \begin{scope}[shift={(0.25,0.25)}]
        \node[above] at (0.5,0) {};
        \node[above] at (1.5,0) {};
        \node[above] at (2.5,0) {};
        \node[above] at (0.5,1) {};
        \node[above] at (1.5,1) {};
        \node[above] at (2.5,1) {};

        \node[right] at (0,0.5) {};
        \node[right] at (1,0.5) {};
        \node[right] at (2,0.5) {};
        \node[right] at (0,1.5) {};
        \node[right] at (1,1.5) {};
        \node[right] at (2,1.5) {};
      \end{scope}
    \end{tikzpicture}
  }
  \qquad
  \subcaptionbox{\label{fig:dia-quiver-b}}{
    \begin{tikzpicture}[scale=1.2]
      \draw[dashed] (0,0) rectangle (3,2);

      \begin{scope}[shift={(0.25,0)}]
        \node[fnode] (i1) at (0.5,0) {};
        \node[fnode] (i2) at (1.5,0) {};
        \node[fnode] (i3) at (2.5,0) {};
        \node[gnode] (j1) at (0.5,1) {};
        \node[gnode] (j2) at (1.5,1) {};
        \node[gnode] (j3) at (2.5,1) {};
        \node[fnode] (h1) at (0.5,2) {};
        \node[fnode] (h2) at (1.5,2) {};
        \node[fnode] (h3) at (2.5,2) {};

        \node[gnode] (k1) at (0,0.5) {};
        \node[gnode] (k2) at (0,1.5) {};
        \node[gnode] (l1) at (1,0.5) {};
        \node[gnode] (l2) at (1,1.5) {};
        \node[gnode] (m1) at (2,0.5) {};
        \node[gnode] (m2) at (2,1.5) {};
      \end{scope}

      \draw[q->] (i1) -- (l1);
      \draw[q->] (i2) -- (m1);
      \draw[q-] (i3) -- ++(45:{sqrt(2)/4} );
      \draw[q->] (k1) -- (i1);
      \draw[q<-] (k1) -- ++(-135:{sqrt(2)/4} );
      \draw[q-] (k1) -- ++(135:{sqrt(2)/4} );
      \draw[q->] (l1) -- (i2);
      \draw[q->] (l1) -- (j1);
      \draw[q->] (m1) -- (i3);
      \draw[q->] (m1) -- (j2);
      \draw[q->] (j1) -- (k1);
      \draw[q->] (j1) -- (l2);
      \draw[q->] (j2) -- (m2);
      \draw[q->] (j2) -- (l1);
      \draw[q->] (j3) -- (m1);
      \draw[q-] (j3) -- ++(45:{sqrt(2)/4} );
      \draw[q<-] (j3)  -- ++(-45:{sqrt(2)/4} );
      \draw[q->] (k2) -- (j1);
      \draw[q<-] (k2) -- (h1);
      \draw[q-] (k2) -- ++(135:{sqrt(2)/4} );
      \draw[q<-] (k2) -- ++(-135:{sqrt(2)/4} );
      \draw[q->] (l2) -- (j2);
      \draw[q->] (m2) -- (h2);
      \draw[q<-] (l2) -- (h2);
      \draw[q->] (l2) -- (h1);
      \draw[q->] (m2) -- (j3);
      \draw[q<-] (m2) -- (h3);
      \draw[q<-] (h3) -- ++(-45:{sqrt(2)/4} );
    \end{tikzpicture}
  }
  \caption{(a) A $2 \times 3$ lattice on a cylinder constructed from
    the crossing~\protect\eqref{eq:RD}.  The horizontal direction is
    periodic, while the vertical direction is a finite interval.  (b)
    The quiver of the corresponding gauge theory.}
  \label{fig:dia-quiver}
\end{figure}

To circumvent this difficulty, we take a pair of zigzag paths with
opposite orientation and regard them as a single thick line:
\begin{equation}
  \label{eq:solid-line}
  \begin{tikzpicture}[baseline=(x.base)]
    \node (x) at (0,0) {\vphantom{x}};

    \draw[r->] (0,0) node[left] {$(a,b)$} -- (1,0);
  \end{tikzpicture}
  \ =
  \begin{tikzpicture}[baseline=(x.base)]
    \node (x) at (0,1/6) {\vphantom{x}};

    \draw[z<-] (0,0) -- (1,0) node[right] {$b$};
    \draw[z->] (0,1/3) node[left] {$a$} -- (1,1/3);
  \end{tikzpicture}
  \,.
\end{equation}
If we make a lattice using this line in an $(N,-1)$ 5-brane
background, undesirable regions do not arise.  Indeed, a crossing of
two lines gives a diamond of arrows:
\begin{equation}
  \label{eq:RD}
  \begin{tikzpicture}[baseline=(x.base), scale=0.8]
    \node (x) at (0,1) {\vphantom{x}};

    \fill[lshaded] (0,0) rectangle (2,2);
    \draw[r->] (0,1) node[left] {$(a_i, b_i)$} -- (2,1);
    \draw[r->] (1,0) node[below] {$(a_j, b_j)$}-- (1,2);
  \end{tikzpicture}
  \ =
  \begin{tikzpicture}[baseline=(x.base), scale=0.8]
    \node (x) at (0,1) {\vphantom{x}};
    
    \fill[lshaded] (0,0) rectangle (2/3,2/3);
    \fill[lshaded] (4/3,0) rectangle (2,2/3);
    \fill[lshaded] (0,4/3) rectangle (2/3,2);
    \fill[lshaded] (4/3,4/3) rectangle (2,2);
    \fill[dshaded] (2/3,2/3) rectangle (4/3,4/3);

    \draw[z->] (0,4/3) node[left] {$a_i$} -- (2,4/3);
    \draw[z<-] (0,2/3) -- (2,2/3) node[right] {$b_i$};
    \draw[z->] (2/3,0) node[below] {$a_j$} -- (2/3,2);
    \draw[z<-] (4/3,0) -- (4/3,2) node[above] {$b_j$};

    \node (zj) at (1,1/3) {$z_j$};
    \node (zi) at (1/3,1) {$z_i$};
    \node (wj) at (1,5/3) {$w_j$};
    \node (wi) at (5/3,1) {$w_i$};
  \end{tikzpicture}
  \ =
  \begin{tikzpicture}[baseline=(x.base), scale=0.8]
    \node (x) at (0,1) {\vphantom{x}};

    \node[fnode, above] (zj) at (1,0) {$z_j$};
    \node[fnode, right] (zi) at (0,1) {$z_i$};
    \node[fnode, below] (wj) at (1,2) {$w_j$};
    \node[fnode, left] (wi) at (2,1) {$w_i$};

    \draw[q->] (zj) --
    node[below left=-4pt] {$\sqrt{pq} b_i/a_j$} (zi);
    \draw[q->] (zi) --
    node[above left=-4pt] {$a_j/a_i$} (wj);
    \draw[q->] (wj) --
    node[above right=-4pt] {$\sqrt{pq} a_i/b_j$} (wi);
    \draw[q->] (wi) --
    node[below right=-4pt] {$b_j/b_i$} (zj);
  \end{tikzpicture}
  \,.
\end{equation}
An example of a quiver constructed from crossings of this type is
shown in Fig.~\ref{fig:dia-quiver}.  Note that our R-charge
assignment satisfies the two constraints described above.

The R-operator
$\Rdia_{ij}\colon \Vdia_i \otimes \Vdia_j \to \Vdia_j \otimes \Vdia_i$
for the corresponding lattice model is given by the supersymmetric
index of the quiver~\eqref{eq:RD}, and depends on two pairs of spectral
parameters.  This is in fact the lattice model discovered by Bazhanov
and Sergeev in~\cite{Bazhanov:2011mz}.  The vector space $\Vdia$
supported on a line is the space of symmetric meromorphic functions
$f(z)$ of $N$ complex variables $z = \{z_1, \dotsc, z_N\}$ satisfying
the constraint $z_1 \dotsm z_N = 1$.  The variables $z$ are to be
identified with the fugacities for the $\SU(N)$ node on that line.
For example, $\Vdia_i$ is the space of symmetric meromorphic functions
of the variables $z_i$ or $w_i$ in the above diagram.

For $X \in \End(\Vdia)$, we define its matrix elements $X^w_z$ by
\begin{equation}
  (Xf)(w)
  =
  \int_{\T^{N-1}} \prod_{I=1}^{N-1} \frac{\rmd z_I}{2\pi i z_I}
  X^w_z \IV(z) f(z)
  \,.
\end{equation}
Then the matrix elements of $\Rdia_{ij}$ are%
\footnote{Apart from the normalization, this definition differs from
  the R-operator (4.5) in~\cite{Yagi:2015lha} by the vector multiplet
  factors $\IV(z_i) \IV(z_j)$.  The difference is due to the fact
  that the factor $\IV(z)$ was not included in the definition of
  matrix elements in that paper.}
\begin{multline}
  \label{eq:Rdia}
  \Rdia_{ij}\bigl((a_i,b_i), (a_j,b_j)\bigr)_{z_i z_j}^{w_j w_i}
  \\
  =
  \IBt\biggl(z_i, w_j; \frac{a_j}{a_i}\biggr) 
  \IB\biggl(w_j, w_i; \sqrt{pq} \frac{a_i}{b_j}\biggr)
  \IBt\biggl(w_i, z_j; \frac{b_j}{b_i}\biggr)
  \IB\biggl(z_j, z_i; \sqrt{pq} \frac{b_i}{a_j}\biggr)
  \,.
\end{multline}
Here $\IBt$ is a normalized index of a bifundamental chiral
multiplet:
\begin{equation}
  \label{eq:It_B}
  \IBt(z, w; a)
  =
  \frac{\IB(z, w; a)}{\Gamma(a^N)}
  \,.
\end{equation}
We choose this normalization for contributions from arrows with $R =
0$ so that the R-operator satisfies the unitarity relation
\eqref{eq:invert}.

Plugging this R-operator into the Yang--Baxter equation \eqref{eq:YBE-G},
we see that the integrability of the lattice model is the statement
that the indices of two quivers are equal:
\begin{equation}
  \label{eq:YBE-BT}
  \begin{tikzpicture}[scale=0.8, baseline=(x.base)]
    \node (x) at (150:1) {\vphantom{x}};

    \node[gnode] (j2) at (0,0) {};
    \node[gnode] (k2) at (150:1) {};
    \node[gnode] (i2) at (90:1) {};

    \node[fnode] (i3) at (0:1) {};
    \node[fnode] (j3) at ($(i2)+(0:1)$) {};

    \node[fnode] (k1) at ($(j2) + (-120:1)$) {};
    \node[fnode] (j1) at ($(k2) + (-120:1)$) {};

    \node[fnode] (k3) at ($(i2) + (120:1)$) {};
    \node[fnode] (i1) at ($(k2) + (120:1)$) {};

    \draw[q->] (j2) -- (k1);
    \draw[q->] (k1) -- (j1);
    \draw[q->] (j1) -- (k2);
    \draw[q->] (k2) -- (j2);

    \draw[q->] (k2) -- (i1);
    \draw[q->] (i1) -- (k3);
    \draw[q->] (k3) -- (i2);
    \draw[q->] (i2) -- (k2);

    \draw[q->] (i2) -- (j3);
    \draw[q->] (j3) -- (i3);
    \draw[q->] (i3) -- (j2);
    \draw[q->] (j2) -- (i2);
  \end{tikzpicture}
  \ = \
  \begin{tikzpicture}[rotate=-60, scale=0.8, baseline=(x.base)]
    \node (x) at (90:1) {\vphantom{x}};

    \node[gnode] (i2) at (0,0) {};
    \node[gnode] (j2) at (150:1) {};
    \node[gnode] (k2) at (90:1) {};

    \node[fnode] (k1) at (0:1) {};
    \node[fnode] (i3) at ($(k2)+(0:1)$) {};

    \node[fnode] (j1) at ($(i2) + (-120:1)$) {};
    \node[fnode] (i1) at ($(j2) + (-120:1)$) {};

    \node[fnode] (j3) at ($(k2) + (120:1)$) {};
    \node[fnode] (k3) at ($(j2) + (120:1)$) {};

    \draw[q->] (j2) -- (k3);
    \draw[q->] (k3) -- (j3);
    \draw[q->] (j3) -- (k2);
    \draw[q->] (k2) -- (j2);

    \draw[q->] (k2) -- (i3);
    \draw[q->] (i3) -- (k1);
    \draw[q->] (k1) -- (i2);
    \draw[q->] (i2) -- (k2);

    \draw[q->] (i2) -- (j1);
    \draw[q->] (j1) -- (i1);
    \draw[q->] (i1) -- (j2);
    \draw[q->] (j2) -- (i2);
  \end{tikzpicture}
  \ .
\end{equation}
This is indeed true, as the two quivers describe theories that are
dual in the infrared; repeated application of the basic Seiberg
duality transformation~\cite{Seiberg:1994pq} cyclically four times to
the three gauge nodes turns the quiver on the left-hand side to the
one on the right-hand side~\cite{Yagi:2015lha}.

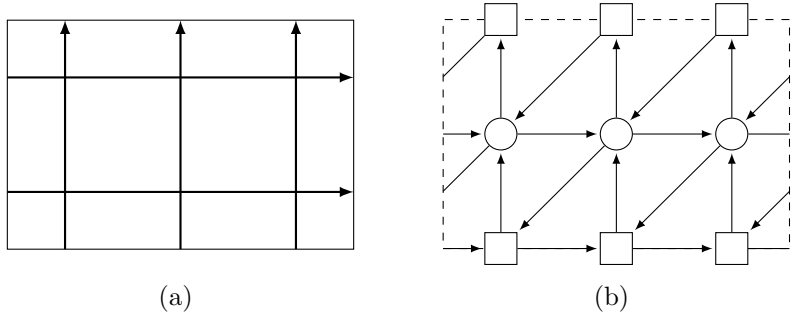
\begin{figure}
  \centering
  \subcaptionbox{\label{fig:tri-quiver-a}}{
    \begin{tikzpicture}[scale=1.2]
      \node[fnode, draw=none, fill=none] at (1,0) {};

      \draw (0,0) rectangle (3, 2);
      
      \begin{scope}[shift={(0.5,0)}]
        \draw[r->] (0, 0) -- (0, 2);
        \draw[r->] (1, 0) -- (1, 2);
        \draw[r->] (2, 0) -- (2, 2);
      \end{scope}
      
      \begin{scope}[shift={(0,0.5)}]
        \draw[r->] (0, 0) -- (3, 0);
        \draw[r->] (0, 1) -- (3, 1);
      \end{scope}
    \end{tikzpicture}
  }
  \qquad
  \subcaptionbox{\label{fig:tri-quiver-b}}{
    \begin{tikzpicture}[scale=1.2]
      \draw[dashed] (0,0) rectangle (3,2);
      
      \begin{scope}[shift={(0.5,0)}]
        \node[fnode] (a) at (0,0) {};
        \node[fnode] (b) at (1,0) {};
        \node[fnode] (c) at (2,0) {};
        \node[gnode] (d) at (0,1) {};
        \node[gnode] (e) at (1,1) {};
        \node[gnode] (f) at (2,1) {};
        \node[fnode] (g) at (0,2) {};
        \node[fnode] (h) at (1,2) {};
        \node[fnode] (q) at (2,2) {};
      \end{scope}
      
      \draw[q->] (0,0) -- (a);
      \draw[q->] (a) -- (b);
      \draw[q->] (b) -- (c);
      \draw[q-] (c) -- (3,0);

      \draw[q->] (0,1) -- (d);
      \draw[q->] (d) -- (e);
      \draw[q->] (e) -- (f);
      \draw[q-] (f) -- (3,1);
      
      \draw[q->] (a) -- (d);
      \draw[q->] (b) -- (e);
      \draw[q->] (c) -- (f);
      \draw[q->] (d) -- (g);
      \draw[q->] (e) -- (h);
      \draw[q->] (f) -- (q);
      
      \draw[q-] (d) -- ++(-135:{sqrt(2)/2} );
      \draw[q->] (e) -- (a);
      \draw[q->] (f) -- (b);
      \draw[q<-] (c) -- ++(45:{sqrt(2)/2} );

      \draw[q-] (g) -- ++(-135:{sqrt(2)/2} );
      \draw[q->] (h) -- (d);
      \draw[q->] (q) -- (e);
      \draw[q<-] (f) -- ++(45:{sqrt(2)/2} );
    \end{tikzpicture}
  }
  \caption{(a) A $2 \times 3$ lattice on a cylinder constructed from
    the crossing~\eqref{eq:RT}.  (b) The corresponding quiver.}
  \label{fig:tri-quiver}
\end{figure}

There is another R-operator
$\Rtri_{ij}\colon \Vtri_i \otimes \Vtri_j \to \Vtri_j \otimes \Vtri_i$
that can be constructed from the
line~\eqref{eq:solid-line}~\cite{Yagi:2015lha}.  When two lines meet,
we can let them exchange their constituent zigzag paths:
\begin{equation}
  \label{eq:RT}
  \begin{tikzpicture}[baseline=(x.base), scale=0.8]
    \node (x) at (0,1) {\vphantom{x}};

    \draw[r->] (0,1) node[left] {$(a_i, b_i)$}
    -- (2,1) node[right] {$(a_j, b_i)$};
    \draw[r->] (1,0) node[below] {$(a_j, b_j)$}
    -- (1,2) node[above] {$(a_i, b_j)$};
  \end{tikzpicture}
  =
  \begin{tikzpicture}[baseline=(x.base), scale=0.8]
    \node (x) at (0,1) {\vphantom{x}};

    \node (zi) at (0.35,1.65) {$z_i$};
    \node (zj) at (0.35,0.35) {$z_j$};
    \node (wi) at (1.65,0.35) {$w_i$};
    \node (wj) at (1.65,1.65) {$w_j$};

    \fill[lshaded] (1,0.7) rectangle (1.3,1);
    \fill[dshaded] (0.7,0) rectangle (1,0.7);
    \fill[dshaded] (1.3,1) rectangle (2,1.3);
    \fill[dshaded] (0,1) -- (1,1) -- (1,2) -- (0.7,2) -- (0.7,1.3)
    -- (0,1.3) -- cycle;

    \draw[z->] (1,2) node[above] {$b_j$} -- (1,0);
    \draw[z->] (2,1) node[right] {$b_i$} -- (0,1);

    \draw[z->] (0.7,0) node[below] {$a_j$}
    -- (0.7,0.7) -- (1.3,0.7) -- (1.3,1.3) -- (2,1.3);
    \draw[z->] (0,1.3) node[left] {$a_i$}
    -- (0.7,1.3) -- (0.7,2);
  \end{tikzpicture}
  =
  \begin{tikzpicture}[baseline=(x.base), scale=0.8]
    \node (x) at (0,1) {\vphantom{x}};

    \node[fnode] (zj) at (0.35,0.35) {$z_j$};
    \node[fnode] (wi) at (1.65,0.35) {$w_i$};
    \node[fnode] (wj) at (1.65,1.65) {$w_j$};

    \draw[q->] (zj) -- node[below=4pt] {$\sqrt{pq} a_j/b_j$} (wi);
    \draw[q->] (wi) -- node[right] {$\sqrt{pq} b_i/a_j$} (wj);
    \draw[q->] (wj) -- node[above left=-4pt] {$b_j/b_i$} (zj);
  \end{tikzpicture}
  \,.
\end{equation}
A lattice constructed from crossings of this type consists of
triangles of arrows, as shown in Fig.~\ref{fig:tri-quiver}.  The
lattice model thus obtained is an interaction-round-a-face (IRF)
model, for which spins are assigned on the faces of the lattice.  In
the open string picture \eqref{open-string-pic}, all physical degrees
of freedom are localized at the ends of strings as ``Chan--Paton
factors.''

Formally, we can reformulate this model as a vertex model.  To do so,
we take the vector space $\Vtri$ supported on a line to be the tensor
product of the Chan--Paton spaces from both ends,
\begin{equation}
  \Vtri \iso \Vdia \otimes \Vdia
  \,,
\end{equation}
and include in the definition of the R-operator, delta functions that
ensure that the Chan--Paton factors match correctly.  In the present
case, the matrix elements of $\Rtri_{ij}$ are
\begin{multline}
  \Rtri_{ij}\bigl((a_i, b_i), (a_j,b_j)\bigr)
  _{z_i z_i', z_j z_j'}^{w_j' w_j, w_i' w_i}
  \\
  =
  \IB\biggl(z_j, w_i; \sqrt{pq} \frac{a_j}{b_j}\biggr)
  \IB\biggl(w_i, w_j; \sqrt{pq} \frac{b_i}{a_j}\biggr)
  \IBt\biggl(w_j, z_j; \frac{b_j}{b_i}\biggr)
  \\
  \times
  \delta(z_i, w_j') 
  \delta(z_j, z_i') 
  \delta(w_i, z_j') 
  \delta(w_j, w_i')
  \,.
\end{multline}
Strictly speaking, this reformulation is a little problematic in the
present case where the spin variables are continuous, as the
integration over each set of gauge fugacities $z$ gets accompanied by
a factor $\delta(z,z)$.  It is understood that this factor is to be
dropped.

The Yang--Baxter equation is simpler for this R-operator.  After
canceling some factors and using the identity \eqref{eq:II=0}, we find
that the equation reduces to the following form:
\begin{equation}
  \label{eq:Seiberg}
  \begin{tikzpicture}[baseline=(x.base)]
    \node (x) at (0,0) {\vphantom{x}};

    \node[fnode] (a) at (240:1){};
    \node[fnode] (c) at (0:1) {};
    \node[fnode] (d) at (60:1) {};
    \node[fnode] (f) at (180:1) {};
    \node[gnode] (g) at (0,0) {};

    \draw[q->] (a) -- (g);
    \draw[q->] (g) -- (c);
    \draw[q->] (d) -- (g);    
    \draw[q->] (g) -- (f);
  \end{tikzpicture}
  \
  =
  \
  \begin{tikzpicture}[baseline=(x.base)]
    \node (x) at (0,0) {\vphantom{x}};

    \node[fnode] (a) at (240:1){};
    \node[fnode] (c) at (0:1) {};
    \node[fnode] (d) at (60:1) {};
    \node[fnode] (f) at (180:1) {};
    \node[gnode] (g) at (0,0) {};
    
    \draw[q->] (g) -- (a);
    \draw[q->] (c) -- (g);
    \draw[q->] (f) -- (g);
    \draw[q->] (g) -- (d);
    \draw[q->] (d) -- (f);

    \draw[q<-] (c) -- (a);
    \draw[q<-] (c) -- (d);
    \draw[q<-] (f) -- (a);
  \end{tikzpicture}
  \ .
\end{equation}
The two quivers are related by Seiberg duality for $\SU(N)$ SQCD with
$2N$ flavors, hence their indices are indeed equal.  Mathematically,
it is a consequence of an integral identity~\cite{MR2044635,
  MR2630038} obeyed by the elliptic gamma function, as pointed out
in~\cite{Dolan:2008qi}.  Note that we obtained Seiberg duality for
$2N$ flavors from the Yang--Baxter equation, even though the quiver
for the lattice model has $3N$ flavors for each gauge group.  This is
necessary: the duality transformation would change the rank of the
gauge group if the number of flavors were different from $2N$.

\section{Surface defects as transfer matrices}
\label{sec:SD}

Having understood how the supersymmetric indices of brane tiling
models give rise to integrable lattice models, we now discuss the
lattice model realization of a class of half-BPS surface defects in
the 4d theories.  We will see that these surface defects are mapped to
transfer matrices constructed from L-operators.  In the simplest case,
we will identify the concrete form of the relevant L-operator.

\subsection{Surface defects and L-operators}

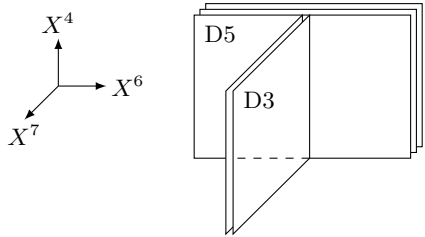
\begin{figure}
  \centering
  \begin{tikzpicture}[scale=0.5, align at top]
    \draw[->] (0,0) -- (1,0) node[anchor=west, right=-2pt] {$X^6$};
    \draw[->] (0,0) -- (0,1) node[anchor=south, above=-2pt] {$X^4$};
    \draw[->] (0,0) -- (-135:1) node[anchor=north, below=-2pt] {$X^7$};
  \end{tikzpicture}
  \quad
  \begin{tikzpicture}[scale=3/4, align at top]
    \draw[fill=white, shift={(0.16,0.16)}] (0,0) rectangle (3,2);
    \draw[fill=white, shift={(0.08,0.08)}] (0,0) rectangle (3,2);
    \draw[fill=white] (0,0) rectangle (3,2);
    \node[below right] at (0,2) {D5};

    \draw[fill=white] (1.5,0) -- ++(90:2) -- ++ (-135:1.5) -- ++(-90:2)
    -- cycle;

    \begin{scope}[shift={(0.1, 0)}]
      \draw[fill=white] (1.5,0) -- ++(90:2) -- ++ (-135:1.5) 
      node[below right, shift={(0,0.1)}] {D3}  -- ++(-90:2) -- cycle;

      \draw[dashed] (1.5,0) -- ++(180:{1.5/sqrt(2)} );
    \end{scope}
  \end{tikzpicture}
  \caption{D3-branes ending on the D5-branes create a surface defect.}
  \label{fig:D3-D5}
\end{figure}

For the sake of clarity, let us go back to the brane box
configuration \eqref{eq:BB} and explain the construction of these
surface defects in this situation; adapting the construction to brane
tilings is straightforward.  To this configuration we add $r$
D3-branes:
\begin{equation}
  \label{eq:SD-table}
  \begin{tabular}{|l|cccccccccc|}
    \hline
    & 0 & 1 & 2 & 3 & 4 & 5 & 6 & 7 & 8 & 9
    \\ \hline
    D5 & $\times$ & $\times$ & $\times$ & $\times$ & $\times$ & & $\times$
    &&&
    \\
    NS5 & $\times$ & $\times$ & $\times$ & $\times$ & $\times$ & $\times$
    &&&&
    \\
    NS5 & $\times$ & $\times$ & $\times$ & $\times$ & & & $\times$ & $\times$
    &&
    \\
    D3 & $\times$ & $\times$ & & & $\times$ & & & $\times$
    &&
    \\ \hline
  \end{tabular}
\end{equation}
The D3-branes come from $X^7 = +\infty$ and end on the D5-branes
located at, say $X^7 = 0$ (Fig.~\ref{fig:D3-D5}).  Out of the $4$
supercharges preserved by the other branes, they preserve the half
that generate $\CN = (0,2)$ supersymmetry on the $01$-plane.  From the
point of view of the 6d theory on the D5-branes, they create a
codimension-$3$ defect.  In the 4d theory obtained by compactifying
the $46$-plane, this is a codimension-$2$ or surface defect.

In the absence of NS5-branes extending along the $012367$ directions,
this brane configuration is related, via T-duality along the
$X^4$-direction, to the familiar configuration of D2-branes creating a
surface defect~\cite{Alday:2009fs} in a 4d $\CN = 2$ gauge theory
realized by a D4--NS5 system~\cite{Witten:1997sc}.  If those NS5-branes
are present, the T-duality converts them to an orbifold which breaks
the $\CN = 2$ supersymmetry to $\CN = 1$.  Still, our setup is locally
identical to the $\CN = 2$ case, and we can rely on various results
that have been obtained in that context.
 
\begin{figure}
  \centering
  \subcaptionbox*{}[\qsep]{
    \begin{tikzpicture}[scale=0.5]
      \draw[<->] (1,0) node[right] {$X^7$}
      -- (0,0) -- (0,1) node[above] {$X^6$};
      \node at (0,-2) {};
    \end{tikzpicture}
  }
  \subcaptionbox{\label{fig:SD-sym}}{
    \begin{tikzpicture}
      \draw (0,-1) -- (0,1) node[below left] {$N$ D5};
      \draw (0.1,-1) -- (0.1,1);
      \draw (0.2,-1) -- (0.2,1);
      \draw (0.3,-1) -- (0.3,1);

     \draw (0,0.15) -- (2,0.15);
     \draw (0,0.05) -- (2,0.05);
     \draw (0,-0.05) -- (2,-0.05);
     \draw (0.1,-0.15) -- (2,-0.15);

     \node[anchor=south, above] at (1,0.15) {$r$ D3};

     \draw[fill=white] (2,0) circle [radius=0.2];
     \draw ($(2,0) + (0:0.2)$) -- ++(180:0.4);
     \draw ($(2,0) + (-90:0.2)$) -- ++(90:0.4);
     
     \node[anchor=north west, below] at (2,-0.25) {NS5 ($012345$)};

    \end{tikzpicture}
  }
  \qquad
  \subcaptionbox{\label{fig:SD-antisym}}{
    \begin{tikzpicture}
      \draw (0,-1) -- (0,1) node[below left] {$N$ D5};
      \draw (0.1,-1) -- (0.1,1);
      \draw (0.2,-1) -- (0.2,1);
      \draw (0.3,-1) -- (0.3,1);

      \draw (0,0.1) -- (2,0.1);
      \draw (0.1,0) -- (2,0);
      \draw (0.2,-0.1) -- (2,-0.1);

      \node[anchor=south, above] at (1,0.1) {$r$ D3};

      \draw[fill=white] (2,0) circle [radius=0.2];
      \draw ($(2,0) + (45:0.2)$) -- ++(-135:0.4);
      \draw ($(2,0) + (-45:0.2)$) -- ++(135:0.4);
     
      \node[anchor=north west, below] at (2,-0.25) {NS5 ($014589$)};
    \end{tikzpicture}
  }
  \caption{Brane configurations for surface defects labeled with (a) the
    $r$th symmetric representation and (b) the $r$th antisymmetric
    representation of $\SU(N)$.}
  \label{fig:SD}
\end{figure}
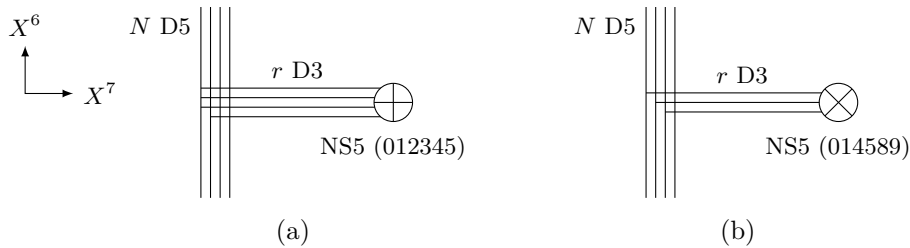

Instead of letting the D3-branes extend indefinitely along the
$X^7$-axis, we can make them end on an NS5-brane that spans the
$012345$ directions and is located at some $X^7 > 0$
(Fig.~\ref{fig:SD-sym}).  The number $r$ of D3-branes can be any
integer, so this configuration may be thought of as corresponding to a
symmetric representation of $\SU(N)$.  Hence, we label the surface
defect created by this configuration of D3-branes with the $r$th
symmetric representation~\cite{Gadde:2013ftv}.

Rather than the above NS5-brane, we may also introduce an NS5-brane
extending along the $014589$ directions and have the D3-branes end on
it (Fig.~\ref{fig:SD-antisym}).  In this case, due to the fermionic
nature of D-branes, the other ends of the D3-branes must attach to
separate D5-branes (``s-rule'').  Thus, $r$ cannot exceed $N$.
Furthermore, if we pass the NS5-brane to the other side of the
D5-branes, by the Hanany--Witten transition we obtain a similar
configuration with $N-r$ D3-branes.  This means that we can label the
corresponding surface defect with the $r$th antisymmetric
representation of $\SU(N)$.  This configuration is dual to that for a
Wilson line in the antisymmetric representation in 4d $\CN = 4$ super
Yang--Mills theory~\cite{Yamaguchi:2006tq, Gomis:2006sb,
  Gomis:2006im}.

\begin{figure}
  \centering
  \subcaptionbox{}{
    \begin{tikzpicture}
      \draw (0.15,-1) -- (0.15,1);
      \draw (0.05,-1) -- (0.05,1);
      \draw (-0.05,-1) -- (-0.05,1);
      \draw (-0.15,-1) -- (-0.15,1);
      
      \begin{scope}[shift={(0,-0.5)}]
        \node[below] at (0.5,-0.15) {$r_1$};
        \draw (-0.15,0.15) -- (1,0.15);
        \draw (-0.05,0.05) -- (1,0.05);
        \draw (0.05,-0.05) -- (1,-0.05);
        \draw (0.15,-0.15) -- (1,-0.15);
        \draw[fill=white] (1,0) circle [radius=0.2];
        \draw ($(1,0) + (45:0.2)$) -- ++(-135:0.4);
        \draw ($(1,0) + (-45:0.2)$) -- ++(135:0.4);
      \end{scope}

      \begin{scope}[shift={(0,0)}]
        \node[below] at (1.75,-0.1) {$r_2$};
        \draw (-0.15,0.1) -- (2.25,0.1);
        \draw (-0.05,0) -- (2.25,0);
        \draw (0.15,-0.1) -- (2.25,-0.1);
        \draw[fill=white] (2.25,0) circle [radius=0.2];
        \draw ($(2.25,0) + (45:0.2)$) -- ++(-135:0.4);
        \draw ($(2.25,0) + (-45:0.2)$) -- ++(135:0.4);
      \end{scope}

      \node at (0.5,0.5) {$\vdots$};

      \begin{scope}[shift={(0,0.75)}]
        \node[below] at (2.5,0) {$r_n$};
        \draw (0.05,0) -- (3,0);
        \draw[fill=white] (3,0) circle [radius=0.2];
        \draw ($(3,0) + (45:0.2)$) -- ++(-135:0.4);
        \draw ($(3,0) + (-45:0.2)$) -- ++(135:0.4);
      \end{scope}
    \end{tikzpicture}
  }
  \qquad
  \subcaptionbox{}{
    \begin{tikzpicture}[scale=1/3]
      \draw (0,0) -- (2,0);
      \draw (0,-1) -- (2,-1);
      \draw (0,-2) -- (2,-2);
      \draw (0,-3) -- (2,-3);
      \draw (0,-4) -- (1,-4);
      \draw (0,0) -- (0,-4);
      \draw (1,0) -- (1,-4);
      \draw (2,0) -- (2,-3); 
      
      \node at (2.7,-0.5) {$\dots$};
      
      \begin{scope}[shift={(0.4,0)}]
        \draw (3,0) rectangle (4,-1);
      \end{scope}
      
      \node[anchor=north west] at (-0.1,-4) {$r_1$};
      \node[anchor=north west] at (0.9,-3) {$r_2$};
      \node[anchor=north west] at (3.3,-1) {$r_n$};
    \end{tikzpicture}
  }
  \caption{(a) A surface defect labeled with a general representation
    $R$ is constructed from $r_1$, $\dotsc$, $r_n$ D3-branes suspended
    between the D5-branes and an NS5-brane, with
    $r_1 \geq \dotsb \geq r_n$.  The NS5-branes are separated
    vertically in this picture, but should really be positioned at the
    same level.  (b) The Young diagram for $R$ has $n$ columns with
    $r_1$, $\dotsc$, $r_n$ boxes.}
  \label{fig:SD-genrep}
\end{figure}
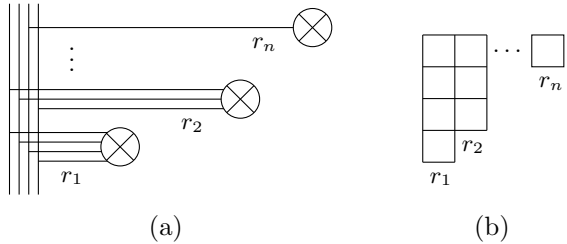

We can generate many more surface defects by taking products of these
basic ones.  It is known that they form a class of surface defects
classified by irreducible representations of $\SU(N)$.  The brane
configuration for a surface defect labeled with a general
representation~$R$, after a slight deformation, looks as in
Fig.~\ref{fig:SD-genrep}~\cite{Gomis:2006sb, Gomis:2006im,
  Gomis:2014eya}.%
\footnote{For a given $n$-tuple $(r_1, \dotsc, r_n)$, there are
  distinct brane configurations that differ in the choice of the
  D5-brane on which each D3-brane ends.  The surface defect under
  consideration is the superposition of all possible such choices.
  The inequivalent choices are in one-to-one correspondence with the
  semistandard Young tableaux obtained from the Young diagram for $R$,
  or equivalently, by the weights of $R$.  This structure is visible
  in the supersymmetric index in the presence of the surface defect
 ~\cite{Alday:2013kda, Bullimore:2014nla}.}
Dual configurations realizing line operators in three dimensions were
considered in~\cite{Assel:2015oxa}.

In the above construction, we may replace the $46$-plane with any
Riemann surface $\Sigma$ and let the D3-branes end on the D5-branes
along curves in $\Sigma$.  The NS5-branes can also take more general
configurations representing a brane tiling.

Now that we have a 4d $\CN = 1$ theory with a half-BPS surface defect,
we can place it on $S^3 \times S^1$ and compute its supersymmetric
index.  Assuming that the system flows to a conformal fixed point, we
can do this by conformally mapping the Euclidean spacetime $\R^4$
(minus the origin) to $S^3 \times \R$ and then compactifying the
radial direction $\R$.  After the mapping, the surface defect wraps
$S^1 \times S^1 \subset S^3 \times S^1$.  The first $S^1$ factor may
be taken to be either $\{\zeta_1 = 0\}$ or $\{\zeta_2 = 0\}$ in the
parameterization $|\zeta_1|^2 + |\zeta_2|^2 = 1$ of $S^3$.  These are
the only circles in $S^3$ that are left invariant under the action of
the isometry group $\U(1)_p \times \U(1)_q$, for the orbit of a point
outside these circles is two-dimensional.

The index in the presence of the surface defect is again given by a
correlation function of line operators in a 2d TQFT on $\Sigma$.  The
difference is that this time, the correlator contains a new line
operator created by the D3-branes ending on the D5-branes.  We
represent it by a dashed arrow:
\begin{equation}
  \begin{tikzpicture}[baseline=(x.base)]
    \node (x) at (0,0) {\vphantom{x}};

    \draw[dz->] (0,0) -- (1,0);
  \end{tikzpicture}
  \ .
\end{equation}
As we have seen, this line operator is specified by a representation
of $\SU(N)$.  In fact, it is labeled with a \emph{pair} of
representations $(R_1, R_2)$ since in general we can take
superposition of two surface defects, each wrapped around either
circle in $S^3$.

In any case, the correlation function equals the partition function of
a lattice model whose lattice is made of two kinds of lines, zigzag
paths coming from NS5-branes and the dashed line coming from the
D3-branes.  An extra dimension emerges as the M-theory circle if the
brane system is embedded in the M-theory via T-duality along the
second $S^1$ factor.  Under this embedding, the D3-branes are mapped
to M2-branes supported at points on the M-theory circle.  Thus, the
inclusion of the dashed line does not spoil the integrability of the
lattice model.

The position of the M2-branes on the M-theory circle (which is also
the position of the M5-brane on which they have one end) provides a
spectral parameter for the dashed line.  From the viewpoint of the
theory on the D3-branes, this is the holonomy around the second $S^1$
of the dual gauge field for the diagonal $\U(1)$ subgroup of the
$\U(r)$ gauge group.  For the theory on the D2-branes obtained by
T-duality along the $X^4$-direction, it is the holonomy of the $\U(1)$
gauge field dual to a periodic scalar.

We denote by $\W_{(R_1, R_2)}$ the vector space for a dashed line
labeled $(R_1, R_2)$.  At least when one of the representations is
trivial, $(R_1, R_2) = (R, \emptyset)$, it is natural to expect that
this space is isomorphic to the representation space $V_R$ of $R$ for
the following reason.  Under the M-theory embedding, the $N$ D5-branes
become M5-branes and support the 6d $\CN = (2,0)$ superconformal
theory of type $A_{N-1}$, placed on $S^3 \times \Sigma \times S^1$.
It is known that a BPS sector of the 6d theory compactified on $S^3$
is equivalent to Chern--Simons theory with gauge group
$\SL(N,\C)$~\cite{Terashima:2011qi, Dimofte:2011jd, Dimofte:2011ju,
  Dimofte:2011py, Cecotti:2011iy, Cordova:2013cea}.  The D3-branes, on
the other hand, become M2-branes and create a half-BPS codimension-$4$
defect supported on $S^1 \times C \times \{p\}$, where
$C \subset \Sigma$ is the curve along which the D3-branes are attached
on the D5-branes, and $p$ is a point on the M-theory circle.  In the
Chern--Simons theory, this defect reduces to a line operator labeled
$R$.  This is a Wilson line operator in the representation $R$, which
may be thought of as the worldline of a heavy charged particle whose
Hilbert space is $V_R$.  Thus we expect
\begin{equation}
 \W_{(R,\emptyset)} \iso V_R
 \,.
\end{equation}
In particular, $\W_{(R,\emptyset)}$ is a finite-dimensional space.
For general $(R_1, R_2)$, it is likely that $\W_{(R_1, R_2)}$ is
isomorphic to the tensor product $V_{R_1} \otimes V_{R_2}$.  To avoid
clutter, in what follows we fix the representations and simply write
$\W$ for $\W_{(R_1, R_2)}$.

Let us ask how the introduction of the surface defect is represented
on the lattice model side.  Consider a general brane tiling
configuration (which may or may not have a quiver description), and
suppose that the D3-branes end on the D5-branes along a loop
in~$\Sigma$.  Due to the periodic boundary condition, the dashed line
crosses zigzag paths coming from the right as many times as those
coming from the left.  By deforming these zigzag paths near the dashed
line, we can always make the two cases occur alternately.  Then, the
neighborhood of the dashed line looks like
\begin{equation}
  \label{eq:TM-L-num}
  \begin{tikzpicture}[scale=0.5]
    \draw[dz->, right hook->] (-0.1,0) -- (1.1,0);
    \draw[dz->, >=left hook] (1,0) -- (4.1,0);
    \node[fill=white, inner sep=1pt] at (2.5,0) {$\,\dots$};

    \begin{scope}[shift={(0,-0.5)}]
      \draw[r->] (0.5,0)
      node[below] {$1$} -- (0.5,1.2);
      \draw[r->] (1.5,0)
      node[below] {$2$} -- (1.5,1.2);
      \draw[r->] (3.5,0)
      node[below] {$n$} -- (3.5,1.2);
    \end{scope}
  \end{tikzpicture}
\end{equation}
in some $(N,q)$ 5-brane background.  Each crossing of a solid line and
the dashed one gives an R-operator
$\LM_i\colon \W \otimes \V_i \to \V_i \otimes \W$.  We call it an
\emph{L-operator}.  In this terminology, the
object~\eqref{eq:TM-L-num} created by the surface defect is the
transfer matrix
\begin{equation}
  \Tr_\W\bigl(\LM_n \circ_\W \dotsb \circ_\W \LM_2 \circ_\W
  \LM_1\bigr)
  \,.
\end{equation}
We conclude that the surface defect is represented in the lattice
model by the insertion of a transfer matrix constructed from
L-operators.

We can also use two dashed lines to make an R-operator
$\CRM_{ij}\colon \W_i \otimes \W_j \to \W_j \otimes \W_i$, which is
the Boltzmann weight for the lattice model constructed from dashed
lines.  In total, we have three R-operators:
\begin{equation}
  \RM_{ij}
  =
  \begin{tikzpicture}[scale=0.5, baseline=(x.base)]
    \node (x) at (0,1) {\vphantom{x}};
    \draw[r->] (0,1) node[left] {$i$} -- (2,1);
    \draw[r->] (1,0) node[below] {$j$} -- (1,2);
  \end{tikzpicture}
  \ ,
  \quad
  \LM_{ij}
  =
  \begin{tikzpicture}[scale=0.5, baseline=(x.base)]
    \node (x) at (0,1) {\vphantom{x}};

    \draw[dz->] (0,1) node[left] {$i$} -- (2,1);
    \draw[r->] (1,0) node[below] {$j$} -- (1,2);
  \end{tikzpicture}
  \ ,
  \quad
  \CRM_{ij}
  =
  \begin{tikzpicture}[scale=0.5, baseline=(x.base)]
    \node (x) at (0,1) {\vphantom{x}};

    \draw[dz->] (0,1) node[left] {$i$} -- (2,1);
    \draw[dz->] (1,0) node[below] {$j$} -- (1,2);
  \end{tikzpicture}
  \ .
\end{equation}
Correspondingly, we have four Yang--Baxter equations, involving $0$,
$1$, $2$, or $3$ dashed lines.  Those that contain dashed lines,
\begin{equation}
  \label{eq:RLL-1}
  \begin{tikzpicture}[scale=0.5, baseline=(x.base)]
    \node (x) at (30:2) {\vphantom{x}};

    \draw[r->] (0,0) node[left] {$1$} -- ++(30:3);
    \draw[dz->] (0,2) node[left] {$$} -- ++(-30:3);
    \draw[r->] (-30:1) node[below] {$2$} -- ++(0,3);
  \end{tikzpicture}
  \ =
  \begin{tikzpicture}[scale=0.5, baseline=(x.base)]
    \node (x) at (30:1) {\vphantom{x}};

    \draw[r->] (0,0) node[left] {$1$} -- ++(30:3);
    \draw[dz->] (0,1) node[left] {$$} -- ++(-30:3);
    \draw[r->] (-30:2) node[below] {$2$} -- ++(0,3);
  \end{tikzpicture}
  \ \Longleftrightarrow \
 \LM_1 \LM_2 \RM_{12} = \RM_{12} \LM_2 \LM_1
\end{equation}
and
\begin{equation}
  \label{eq:RLL-2}
  \begin{tikzpicture}[scale=0.5, baseline=(x.base)]
    \node (x) at (30:2) {\vphantom{x}};

    \draw[dz->] (0,0) node[left] {$2$} -- ++(30:3);
    \draw[dz->] (0,2) node[left] {$1$} -- ++(-30:3);
    \draw[r->] (-30:1) node[below] {$$} -- ++(0,3);
  \end{tikzpicture}
  \ =
  \begin{tikzpicture}[scale=0.5, baseline=(x.base)]
    \node (x) at (30:1) {\vphantom{x}};

    \draw[dz->] (0,0) node[left] {$2$} -- ++(30:3);
    \draw[dz->] (0,1) node[left] {$1$} -- ++(-30:3);
    \draw[r->] (-30:2) node[below] {$$} -- ++(0,3);
  \end{tikzpicture}
  \ \Longleftrightarrow \
  \CRM_{12} \LM_1 \LM_2 = \LM_2 \LM_1 \CRM_{12}
  \,,
\end{equation}
take the form of so-called RLL relations.  These relations, together
with the Yang--Baxter equation with solid lines only, imply that
transfer matrices~\eqref{eq:TM} and \eqref{eq:TM-L-num} commute among
themselves.  The last Yang--Baxter equation,
\begin{equation}
  \label{eq:YBE-CR}
  \begin{tikzpicture}[scale=0.5, baseline=(x.base)]
    \node (x) at (30:2) {\vphantom{x}};

    \draw[dz->] (0,0) node[left] {$2$} -- ++(30:3);
    \draw[dz->] (0,2) node[left] {$1$} -- ++(-30:3);
    \draw[dz->] (-30:1) node[below] {$3$} -- ++(0,3);
  \end{tikzpicture}
  \ =
  \begin{tikzpicture}[scale=0.5, baseline=(x.base)]
    \node (x) at (30:1) {\vphantom{x}};

    \draw[dz->] (0,0) node[left] {$2$} -- ++(30:3);
    \draw[dz->] (0,1) node[left] {$1$} -- ++(-30:3);
    \draw[dz->] (-30:2) node[below] {$3$} -- ++(0,3);
  \end{tikzpicture}
  \ \Longleftrightarrow \
  \CRM_{12} \CRM_{13} \CRM_{23}
  =
  \CRM_{23} \CRM_{13} \CRM_{12}
  \,,
\end{equation}
implies integrability of the lattice model on dashed lines.

We emphasize that the surface defect is represented by an object that
is defined locally near the dashed line.  Hence, the same dashed line
acts on the indices of any brane tiling models in the same way, as
long as the neighborhoods of the dashed line in the respective models
are topologically equivalent and the spectral parameters match.  This
locality holds even when we couple a brane tiling model to an
arbitrary 4d $\CN = 1$ theory by gauging appropriate flavor groups: to
compute the index of the combined theory in the presence of a surface
defect, we can first let the corresponding transfer matrix act on the
index of the brane tiling model, and then couple the result to the
index of the other theory by formula~\eqref{eq:gauging-I}.  This
property is a consequence of ``associativity'' of the gauging
operation.  The surface defect considered here may be thought of as a
2d $\CN = (0,2)$ theory coupled to a 4d $\CN = 1$ theory.  To insert
it in the combined 4d theory, we may first couple the 2d theory to a
brane tiling model and then couple the resulting 2d--4d system to the
other 4d theory.

\subsection{\texorpdfstring{Fundamental representation of
    $\SU(2)$}{Fundamental representation of SU(2)}}

Let us consider the simplest interesting setup where we have two
D5-branes and a single D3-brane, and identify the concrete form of the
transfer matrix~\eqref{eq:TM-L-num} in this case.  For $N=2$, $(N,q)$
and $(N,q+2)$ 5-branes are related by an $\SL(2; \Z)$ transformation
of type IIB string theory.  Therefore, we can go to a duality frame in
which the transfer matrix only involves either $(N,0)$ and $(N,-1)$
regions, or $(N,0)$ and $(N,1)$ regions.  The two cases are on an
equal footing, and in fact related in a simple way, as we will see.
We first consider the transfer matrix in the $(N,-1)$ background.

We denote the L-operator in this case by $\Ldia$ since it is the
operator that arises when a dashed line is inserted in a brane tiling
model described by the diamond quiver constructed from the
R-operator~\eqref{eq:RD}.  In the situation under consideration, the
gauge group of a brane tiling model is a product of $\SU(2)$ groups,
and the surface defect is labeled $(R_1, R_2) = (\emptyset, \yng(1))$;
the D3-brane wraps the circle $\{\zeta_2 = 0\}$ in $S^3$.  Thus,
$\Vdia$ is the space of meromorphic functions $f(z)$ such that
$f(z) = f(1/z)$, and $\W = \C^2$.  Accordingly, we can represent
$\Ldia\colon \W \otimes \Vdia \to \Vdia \otimes \W$ as a $2 \times 2$
matrix whose entries are operators acting on functions in $\Vdia$.
The R-operator
$\CRdia_{ij}\colon \W_i \otimes \W_j \to \W_j \otimes \W_i$ is a
$4 \times 4$ matrix.  These operators, together with the R-operator
$\Rdia$, satisfy the Yang--Baxter equations~\eqref{eq:YBE-G},
\eqref{eq:RLL-1}, \eqref{eq:RLL-2}, and~\eqref{eq:YBE-CR}.

Sklyanin constructed~\cite{Sklyanin:1983ig} an L-operator
$\LSkl\colon \W \otimes \Vdia \to \Vdia \otimes \W$ that solves the
RLL relation
\begin{equation}
  \RBax_{12}(u_1, u_2)
  \LSkl_1\bigl(u_1, (\nu,\ell)\bigr) \LSkl_2\bigl(u_2, (\nu,\ell)\bigr)
  =
   \LSkl_2\bigl(u_2, (\nu, \ell)\bigr) \LSkl_1\bigl(u_1, (\nu,\ell)\bigr)
   \RBax_{12}(u_1, u_2)
   \,,
\end{equation}
with Baxter's R-operator
$\RBax_{ij}\colon \W_i \otimes \W_j \to \W_j \otimes \W_i$ for the
eight-vertex model~\cite{Baxter:1971cr, Baxter:1972hz}.  Here $u_i$ is
a complex spectral parameter for $\W_i$, and $(\nu, \ell)$ is a pair
of complex spectral parameters for $\Vdia$.  Baxter's R-operator
satisfies the Yang--Baxter equation
\begin{equation}
  \RBax_{12}(u_1, u_2) \RBax_{13}(u_1, u_3) \RBax_{23}(u_2, u_3) 
  =
  \RBax_{23}(u_2, u_3) \RBax_{13}(u_1, u_3) \RBax_{12}(u_1, u_2)
  \,.
\end{equation}
If we write $\sigma_0$ for the $2\times 2$ unit matrix and $\sigma_1$,
$\sigma_2$, $\sigma_3$ for the Pauli matrices, then
\begin{equation}
  \RBax_{ij}(u_i, u_j)
  =
  P
  \sum_{a=0}^3 w_a(u_i - u_j) \sigma_a \otimes \sigma_a
  \,;
  \quad
  w_a(u) = \frac{\theta_{a+1}(u+\eta)}{\theta_{a+1}(\eta)}
  \,,
\end{equation}
where $P\colon \W_i \otimes \W_j \to \W_j \otimes \W_i$ is the
permutation operator, $\theta_{a+1}(\zeta) = \theta_{a+1}(\zeta|\tau)$
are the Jacobi theta functions, and $\tau$, $\eta$ are complex
parameters of the eight-vertex model.  Sklyanin's L-operator is
defined by
\begin{equation}
  \LSkl\bigl(u, (\nu,\ell)\bigr)
  =
  P
  \sum_{a = 0}^3 w_a(u + \nu) \sigma_a \otimes \mathbf{S}_a^{(\ell)}
  \,.
\end{equation}
The operators $\mathbf{S}_a^{(\ell)}$ act on meromorphic functions
$f(\zeta)$ as difference operators:
\begin{equation}
  \bigl(\mathbf{S}_a^{(\ell)} f\bigr)(\zeta)
  =
  i^{\delta_{a,2}} \frac{\theta_{a+1}(\eta)}{\theta_{a+1} (2\zeta)}
  \bigl(\theta_{a+1}(2\zeta - 2\eta\ell) f(\zeta+\eta)
  - \theta_{a+1}(-2\zeta - 2\eta\ell) f(\zeta-\eta)\bigr)
  \,.
\end{equation}
They generate the so-called Sklyanin algebra~\cite{Sklyanin:1982tf,
  Sklyanin:1983ig}.

In~\cite{Derkachov:2012iv}, Derkachov and Spiridonov constructed an
R-operator
$\RDS_{ij}\colon \Vdia_i \otimes \Vdia_j \to \Vdia_j \otimes \Vdia_i$
that satisfies the RLL relation%
\footnote{This RLL relation was considered for $u = 0$ in
 ~\cite{Derkachov:2012iv}.  The relation for general $u$ readily
  follows from this special case, since $\LSkl(u, (\nu, \ell)) =
  \LSkl(0, (\nu + u, \ell))$ and $\RDS_{12}$ is invariant under
  the overall shift $\nu_i \to \nu_i + u$.}
\begin{multline}
 \LDS_1\bigl(u, (\nu_1,\ell_1)\bigr) \LDS_2\bigl(u, (\nu_2,\ell_2)\bigr)
 \RDS_{12}\bigl((\nu_1,\ell_1), (\nu_2,\ell_2)\bigr)
 \\
 =
 \RDS_{12}\bigl((\nu_1,\ell_1), (\nu_2,\ell_2)\bigr)
 \LDS_2\bigl(u, (\nu_2,\ell_2)\bigr) \LDS_1\bigl(u,
 (\nu_1,\ell_1)\bigr)
 \,.
\end{multline}
The L-operator $\LDS\colon \W \otimes \Vdia \to \Vdia \otimes
\W$ is essentially Sklyanin's L-operator, differing only by an
automorphism of the Sklyanin algebra:
\begin{equation}
  \LDS\bigl(u, (\nu,\ell)\bigr)
  =
  \varphi \sigma_3 \LSkl\bigl(u, (\nu,\ell)\bigr) \varphi^{-1}
  \,;
  \quad
  (\varphi f)(\zeta) = \exp(\pi i\zeta^2/\eta) f(\zeta)
  \,.
\end{equation}
This L-operator also satisfies the RLL relation with $\RBax$.  We will
say more about the construction of Derkachov and Spiridonov in
section~\ref{sec:BS}.  At this point, what is important for us is that
their R-operator is precisely the R-operator~\eqref{eq:Rdia} for the
diamond quiver.  It turns out that%
\footnote{In the notation of~\cite{Derkachov:2012iv}, we have
  $\R_{12}(u_1,u_2|v_1,v_2) = \Rdia_{21}((a_2,b_2), (a_1,b_1)) P$, with
  $(a_1, b_1) = (\exp(-2\pi iu_2), \exp(-2\pi iu_1))$ and $(a_2, b_2) =
  (\exp(-2\pi iv_2), \exp(-2\pi iv_1))$.}
\begin{equation}
  \RDS_{ij}\bigl((\nu_i,\ell_i), (\nu_j,\ell_j)\bigr)
  =
  \Rdia_{ij}\bigl((a_i,b_i), (a_j,b_j)\bigr)
  \,,
\end{equation}
with the variables $\zeta$ and $z$ related by $z = \exp(2\pi i\zeta)$
and the parameters matched as
\begin{equation}
  a_i b_i = \exp(-2\pi  i\nu_i)
  \,,
  \quad
  \frac{a_i}{b_i} = \exp(2\pi i\eta(2\ell_i + 1))
  \,;
  \quad
  (p, q) = (\exp(2\pi i\tau), \exp(4\pi i\eta))
  \,.
\end{equation}

Based on this observation, we propose that the L-operator for the
diamond quiver
\begin{equation}
  \Ldia\bigl(c, (a,b)\bigr)
  =
  \begin{tikzpicture}[scale=0.5, baseline=(x.base)]
    \node (x) at (0,1) {\vphantom{x}};

    \fill[lshaded] (0,0) rectangle (2,2);
    \draw[dz->] (0,1) node[left] {$c$} -- (2,1);
    \draw[r->] (1,0) node[below] {$(a,b)$} -- (1,2);
  \end{tikzpicture}
\end{equation}
is the L-operator of Derkachov and Spiridonov:
\begin{equation}
  \Ldia\bigl(c, (a, b)\bigr) 
  =
  \LDS\bigl(u, (\nu, \ell)\bigr)
  \,.
\end{equation}
Requiring $\Ldia\bigl(c, (a, b)\bigr) = \Ldia\bigl(1, (a /c, b
/c)\bigr)$ fixes the relation between the two spectral parameters
for the dashed line to be
\begin{equation}
    c = \exp(\pi i u)
    \,.
\end{equation}
This requirement is natural from the point of view of the brane
construction.  In the M-theory picture, the spectral parameters are
identified with the values of $\exp(iX^{10})$ of the relevant branes.
Translation invariance implies that $\Ldia$ depends only on the
differences of the $X^{10}$ coordinates, and hence on the ratios of
the spectral parameters $a$, $b$, and $c$.

For the computation of the transfer matrix, we exploit the fact that
$\Ldia$ really consists of three parts separated by zigzag paths:
\begin{equation}
  \label{eq:Ldia-zz}
  \Ldia_i\bigl(c, (a_i,b_i)\bigr)
  =
  \begin{tikzpicture}[scale=0.5, baseline=(x.base)]
    \node (x) at (0,1) {\vphantom{x}};

    \fill[lshaded] (0,0) rectangle (2/3,2);
    \fill[lshaded] (4/3,0) rectangle (2,2);
    \draw[dz->] (0,1) node[left] {$c$} -- (2,1);
    \draw[z->] (2/3,0) node[below] {$a_i$}-- (2/3,2);
    \draw[z<-]   (4/3,0) -- (4/3,2) node[above] {$b_i$};
    \node at (1,1.5) {$z_i$};
  \end{tikzpicture}
  \ .
\end{equation}
Reflecting this structure, $\Ldia$ can be expressed in the following
factorized form:
\begin{equation}
  \label{eq:Ldia-fact}
  \Ldia_i\bigl(c, (a_i, b_i)\bigr)
  =
  B\biggl(z_i; \frac{b_i}{c}\biggr)
  \cdot
  \varphi(z_i)
  \frac{1}{\theta_1(z_i^2)}
  \Biggl(
  \begin{array}{cc}
    \Delta_i^{1/2} & 0 \\
    0 & \Delta_i^{-1/2}
  \end{array}
  \Biggr)
  \varphi^{-1}(z_i)
  \cdot
  A\biggl(z_i; \frac{a_i}{c}\biggr)
  \,.
\end{equation}
In this expression, $\Delta_i^{\pm 1/2}$ are difference operators
acting on functions of $z_i$ as
$(\Delta_i^{\pm 1/2} f)(z_i) = f(q^{\pm 1/2} z_i)$ and
\begin{equation}
  A(z;a)
  =
  \biggl(
  \begin{array}{cc}
    \thetab_4(a/z)
    & \thetab_3(a/z) \\
    \thetab_4(az)
    & \thetab_3(az)
  \end{array}
  \biggr),
  \quad
  B(z;b)
  =
  \biggl(
  \begin{array}{cc}
    \thetab_3(bz)
    & -\thetab_3(b/z) \\
    \thetab_4(bz)
    & -\thetab_4(b/z)
  \end{array}
  \biggr)
  \,,
\end{equation}
where $\thetab_a(z) = \theta_a(z; \sqrt{p})$ and we used the
multiplicative notation for the theta functions.  Roughly speaking, we
may think of the three matrices in the expression~\eqref{eq:Ldia-fact}
as corresponding to the left, middle and right parts of the above
diagram.

The transfer matrix~\eqref{eq:TM-L-num} is obtained by concatenating
$n$ copies of the object~\eqref{eq:Ldia-zz} along a loop:
\begin{equation}
  \begin{tikzpicture}[scale=0.5, baseline=(x.base)]
    \node (x) at (0,0) {\vphantom{x}};

    \begin{scope}[shift={(0,-1)}]
      \fill[lshaded] (-0.1,0) rectangle (0.5,2);
      \fill[lshaded] (1.5,0) rectangle (2.5,2);
      \fill[lshaded] (3.5,0) rectangle (4.05,2);
      \fill[lshaded] (4.95,0) rectangle (5.5,2);
      \fill[lshaded] (6.5,0) rectangle (7.1,2);

      \draw[z->] (0.5,0) node[below] {$a_1$} -- (0.5,2);
      \draw[z<-] (1.5,0) -- (1.5,2)
      node[above] {$b_1$};
      \node at (1,1.5) {$z_1$};

      \draw[z->] (2.5,0) node[below] {$a_2$} -- (2.5,2);
      \draw[z<-] (3.5,0) -- (3.5,2)
      node[above] {$b_2$};
      \node at (3,1.5) {$z_2$};

      \draw[z->] (5.5,0) node[below] {$a_n$} -- (5.5,2);
      \draw[z<-] (6.5,0) -- (6.5,2)
      node[above] {$b_n$};
      \node at (6,1.5) {$z_n$};
    \end{scope}

    \draw[dz->, right hook->] (-0.1,0)
    node[left] {$c$} -- (2.1,0);
    \draw[dz->, >=left hook] (2,0) -- (7.1,0);
    \node[fill=white, inner sep=1pt] at (4.5,0) {$\,\dots$};
  \end{tikzpicture}
  \ .
\end{equation}
Alternatively, we may place $n$ copies of
\begin{equation}
  \begin{tikzpicture}[scale=0.5, baseline=(c.base)]
    \fill[lshaded] (0,0) rectangle (1,2);
    \draw[dz->] (0,1) -- (2,1);
    \node[left] (c) at (0,1) {$c$};
    \draw[z<-] (0,0) -- (0,2) node[above] {$b_{i-1}$};
    \draw[z->] (1,0) node[below] {$a_i$} -- (1,2);
    \node at (1.5,1.5) {$z_i$};
  \end{tikzpicture}
  =
  \varphi(z_i)
  \frac{1}{\theta_1(z_i^2)}
  \biggl(
  \begin{array}{cc}
    \Delta_i^{1/2} & 0 \\
    0 & \Delta_i^{-1/2}
  \end{array}
  \biggr)
  \varphi^{-1}(z_i)
  \cdot
  A\biggl(z_i; \frac{a_i}{c}\biggr)
  \cdot
  B\biggl(z_{i-1}; \frac{b_{i-1}}{c}\biggr)
  \,.
\end{equation}
Using formulas in the appendix, we calculate its matrix elements and
find
\begin{equation}
  \makeatletter
  \newcommand{\biggg}{\bBigg@{5}}
  \makeatother
  \mathopen{\biggg(}
  \begin{tikzpicture}[scale=0.5, baseline=(c.base)]
    \fill[lshaded] (0,0) rectangle (1,2);
    \draw[dz->] (0,1) -- (2,1);
    \node[left] (c) at (0,1) {$c$};
    \draw[z<-] (0,0) -- (0,2) node[above] {$b_{i-1}$};
    \draw[z->] (1,0) node[below] {$a_i$} -- (1,2);
    \node at (1.5,1.5) {$z_i$};
  \end{tikzpicture}
  \mathclose{\biggg)}_{s_i s_{i-1}}
  =
  2i^{s_i + s_{i-1} + 1}
  \frac{(p;p)_\infty}{p^{1/8} q^{1/4}}
  \check l\biggl(z_{i-1}^{s_{i-1}}, z_i^{s_i}; \frac{b_{i-1}}{c}, \frac{a_i}{c}\biggr)
  \Delta_i^{s_i/2}
  \,,
\end{equation}
where $s_{i-1}$, $s_i$ take $\pm 1$ and we defined
\begin{equation}
  \check l\bigl(z_{i-1}, z_i; b_{i-1}, a_i\bigr)
  =
  \frac{1}{\theta(z_i^2)}
  \theta\biggl(\sqrt{\frac{p}{q}} b_{i-1} a_i \frac{z_{i-1}}{z_i}\biggr)
  \theta\biggl(\sqrt{\frac{p}{q}} \frac{a_i}{b_{i-1}} \frac{1}{z_{i-1}
    z_i}\biggr)
  \,.
\end{equation}
Multiplying $n$ copies of this matrix and then setting $z_0 = z_n$
and $b_0 = b_n$, we obtain the following formula for the transfer
matrix:
\begin{multline}
  \label{eq:TM-Ldia}
  \Tr_\W\Bigl(\Ldia_n\bigl(c, (a_n,b_n)\bigr) \circ_\W
  \dotsb \circ_\W \Ldia_1\bigl(c, (a_1,b_1)\bigr)\Bigr)
  \\
  =
  \sum_{s_1 = \pm 1}
  \dotso
  \sum_{s_n = \pm 1}
  \prod_{i=1}^n 
  \check l\biggl(z_{i-1}^{s_{i-1}}, z_i^{s_i}; \frac{b_{i-1}}{c}, \frac{a_i}{c}\biggr)
  \prod_{j=1}^n 
  \Delta_j^{s_j/2}
  \,.
\end{multline}
In this formula we have dropped an overall constant independent of the
spectral parameters.  At any rate, the overall normalization of the
L-operator cannot be determined by the RLL relations.

The RLL relations actually admit more degrees of freedom than just the
overall normalization.  For example, we can multiply $\Ldia(c, (a,b))$
by a function $f(c,(a,b))$ of its spectral parameters, and the
result still solves the RLL relations.  In sections~\ref{sec:class-S}
and~\ref{sec:class-Sk} we will check our proposal by comparing it with
independent computations from gauge theory.

With the knowledge of the transfer matrix in the $(N,-1)$ 5-brane
background, we can identify the transfer matrix in the $(N,1)$
background from the relation
\begin{equation}
  \begin{tikzpicture}[baseline=(x.base), xscale=1/3, yscale=2/3]
    \node (x) at (0,1) {\vphantom{x}};

    \def\lineA{(0,0) to[out=90, in=270] (1,2)}
    \def\lineB{(0,2) to[out=270, in=90] (1,0)}

    \begin{scope}[shift={(1,0)}]
      \begin{scope}
        \clip (0,0) rectangle (1,1);
        \fill[lshaded] \lineA -- (0,2) -- \lineB -- cycle;
      \end{scope}

      \begin{scope}
        \clip (0,1) rectangle (1,2);
        \fill[dshaded] \lineA -- (0,2) -- \lineB -- cycle;
      \end{scope}

      \draw[z<-] \lineA;
      \draw[z<-] \lineB;
      \node[above] at (1,2) {$b_1$};
      \node[below] at (1,0) {$a_2$};
    \end{scope}

    \begin{scope}[shift={(3,0)}]
      \begin{scope}
        \clip (0,0) rectangle (1,1);
        \fill[lshaded] \lineA -- (0,2) -- \lineB -- cycle;
      \end{scope}

      \begin{scope}
        \clip (0,1) rectangle (1,2);
        \fill[dshaded] \lineA -- (0,2) -- \lineB -- cycle;
      \end{scope}

      \draw[z<-] \lineA;
      \draw[z<-] \lineB;
      \node[above] at (1,2) {$b_2$};
      \node[below] at (1,0) {$a_3$};
    \end{scope}

    \begin{scope}[shift={(6,0)}]
      \begin{scope}
        \clip (0,0) rectangle (1,1);
        \fill[lshaded] \lineA -- (0,2) -- \lineB -- cycle;
      \end{scope}

      \begin{scope}
        \clip (0,1) rectangle (1,2);
        \fill[dshaded] \lineA -- (0,2) -- \lineB -- cycle;
      \end{scope}

      \draw[z<-] \lineA;
      \draw[z<-] \lineB;
      \node[above] at (1,2) {$b_n$};
      \node[below] at (1,0) {$a_1$};
    \end{scope}

    \draw[dz->, right hook->] (-0.1,0.5) node[left] {$c$} -- (2.6,0.5);
    \draw[dz->, >=left hook] (2.5,0.5) -- (8.1,0.5);
    \node[fill=white, inner sep=1pt] at (5,0.5) {$\,\dots$};
  \end{tikzpicture}
  \
  =
  \
  \begin{tikzpicture}[baseline=(x.base), xscale=1/3, yscale=2/3]
    \node (x) at (0,1) {\vphantom{x}};

    \def\lineA{(0,0) to[out=90, in=270] (1,2)}
    \def\lineB{(0,2) to[out=270, in=90] (1,0)}

    \begin{scope}[shift={(1,0)}]
      \begin{scope}
        \clip (0,0) rectangle (1,1);
        \fill[lshaded] \lineA -- (0,2) -- \lineB -- cycle;
      \end{scope}

      \begin{scope}
        \clip (0,1) rectangle (1,2);
        \fill[dshaded] \lineA -- (0,2) -- \lineB -- cycle;
      \end{scope}

      \draw[z<-] \lineA;
      \draw[z<-] \lineB;
      \node[above] at (1,2) {$b_1$};
      \node[below] at (1,0) {$a_2$};
    \end{scope}

    \begin{scope}[shift={(3,0)}]
      \begin{scope}
        \clip (0,0) rectangle (1,1);
        \fill[lshaded] \lineA -- (0,2) -- \lineB -- cycle;
      \end{scope}

      \begin{scope}
        \clip (0,1) rectangle (1,2);
        \fill[dshaded] \lineA -- (0,2) -- \lineB -- cycle;
      \end{scope}

      \draw[z<-] \lineA;
      \draw[z<-] \lineB;
      \node[above] at (1,2) {$b_2$};
      \node[below] at (1,0) {$a_3$};
    \end{scope}

    \begin{scope}[shift={(6,0)}]
      \begin{scope}
        \clip (0,0) rectangle (1,1);
        \fill[lshaded] \lineA -- (0,2) -- \lineB -- cycle;
      \end{scope}

      \begin{scope}
        \clip (0,1) rectangle (1,2);
        \fill[dshaded] \lineA -- (0,2) -- \lineB -- cycle;
      \end{scope}

      \draw[z<-] \lineA;
      \draw[z<-] \lineB;
      \node[above] at (1,2) {$b_n$};
      \node[below] at (1,0) {$a_1$};
    \end{scope}

    \draw[dz->, right hook->] (-0.1,1.5) node[left] {$c$} -- (2.6,1.5);
    \draw[dz->, >=left hook] (2.5,1.5) -- (8.1,1.5);
    \node[fill=white, inner sep=1pt] at (5,1.5) {$\,\dots$};
  \end{tikzpicture}
  \ ,
\end{equation}
which should hold according to our extra dimension argument.  This
relation says that the transfer matrices in the two backgrounds are
related by conjugation with a loop of bifundamental chiral multiplets.
Let us assign R-charge $R = 1$ to these multiplets.  A short
calculation shows
\begin{multline}
  \IB\biggl(z_i, z_{i-1}; \sqrt{pq} \frac{b_{i-1}}{a_i}\biggr)
  \check l\biggl(z_{i-1}^{s_{i-1}}, z_i^{s_i}; \frac{b_{i-1}}{c}, \frac{a_i}{c}\biggr)
  \Delta_{i-1}^{s_{i-1}/2}
  \Delta_i^{s_i/2}
  \\
  =
  \check l\biggl(z_{i-1}^{s_{i-1}}, z_i^{s_i}; \frac{a_i}{c}, \frac{b_{i-1}}{c}\biggr)
  \Delta_{i-1}^{s_{i-1}/2}
  \Delta_i^{s_i/2}
  \IB\biggl(z_i, z_{i-1}; \sqrt{pq} \frac{b_{i-1}}{a_i}\biggr)
  \,.
\end{multline}
From this equation we see that the conjugation just exchanges
$b_{i-1}$ and $a_i$ in the transfer matrix~\eqref{eq:TM-Ldia}.  Thus,
we conclude
\begin{equation}
  \label{eq:TM-Ldiabar}
  \begin{tikzpicture}[scale=0.5, baseline=(x.base)]
    \node (x) at (0,0) {\vphantom{x}};

    \begin{scope}[shift={(0,-1)}]
      \fill[dshaded] (-0.1,0) rectangle (0.5,2);
      \fill[dshaded] (1.5,0) rectangle (2.5,2);
      \fill[dshaded] (3.5,0) rectangle (4.05,2);
      \fill[dshaded] (4.95,0) rectangle (5.5,2);
      \fill[dshaded] (6.5,0) rectangle (7.1,2);

      \draw[z<-] (0.5,0) -- (0.5,2)
      node[above] {$b_n$};
      \draw[z->] (1.5,0) node[below] {$a_2$} -- (1.5,2);
      \node at (1,1.5) {$z_1$};

      \draw[z<-] (2.5,0) -- (2.5,2)
      node[above] {$b_1$};
      \draw[z->] (3.5,0) node[below] {$a_3$} -- (3.5,2);
      \node at (3,1.5) {$z_2$};

      \draw[z<-] (5.5,0) -- (5.5,2)
      node[above] {$b_{n-1}$};
      \draw[z->] (6.5,0) node[below] {$a_1$} -- (6.5,2);
      \node at (6,1.5) {$z_n$};
    \end{scope}

    \draw[dz->, right hook->] (-0.1,0) node[left] {$c$} -- (2.1,0);
    \draw[dz->, >=left hook] (2,0) -- (7.1,0);
    \node[fill=white, inner sep=1pt] at (4.5,0) {$\,\dots$};
  \end{tikzpicture}
  \ =
  \sum_{(s_i) \in \{\pm 1\}^n}
  \prod_{i=1}^n 
  \check l\biggl(z_{i-1}^{s_{i-1}}, z_i^{s_i}; \frac{a_i}{c}, \frac{b_{i-1}}{c}\biggr)
  \prod_{j=1}^n 
  \Delta_j^{s_j/2}
  \,.
\end{equation}

So far we have treated the surface defect labeled
$(R_1, R_2) = (\emptyset, \yng(1))$.  Of course, we may also consider
the case with $(R_1, R_2) = (\yng(1), \emptyset)$ in the same manner,
by letting surface defects wrap around the other $S^1$ inside $S^3$.
Hence, there are two sets of L-operators related by the symmetry
exchanging $p$ and $q$.  The underlying algebraic structure is the
product of two copies of the Sklyanin algebra, known as the
\emph{elliptic modular double}~\cite{MR2492363}.

\subsection{Relation to the Bazhanov--Sergeev model}
\label{sec:BS}

The reason that we introduced the thick line~\eqref{eq:solid-line} by
pairing up two zigzag paths was that brane tiling diagrams constructed
using this line do not contain regions supporting $(N,q)$ 5-branes
with $|q| > 1$.  If those regions are present, in general we do not
have a description of the 4d theory in terms of a quiver and hence
cannot use the formula~\eqref{eq:I-quiver} for the supersymmetric
index.  Since the Yang--Baxter equation for three zigzag paths always
involves undesirable regions, simply restricting ourselves to the
quiver case is not sufficient for checking the integrability of the
model explicitly.

The situation is different when the number of D5-branes, $N = 2$.  In
this case, any $(N, q)$ 5-brane falls into one of two equivalence
classes under the $\SL(2;\Z)$ duality of type IIB string theory:
either $(N,0)$ or $(N,1)$ 5-brane, which we may visualize as an
unshaded or shaded region.  Every unshaded region generates an
$\SU(2)$ gauge or flavor group.  This fact raises the hope that a
general brane tiling with $N = 2$ leads to a quiver gauge theory.  If
so, there should be the corresponding integrable lattice model whose
R-matrix is made out of the bifundamental factor $\IB$.

For brane tilings on flat surfaces, such an integrable lattice model
was indeed discovered by Bazhanov and Sergeev
in~\cite{Bazhanov:2010kz}.  Given a brane tiling, we can map it to the
Bazhanov--Sergeev model as follows.  First of all, we assume that we
can deform the zigzag paths so that each of them heads either upward
or downward and its slope is never zero (taking the $X^6$-direction as
horizontal and the $X^4$-direction as vertical, say).  With this
assumption, the orientations of zigzag paths are actually irrelevant
for the lattice model, so we omit them from the brane tiling diagram.
Then the diagram consists of two building blocks, and we assign
quivers to them:
\begin{align}
  \label{eq:RBS-1}
  \begin{tikzpicture}[baseline=(z.base)]
    \fill[shaded] (0,0) -- (1,1) -- (0,1) -- (1,0) -- cycle;
    \draw[z-] (0,0) node[below] {$a$} -- (1,1);
    \draw[z-] (1,0) node[below] {$b$} -- (0,1);
    \node (z) at (0.1,0.5) {$z$};
    \node at (0.9,0.5) {$w$};
  \end{tikzpicture}
  &= \
  \begin{tikzpicture}[baseline=(z.base)]
    \node[fnode] (z) at (0,0) {$z$};
    \node[fnode] (w) at (1,0) {$w$};
    \draw[q-] (z) -- node[above=4pt] {$\sqrt{pq} a/b$} (w);
  \end{tikzpicture}
  \ =
  \IB\biggl(z, w; \sqrt{pq} \frac{a}{b}\biggr),
  \\
  \label{eq:RBS-2}
  \begin{tikzpicture}[baseline=(x.base)]
    \node (x) at (0,0.5) {\vphantom{x}};
    \fill[shaded] (0,0) -- (1,1) -- (1,0) -- (0,1) -- cycle;
    \draw[z-] (0,0) node[below] {$a$} -- (1,1);
    \draw[z-] (1,0) node[below] {$b$} -- (0,1);
    \node (z) at (0.5,0.1) {$z$};
    \node at (0.5,0.9) {$w$};
  \end{tikzpicture}
  &= \
  \begin{tikzpicture}[baseline=(x.base)]
    \node (x) at (0,0.5) {\vphantom{x}};
    \node[fnode] (z) at (0,0) {$z$};
    \node[fnode] (w) at (0,1) {$w$};
    \draw[q-] (z) -- node[right] {$b/a$} (w);
  \end{tikzpicture}
  =
  \IBt\biggl(z, w; \frac{b}{a}\biggr)
    \,.
\end{align}
We can also drop orientation from arrows since the fundamental
representation of $\SU(2)$ is pseudoreal.  Finally, we define the
partition function of the lattice model by the supersymmetric index of
the quiver gauge theory obtained in this way.  As usual, we use the
normalized factor~\eqref{eq:It_B} for bifundamental chiral multiplets
with R-charge $R = 0$.  Note that $\IB(z, w; u)$ is symmetric under
exchange of $z$ and $w$, as is consistent with the fact that arrows
are unoriented.  \def\lineA{(0,0) .. controls (1,1) .. (0,2)}
\def\lineB{(1,2) .. controls (0,1) .. (1,0)}

The R-operators defined above satisfy the relations
\begin{equation}
  \begin{tikzpicture}[baseline=(z.base), yscale=0.8]
    \fill[shaded] \lineA -- \lineB -- cycle;

    \draw[z-] \lineA;
    \draw[z-] \lineB;

    \node[below] at (0,0) {$a$};
    \node[below] at (1,0) {$b$};
    \node (z) at (0.1,1) {$z$};
    \node (w) at (0.9,1) {$w$};
  \end{tikzpicture}
  = \
  \begin{tikzpicture}[baseline=(z.base)]
    \node[fnode] (z) at (0,0) {$z$};
    \node[fnode] (w) at (1,0) {$w$};
    \draw[q-] (z) to[bend right] node[below] {$\sqrt{pq} a/b$} (w);
    \draw[q-] (z) to[bend left] node[above] {$\sqrt{pq} b/a$} (w);
  \end{tikzpicture}
  \ = \
  \begin{tikzpicture}[baseline=(z.base)]
    \node[fnode] (z) at (0,0) {$z$};
    \node[fnode] (w) at (1,0) {$w$};
  \end{tikzpicture}
  \ =
  \begin{tikzpicture}[baseline=(z.base), yscale=0.8]
    \fill[shaded] (1/3,0) rectangle (2/3,2);

    \draw[z-] (1/3,0) node[below] {$a$} -- (1/3,2);
    \draw[z-] (2/3,0) node[below] {$b$} -- (2/3,2);

    \node (z) at (0,1) {$z$};
    \node (w) at (1,1) {$w$};
  \end{tikzpicture}  
\end{equation}
and
\begin{equation}
  \begin{tikzpicture}[baseline=(w.base), yscale=0.8]
    \begin{scope}
      \clip \lineA -- (1,2) -- (1,0) -- cycle;
      \fill[shaded] \lineB;
    \end{scope}
    \begin{scope}
      \clip \lineB -- (0,0) -- (0,2) -- cycle;
      \fill[shaded] \lineA;
    \end{scope}

    \draw[z-] \lineA;
    \draw[z-] \lineB;

    \node[below] at (0,0) {$a$};
    \node[below] at (1,0) {$b$};

    \node at (0.5,0.1) {$z$};
    \node (w) at (0.5,1) {$w$};
    \node at (0.5,1.9) {$x$};
  \end{tikzpicture}
  = \
  \begin{tikzpicture}[baseline=(w.base), yscale=0.8]
    \node[fnode] (z) at (0.5,0) {$z$};
    \node[fnode] (w) at (0.5,1) {$w$};
    \node[fnode] (x) at (0.5,2) {$x$};
  
    \draw[q-] (z) -- node[right] {$b/a$} (w);
    \draw[q-] (w) -- node[right] {$a/b$} (x);
  \end{tikzpicture}
  = \
  \begin{tikzpicture}[baseline=(w.base), yscale=0.8]
    \node (w) at (0.5,1) {\vphantom{w}};
    \node[fnode] (z) at (0.5,0) {$z$};
    \node[fnode] (x) at (0.5,2) {$x$};
    \draw[eq-] (z) -- (x);
  \end{tikzpicture}
  \ = \
  \begin{tikzpicture}[baseline=(w.base), yscale=0.8]
    \node (w) at (0.5,1) {\vphantom{w}};
    \fill[shaded] (0,0) rectangle (1/3, 2);
    \fill[shaded] (2/3, 0) rectangle (1,2);

    \draw[z-] (1/3, 0) node[below] {$a$} -- (1/3, 2);
    \draw[z-] (2/3, 0) node[below] {$b$} -- (2/3, 2);

    \node (z) at (0.5,0.1) {$z$};
    \node (x) at (0.5,1.9) {$x$};
  \end{tikzpicture}
  \ .
\end{equation}
Furthermore, they solve the Yang--Baxter equation
\begin{equation}
  \label{eq:STR}
  \def\th{atan(sqrt(3)/5)}
  \begin{tikzpicture}[baseline=(w.base)]
    \fill[shaded] (0,0) circle [radius=1];

    \node[zerosep] (1) at ({-\th}:1) {};
    \node[zerosep] (2) at ({\th}:1) {};
    \node[zerosep] (3) at ({120-\th}:1) {};
    \node[zerosep] (4) at ({120+\th}:1) {};
    \node[zerosep] (5) at ({240-\th}:1) {};
    \node[zerosep] (6) at ({240+\th}:1) {};
    \node[zerosep] (7) at (0:{sqrt(2)*4/15} ) {};
    \node[zerosep] (8) at (120:{sqrt(2)*4/15} ) {};
    \node[zerosep] (9) at (240:{sqrt(2)*4/15} ) {};

    \filldraw[white] (1) arc (-{\th}:{\th}:1) -- (7) -- cycle;
    \filldraw[white] (3) arc ({120-\th}:{120+\th}:1) -- (8) -- cycle;
    \filldraw[white] (5) arc ({240-\th}:{240+\th}:1) -- (9) -- cycle;
    \filldraw[white] (0:{sqrt(2)*4/15} ) -- (120:{sqrt(2)*4/15} ) --
    (240:{sqrt(2)*4/15} ) -- cycle;

    \draw[-] (6) node[below] {$a$} -- (3);
    \draw[-] (1) node[below right=-2pt] {$b$} -- (4);
    \draw[-] (5) node[below left=-2pt] {$c$} -- (2);

    \node (x) at (120:0.8) {$x$};
    \node (y) at (0:0.8) {$y$};
    \node (z) at (240:0.8) {$z$};
    \node (w) at (0,0) {$w$};
  \end{tikzpicture}
  =
  \begin{tikzpicture}[baseline=(w.base), rotate=-60]
    \node[zerosep] (1) at ({-\th}:1) {};
    \node[zerosep] (2) at ({\th}:1) {};
    \node[zerosep] (3) at ({120-\th}:1) {};
    \node[zerosep] (4) at ({120+\th}:1) {};
    \node[zerosep] (5) at ({240-\th}:1) {};
    \node[zerosep] (6) at ({240+\th}:1) {};
    \node[zerosep] (7) at (0:{sqrt(2)*4/15} ) {};
    \node[zerosep] (8) at (120:{sqrt(2)*4/15} ) {};
    \node[zerosep] (9) at (240:{sqrt(2)*4/15} ) {};

    \fill[shaded] (1) arc (-{\th}:{\th}:1) -- (7) -- cycle;
    \fill[shaded] (3) arc ({120-\th}:{120+\th}:1) -- (8) -- cycle;
    \fill[shaded] (5) arc ({240-\th}:{240+\th}:1) -- (9) -- cycle;
    \fill[shaded] (0:{sqrt(2)*4/15} ) -- (120:{sqrt(2)*4/15} ) --
    (240:{sqrt(2)*4/15} ) -- cycle;

    \draw[-] (1) node[below] {$a$} -- (4);
    \draw[-] (2) node[below right=-2pt] {$b$} -- (5);
    \draw[-] (6) node[below left=-2pt] {$c$} -- (3);

    \node (x) at (180:0.8) {$x$};
    \node (y) at (60:0.8) {$y$};
    \node (z) at (300:0.8) {$z$};
    \node (w) at (0,0) {\vphantom{$w$}};
  \end{tikzpicture}
  \,.
\end{equation}
This ``star-triangle'' relation is a consequence of
identity~\eqref{eq:elliptic-beta} and expresses the RG flow from
$\SU(2)$ SQCD with three flavors to the infrared theory:
\begin{equation}
  \def\th{atan(sqrt(3)/5)}
  \begin{tikzpicture}[baseline=(w.base)]
    \node[fnode] (x) at (120:0.8) {$x$};
    \node[fnode] (y) at (0:0.8) {$y$};
    \node[fnode] (z) at (240:0.8) {$z$};
    \node[gnode] (w) at (0,0) {$w$};

    \draw[q-] (w) -- node[below left=-4pt] {$b/a$} (x);
    \draw[q-] (w) -- node[above=4pt] {$\sqrt{pq} c/b$} (y);
    \draw[q-] (w) -- node[below right=-4pt] {$a/c$} (z);
  \end{tikzpicture}
  \ =
  \begin{tikzpicture}[baseline=(w.base), rotate=-60]
    \node[fnode] (x) at (180:0.8) {$x$};
    \node[fnode] (y) at (60:0.8) {$y$};
    \node[fnode] (z) at (300:0.8) {$z$};
    \node (w) at (0,0) {\vphantom{$w$}};

    \draw[q-] (x) -- node[above right=-4pt] {$\sqrt{pq} c/a$} (y);
    \draw[q-] (y) -- node[below right=-4pt] {$\sqrt{pq} a/b$} (z);
    \draw[q-] (z) -- node[left] {$b/c$}  (x);
  \end{tikzpicture}
  \,.
\end{equation}
As expected, the Yang--Baxter equation holds at the level of zigzag
paths.  It implies the Yang--Baxter equation for the
R-operators~\eqref{eq:RD} and~\eqref{eq:RT}.

Similarly, the RLL relation~\eqref{eq:RLL-1} follows from two
Yang--Baxter equations involving a dashed line, namely
\begin{equation}
  \label{eq:YBE-zzd-1}
  \def\lineA{(0,0) to[out=90, in=270] (1,2)}
  \def\lineB{(0,2) to[out=270, in=90] (1,0)}
  \begin{tikzpicture}[baseline=(x.base), scale=2/3]
    \node (x) at (1.5,1) {\vphantom{x}};

    \fill[shaded] (-0.5,0) rectangle (1.5, 2);
    \filldraw[white] \lineA -- (0,2) -- \lineB -- cycle;
    \draw[z-] \lineA;
    \draw[z-] \lineB;
    \draw[dz->] (-0.5,0.5) node[left] {$c$} -- (1.5,0.5);
    \node[below] at (0,0) {$a$};
    \node[below] at (1,0) {$b$};
  \end{tikzpicture}
  \ =
  \begin{tikzpicture}[baseline=(x.base), scale=2/3]
    \node (x) at (1.5,1) {\vphantom{x}};

    \fill[shaded] (-0.5,0) rectangle (1.5, 2);
    \filldraw[white] \lineA -- (0,2) -- \lineB -- cycle;
    \draw[z-] \lineA;
    \draw[z-] \lineB;
    \draw[dz->] (-0.5,1.5) node[left] {$c$} -- (1.5,1.5);
    \node[below] at (0,0) {$a$};
    \node[below] at (1,0) {$b$};
  \end{tikzpicture}
\end{equation}
and another one obtained by flipping the shaded and unshaded regions,
\begin{equation}
  \label{eq:YBE-zzd-2}
  \def\lineA{(0,0) to[out=90, in=270] (1,2)}
  \def\lineB{(0,2) to[out=270, in=90] (1,0)}
  \begin{tikzpicture}[baseline=(x.base), scale=2/3]
    \node (x) at (1.5,1) {\vphantom{x}};

    \fill[shaded] \lineA -- (0,2) -- \lineB -- cycle;
    \draw[z-] \lineA;
    \draw[z-] \lineB;
    \draw[dz->] (-0.5,0.5) node[left] {$c$} -- (1.5,0.5);
    \node[below] at (0,0) {$a$};
    \node[below] at (1,0) {$b$};
  \end{tikzpicture}
  \ =
  \begin{tikzpicture}[baseline=(x.base), scale=2/3]
    \node (x) at (1.5,1) {\vphantom{x}};

    \fill[shaded] \lineA -- (0,2) -- \lineB -- cycle;
    \draw[z-] \lineA;
    \draw[z-] \lineB;
    \draw[dz->] (-0.5,1.5) node[left] {$c$} -- (1.5,1.5);
    \node[below] at (0,0) {$a$};
    \node[below] at (1,0) {$b$};
  \end{tikzpicture}
  \,.
\end{equation}

Following~\cite{Derkachov:2012iv}, let us define an operator $M(a,b)
\in \End(\Vdia)$ by
\begin{equation}
  \bigl(M(a,b) f\bigr)(w)
  =
  \int_\T \frac{\rmd z}{2\pi iz} \IV(z)
  \IBt\biggl(z, w; \frac{b}{a}\biggr) f(z)
  \,.
\end{equation}
Graphically, the action of $M$ concatenates the
crossing~\eqref{eq:RBS-2} to an unshaded region.  Using this operator,
we can write the first Yang--Baxter equation~\eqref{eq:YBE-zzd-1} as
\begin{equation}
  M(a,b) \Ldia\bigl(c, (a,b)\bigr)
  =
  \Ldia\bigl(c, (b,a)\bigr) M(a,b)
  \,.
\end{equation}
We see that $M$ acts on $\Ldia$ by interchange of the spectral
parameters $a$ and $b$.  With a slight modification, the second
Yang--Baxter equation~\eqref{eq:YBE-zzd-2} can also be expressed as an
identity obeyed by $\Ldia$:
\begin{equation}
  \begin{tikzpicture}[baseline=(x.base), xscale=1/3, yscale=2/3]
    \node (x) at (1.5,1) {\vphantom{x}};

    \def\lineA{(0,0) to[out=90, in=270] (1,2)}
    \def\lineB{(0,2) to[out=270, in=90] (1,0)}

    \begin{scope}[shift={(1.75,0)}]
      \fill[shaded] \lineA -- (0,2) -- \lineB -- cycle;
      \draw[z-] \lineA;
      \draw[z-] \lineB;
      \node[below] at (0,0) {$b_1$};
      \node[below] at (1,0) {$a_2$};
    \end{scope}

    \fill[shaded] (0,0) rectangle (0.75,2);
    \draw[z-] (0.75,0) node[below] {$a_1$}-- (0.75,2);

    \begin{scope}[shift={(3.75,0)}]
      \fill[shaded] (0,0) rectangle (0.75,2);
      \draw[z-] (0,0) node[below] {$b_2$}-- (0,2);
    \end{scope}
  
    \draw[dz->] (0,0.5) node[left] {$c$} -- (4.5,0.5);
  \end{tikzpicture}
  \ =
  \begin{tikzpicture}[baseline=(x.base), xscale=1/3, yscale=2/3]
    \node (x) at (1.5,1) {\vphantom{x}};

    \def\lineA{(0,0) to[out=90, in=270] (1,2)}
    \def\lineB{(0,2) to[out=270, in=90] (1,0)}

    \begin{scope}[shift={(1.75,0)}]
      \fill[shaded] \lineA -- (0,2) -- \lineB -- cycle;
      \draw[z-] \lineA;
      \draw[z-] \lineB;
      \node[below] at (0,0) {$b_1$};
      \node[below] at (1,0) {$a_2$};
    \end{scope}

    \fill[shaded] (0,0) rectangle (0.75,2);
    \draw[z-] (0.75,0) node[below] {$a_1$}-- (0.75,2);

    \begin{scope}[shift={(3.75,0)}]
      \fill[shaded] (0,0) rectangle (0.75,2);
      \draw[z-] (0,0) node[below] {$b_2$}-- (0,2);
    \end{scope}
  
    \draw[dz->] (0,1.5) node[left] {$c$} -- (4.5,1.5);
  \end{tikzpicture}
  \ .
\end{equation}
So if we define $S_2(b_1, a_2) \in \End(\Vdia_1 \otimes \Vdia_2)$ by
multiplication with $\IB(z_1, z_2; \sqrt{pq} b_1/a_2)$, or more
graphically,
\begin{equation}
  S_2(b_1, a_2)
  = \
  \begin{tikzpicture}[baseline=(x.base), xscale=1/3, yscale=2/3]
    \node (x) at (1.5,1) {\vphantom{x}};

    \def\lineA{(0,0) to[out=90, in=270] (1,2)}
    \def\lineB{(0,2) to[out=270, in=90] (1,0)}

    \begin{scope}[shift={(1.75,0)}]
      \fill[shaded] \lineA -- (0,2) -- \lineB -- cycle;
      \draw[z-] \lineA;
      \draw[z-] \lineB;
      \node[below] at (0,0) {$b_1$};
      \node[below] at (1,0) {$a_2$};
    \end{scope}

    \fill[shaded] (0,0) rectangle (0.75,2);
    \draw[z-] (0.75,0) node[below] {$a_1$}-- (0.75,2);

    \begin{scope}[shift={(3.75,0)}]
      \fill[shaded] (0,0) rectangle (0.75,2);
      \draw[z-] (0,0) node[below] {$b_2$}-- (0,2);
    \end{scope}
  \end{tikzpicture}
  \ ,
\end{equation}
then its action on $\Ldia_2 \Ldia_1$ interchanges $b_1$ and $a_2$:
\begin{equation}
  S_2(b_1, a_2) \Ldia_2\bigl(c, (a_2, b_2)\bigr) \Ldia_1\bigl(c, (a_1, b_1)\bigr)
  =
  \Ldia_2\bigl(c, (b_1, b_2)\bigr) \Ldia_1\bigl(c, (a_1, a_2)\bigr)
  S_2(b_1, a_2)
  \,. 
\end{equation}

Let us further set $S_1(a_1, b_1) = M_1(a_1, b_1)$ and $S_3(a_2, b_2)
= M_2(a_2, b_2)$, or
\begin{equation}
  S_1(a_1, b_1)
  = \
  \begin{tikzpicture}[baseline=(x.base), xscale=1/3, yscale=2/3]
    \node (x) at (1.5,1) {\vphantom{x}};

    \def\lineA{(0,0) to[out=90, in=270] (1,2)}
    \def\lineB{(0,2) to[out=270, in=90] (1,0)}

    \fill[shaded] (0,0) rectangle (4.5,2);

    \begin{scope}[shift={(0.75,0)}]
      \filldraw[white] \lineA -- (0,2) -- \lineB -- cycle;
      \draw[z-] \lineA;
      \draw[z-] \lineB;
      \node[below] at (0,0) {$a_1$};
      \node[below] at (1,0) {$b_1$};
    \end{scope}

    \begin{scope}[shift={(2.75,0)}]
      \filldraw[white] (0,0) rectangle (1,2);
      \draw[z-] (0,0) node[below] {$a_2$}-- (0,2);
      \draw[z-] (1,0) node[below] {$b_2$}-- (1,2);
    \end{scope}
  \end{tikzpicture}
  \ ,
  \quad
  S_3(a_2, b_2)
  = \
  \begin{tikzpicture}[baseline=(x.base), xscale=1/3, yscale=2/3]
    \node (x) at (1.5,1) {\vphantom{x}};

    \def\lineA{(0,0) to[out=90, in=270] (1,2)}
    \def\lineB{(0,2) to[out=270, in=90] (1,0)}

    \fill[shaded] (0,0) rectangle (4.5,2);

    \begin{scope}[shift={(2.75,0)}]
      \filldraw[white] \lineA -- (0,2) -- \lineB -- cycle;
      \draw[z-] \lineA;
      \draw[z-] \lineB;
      \node[below] at (0,0) {$a_2$};
      \node[below] at (1,0) {$b_2$};
    \end{scope}

    \begin{scope}[shift={(0.75,0)}]
      \filldraw[white] (0,0) rectangle (1,2);
      \draw[z-] (0,0) node[below] {$a_1$}-- (0,2);
      \draw[z-] (1,0) node[below] {$b_1$}-- (1,2);
    \end{scope}
  \end{tikzpicture}
  \ .
\end{equation}
Then, $S_1$, $S_2$, $S_3$ correspond to generators $s_1$, $s_2$, $s_3$
of the symmetric group $\mathfrak{S}_4$ permuting a quadruple of
fugacities $\mathbf{a} = (a_1, b_1, a_2, b_2)$, and by the
Yang--Baxter equations, act as such on $\Ldia_2 \Ldia_1(c, \mathbf{a})
= \Ldia_2(c, (a_2, b_2)) \Ldia_1(c, (a_1, b_1))$:
\begin{align}
  \label{SLL}
  S_1(\mathbf{a}) \Ldia_2 \Ldia_1(c, \mathbf{a})
  &=
  \Ldia_2 \Ldia_1(c, s_1\mathbf{a}) S_1(\mathbf{a})
    \,;
  \quad
  s_1 \mathbf{a} = (b_1, a_1, a_2, b_2)
    \,,
  \\
  S_2(\mathbf{a}) \Ldia_2 \Ldia_1(c, \mathbf{a})
  &=
  \Ldia_2 \Ldia_1(c, s_2\mathbf{a}) S_2(\mathbf{a})
    \,;
  \quad
  s_2 \mathbf{a} = (a_1, a_2, b_1, b_2)
    \,,
  \\
  S_3(\mathbf{a}) \Ldia_2 \Ldia_1(c, \mathbf{a})
  &=
  \Ldia_2 \Ldia_1(c, s_3\mathbf{a}) S_3(\mathbf{a})
    \,;
  \quad
  s_3 \mathbf{a} = (a_1, b_1, b_2, a_2)
    \,.
\end{align}
These permutation operators were used in~\cite{Derkachov:2012iv} to
construct the R-operator $\RDS = \Rdia$, which satisfies the RLL
relation~\eqref{eq:RLL-1}.  In fact, we have
\begin{equation}
  \Rdia_{12}(\mathbf{a})
  =
  \begin{tikzpicture}[baseline=(x.base), scale=2/3]
    \node (x) at (1,0.75) {\vphantom{x}};

    \fill[shaded] (0.5,0) -- (1.25,0.75) -- (0.5,1.5) -- (1.5,1.5) --
    (0.75,0.75) -- (1.5,0) -- cycle;
    \fill[shaded] (0,0) -- (0.75,0.75) -- (0,1.5) -- cycle;
    \fill[shaded] (2,0) -- (1.25,0.75) -- (2,1.5) -- cycle;
    \draw[z-] (0,0) node[below] {$a_1$} -- (1.5,1.5);
    \draw[z-] (0.5,0) node[below] {$b_1$} -- (2,1.5);
    \draw[z-] (1.5,0) node[below] {$a_2$} -- (0,1.5);
    \draw[z-] (2,0) node[below] {$b_2$} -- (0.5,1.5);
  \end{tikzpicture}
  =
  \mathbb{P}_{12}
  S_2(s_1s_3s_2\mathbf{a}) S_1(s_3s_2\mathbf{a})
  S_3(s_2\mathbf{a}) S_2(\mathbf{a})
  \,,
\end{equation}
where $\mathbb{P}_{12}$ acts on a function $f(z_1, z_2)$ as
$(\mathbb{P}_{12} f)(z_1, z_2) = f(z_2, z_1)$.

As mentioned already, this R-operator satisfies the Yang--Baxter
equation thanks to the star-triangle relation~\eqref{eq:STR}.  In the
operator form used here, the last relation arises naturally from the
Bailey lemma proved in~\cite{MR2076912}.  Its higher-rank
generalization~\cite{MR2264067} leads to a web of dualities connecting
4d $\CN = 1$ quiver gauge theories~\cite{Brunner:2016uvv}.

\section{\texorpdfstring{Surface defects in $A_1$ theories of class
    $\CS$}{Surface defects in A1 theories of class S}}
\label{sec:class-S}

Now we aim to check our proposal on surface defects and transfer
matrices by comparing it with independent computations.  In this
section we perform the simplest such check for surface defects in
$A_1$ theories of class $\CS$~\cite{Gaiotto:2009we, Gaiotto:2009hg},
which arise from compactification of the 6d $\CN = (2,0)$ theory of
type $A_1$ on punctured Riemann surfaces.  The action of surface
defects on the supersymmetric indices of class-$\CS$ theories have
been studied before~\cite{Gaiotto:2012xa, Gadde:2013ftv,
  Alday:2013kda, Bullimore:2014nla}.  Here we review the computation
for the surface defect labeled with the fundamental representation of
$\SU(2)$ based on the method developed in~\cite{Gaiotto:2009we}, and
show that the result agrees with the prediction from the transfer
matrix~\eqref{eq:TM-Ldia}.

\subsection{\texorpdfstring{$\CN = 2$ linear and circular quiver
    theories}{N = 2 linear and circular quiver theories}}

Prototypical examples of class-$\CS$ theories are $\CN = 2$ gauge
theories characterized by linear and circular quivers with $\SU(N)$
nodes.  They are actually also examples of brane tiling models
discussed in the previous sections.  As such, they allow us to
translate key notions in class-$\CS$ theories to the language of brane
tilings, and vice versa.  Our first task is to describe these theories
as class-$\CS$ theories as well as brane tiling models, and understand
the relation between the two descriptions.  Although we are mainly
interested in the case with $N = 2$, for now we keep $N$ general.

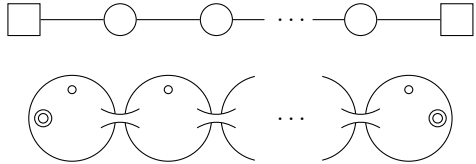
\begin{figure}
  \centering
  \begin{tikzpicture}[baseline=(x)]
    \node[fnode] (a0) at (0,0) {};
    \node[gnode] (a1) at (1,0) {};
    \node[gnode] (a2) at (2,0) {};

    \begin{scope}[shift={(0.5,0)}]
      \node[gnode] (a3) at (3,0) {};
      \node[fnode] (a4) at (4,0) {};
    \end{scope}

    \draw[q-] (a0) -- (a1);
    \draw[q-] (a1) -- (a2);
    \draw[q-] (a3) -- (a4);

    \draw[q-] (a2) -- ($(a2) + (0.5,0)$);
    \draw[q-] ($(a3) + (-0.5,0)$) -- (a3);

    \node (x) at (2.8,0) {$\dots$};
  \end{tikzpicture}

  \bigskip

  \begin{tikzpicture}
    \node[maxp] at (0.2,0) {};
    \node[minp] (min1) at (0.5,0.3) {};
    \node[minp] (min2) at (1.5,0.3) {};
    \node[minp] (min5) at (4,0.3) {};
    \node[maxp] (maxt) at (4.3,0) {};

    \node at (2.8,0) {$\dots$};

    \clip (0,-0.5) rectangle (2.4,0.5) (3.1,-0.5) rectangle (4.5,0.5);
    \draw (0.5,0) circle (0.45);
    \draw (1.5,0) circle (0.45);
    \draw (2.5,0) circle (0.45);
    \draw (3,0) circle (0.45);
    \draw (4,0) circle (0.45);
    \fill[white] (0.8,-0.1) to[bend left] (1.2, -0.1) -- (1.2,0.1) 
    to[bend left]  (0.8,0.1) -- cycle;
    \draw (0.8,-0.1) to[bend left] (1.2, -0.1) (1.2,0.1)  to[bend left]  (0.8,0.1);
    \fill[white] (1.8,-0.1) to[bend left] (2.2, -0.1) -- (2.2,0.1) 
    to[bend left]  (1.8,0.1) -- cycle;
    \draw (1.8,-0.1) to[bend left] (2.2, -0.1) (2.2,0.1)  to[bend left]  (1.8,0.1);
    \fill[white] (3.3,-0.1) to[bend left] (3.7, -0.1) -- (3.7,0.1) 
    to[bend left]  (3.3,0.1) -- cycle;
    \draw (3.3,-0.1) to[bend left] (3.7, -0.1) (3.7,0.1)  to[bend left]  (3.3,0.1);
  \end{tikzpicture}
  \caption{An $\CN = 2$ linear quiver theory associated to a punctured
    sphere.}
  \label{fig:class-S}
\end{figure}

Let us consider the standard type IIA brane configuration for an
$\CN = 2$ linear quiver theory with $m+1$ nodes.  It consists of $N$
D4-branes spanning the $01236$ directions, intersected by $m$
NS5-branes extending along the $012345$ directions:
\begin{equation}
  \label{eq:linear-quiver-IIA}
  \begin{tabular}{|l|cccccccccc|}
    \hline
    & 0 & 1 & 2 & 3 & 4 & 5 & 6 & 7 & 8 & 9
    \\ \hline
    D4 & $\times$ & $\times$ & $\times$ & $\times$ & & & $\times$
    &&&
    \\
    NS5 & $\times$ & $\times$ & $\times$ & $\times$ & $\times$ & $\times$
    &&&&
    \\ \hline
  \end{tabular}
\end{equation}
This brane configuration is lifted in M-theory to $N$ M5-branes,
wrapped on a cylinder with $m$ punctures created by intersecting
M5-branes.  Therefore, the $\CN = 2$ linear quiver theory is obtained
by compactification of the 6d $\CN = (2,0)$ theory of type $A_{N-1}$
on a cylinder with $m$ punctures, or a sphere with $m+2$ punctures.
We distinguish the two punctures coming from the ends of the cylinder
from the $m$ punctures in between.  They are referred to as maximal
and minimal punctures, respectively.  In the class-$\CS$ language, the
$\CN = 2$ linear quiver theory is a class-$\CS$ theory associated to a
sphere with $2$ maximal and $m$ minimal punctures.
Fig.~\ref{fig:class-S} illustrates the correspondence between the
quiver and the sphere.

The R-symmetry of the theory is $\SU(2)_I \times \U(1)_r$, where
$\SU(2)_I$ originates from the rotational symmetry of the $789$-space,
and $\U(1)_r$ from the rotational symmetry of the $45$-plane.  The
$\SU(N)$ flavor node from each end of the quiver is associated to the
maximal puncture on the corresponding side of the sphere.  The $i$th
gauge node is associated to the region between the $i$th and $(i+1)$th
minimal punctures.  To the $i$th minimal puncture is associated a
flavor symmetry $\U(1)_{\alpha_i}$ which acts on the hypermultiplet
charged under the $(i-1)$th and $i$th gauge nodes.

Following the philosophy of class-$\CS$ theories, we decompose this
theory into basic building blocks by decoupling gauge fields.  Roughly
speaking, the gauge coupling of the $i$th gauge node is inversely
proportional to the length between the $i$th and $(i+1)$th minimal
punctures.  To make the gauge couplings small, we take the minimal
punctures far apart from one another.  Then the geometry looks like a
string of $m$ spheres, each containing a single minimal puncture,
connected by long tubes.  The smaller the gauge couplings get, the
longer the tubes become, and eventually these spheres spilt up as the
couplings go to zero.  Each of the spheres represents a bifundamental
hypermultiplet, which is a linear quiver with $m = 1$, so it has one
minimal and two maximal punctures.  The quiver thus breaks into a
collection of three-punctured spheres, or trinions.

Conversely, a sphere with $2$ maximal and $m$ minimal punctures is
obtained by gluing $m$ trinions together, i.e., by replacing pairs of
maximal punctures with tubes.  In general, we can connect two Riemann
surfaces with a tube at maximal punctures.  From the point of view of
gauge theory, gluing corresponds to gauging the diagonal combination
of the $\SU(N)$ flavor symmetries associated to the maximal punctures
involved.  Using trinions with one minimal and two maximal punctures,
we can obtain any linear quiver in this way, and for that matter, also
a circular quiver by further gluing the two ends of a linear quiver
together.  In this sense, these trinions are building blocks for
linear and circular quivers.  As these two kinds of quivers can be
treated essentially in the same manner, we will focus on linear
quivers.

To make contact with brane tilings, we need to describe the $\CN = 2$
linear quiver theory as an $\CN=1$ quiver gauge theory.  In terms of
$\CN = 1$ supermultiplets, the $\CN = 2$ vector multiplet for the
$i$th gauge node decomposes into a vector multiplet and a chiral
multiplet $\Phi_i$ in the adjoint representation with
$(r, I_3) = (-1, 0)$, while the $i$th hypermultiplet consists of two
bifundamental chiral multiplets $Q_i$, $\Qt_i$ with
$(r, I_3) = (0, 1/2)$.  Here $I_3$ is a Cartan generator of
$\SU(2)_I$.  The pair $(Q_i, \Qt_i^\dagger)$ transforms in the doublet
of $\SU(2)_I$ and have $\U(1)_{\alpha_i}$ charge $F_{\alpha_i} = -1$.
From the point of view of $\CN = 1$ supersymmetry, the $\U(1)$
symmetry generated by the combination
\begin{equation}
  \CF = r + I_3
\end{equation}
is a flavor symmetry. We denote the fugacity for $\CF$ by $t$.  For
the standard definition of the $\CN = 2$ index, $r$ and $\CF$ enter
the trace through the combination $(pq)^{-r} t^\CF$.  Then, the
fugacities of $Q_i$, $\Qt_i$ and $\Phi_i$ are $\sqrt{t}/\alpha_i$,
$\sqrt{t} \alpha_i$ and $pq/t$, respectively.

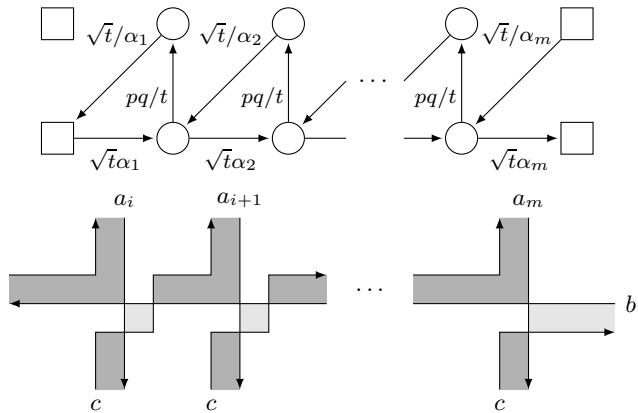
\begin{figure}
  \centering
  \begin{tikzpicture}[scale=1.2]
    \node[fnode] (0) at (0,0) {};
    \node[fnode] (0') at (0,1) {};
    \node[gnode] (1) at (1,0) {};
    \node[gnode] (1') at (1,1) {};
    \node[gnode] (2) at (2,0) {};
    \node[gnode] (2') at (2,1) {};

    \node[gnode] (m) at (3.5,0) {};
    \node[gnode] (m') at (3.5,1) {};
    \node[fnode] (m+1) at (4.5,0) {};
    \node[fnode] (m+1') at (4.5,1) {};

    \draw[q->] (1') --
    node[above=8pt] {$\sqrt{t}/\alpha_1$} (0);
    \draw[q->] (0) --(1)
    node[below, pos=0.5] {$\sqrt{t} \alpha_1$};
    \draw[q->] (1) --(1')
    node[left=-2pt, pos=0.3] {$pq/t$};

    \draw[q->] (2') --
    node[above=9pt] {$\sqrt{t}/\alpha_2$} (1);
    \draw[q->] (1) --(2)
    node[below, pos=0.5] {$\sqrt{t} \alpha_2$};
    \draw[q->] (2) --(2')
    node[left=-2pt, pos=0.3] {$pq/t$};

    \draw[q->] ($(2) + (0.5,0.5)$) -- (2);
    \draw[q-] (2) -- ($(2) + (0.5,0)$);

    \draw[q-] (m') -- ($(m') + (-0.5,-0.5)$);
    \draw[q->] ($(m) + (-0.5,0)$) -- (m);

    \draw[q->] (m) --(m')
    node[left=-2pt, pos=0.3] {$pq/t$};

    \draw[q->] (m+1') --
    node[above=10pt] {$\sqrt{t}/\alpha_m$} (m);
    \draw[q->] (m) -- (m+1)
    node[below, pos=0.5] {$\sqrt{t} \alpha_m$};

    \node at (2.75,0.5) {$\dots$};
  \end{tikzpicture}

  \begin{tikzpicture}[scale=1.2]
    \fill[dshaded] (-1/4,-3/4) rectangle (0,-1/4);
    \fill[dshaded] (3/4,-3/4) rectangle (1,-1/4);

    \fill[dshaded] (-1,1/4) -- (-1/4,1/4) -- (-1/4,3/4) -- (0,3/4) --
    (0,0) -- (-1,0) -- cycle;
    \fill[dshaded] (1/4,1/4) -- (3/4,1/4) -- (3/4,3/4) -- (1,3/4) --
    (1,0) -- (1/4,0) -- cycle;
    \fill[dshaded] (5/4,0) rectangle (7/4,1/4);

    \fill[lshaded] (0,-1/4) rectangle (1/4,0);
    \fill[lshaded] (1,-1/4) rectangle (5/4,0);

    \draw[z->] (0,3/4) node[above] {$a_i$} -- (0,-3/4);

    \draw[z->] (1,3/4) node[above] {$a_{i+1}$} -- (1,-3/4) ;

    \draw[z->] (7/4,0) -- (-1,0)
    node[left] {\vphantom{$b$}};

    \draw[z->] (-1,1/4) -- (-1/4,1/4) -- (-1/4,3/4);
    \draw[z->] (-1/4,-3/4) node[below] {$c$}
    -- (-1/4,-1/4) -- (1/4,-1/4) -- (1/4,1/4) -- (3/4,1/4) -- (3/4,3/4);
    \draw[z->] (3/4,-3/4) node[below] {$c$}
    -- (3/4,-1/4) -- (5/4,-1/4) -- (5/4,1/4) -- (7/4,1/4);

    \begin{scope}[shift={(2.5,0)}]
    \fill[dshaded] (3/4,-3/4) rectangle (1,-1/4);
    \fill[dshaded] (0,1/4) -- (3/4,1/4) -- (3/4,3/4) -- (1,3/4) --
    (1,0) -- (0,0) -- cycle;

    \fill[lshaded] (1,-1/4) rectangle (7/4,0);

    \draw[z->] (0,1/4) -- (3/4,1/4) -- (3/4,3/4);

    \draw[z->] (1,3/4)
    node[above] {$a_m$} -- (1,-3/4);

    \draw[z-] (7/4,0)  node[right] {$b$} -- (0,0) ;

    \draw[z->, shift={(1,0)}] (-1/4,-3/4) node[below] {$c$}
    -- (-1/4,-1/4) -- (3/4,-1/4);
    \end{scope}
    \node at (2.125,0.125) {$\,\dots$};
  \end{tikzpicture}

  \caption{An $\CN = 2$ linear quiver as a brane tiling model.  In the
    quiver, the two nodes in the same column are identified.  In the
    brane tiling diagram, the vertical direction is periodic.}
  \label{fig:linear-quiver}
\end{figure}

It is helpful for us to prepare two copies for each node of the quiver
and impose identification between them.  We draw the arrows in such a
way that $\Phi_i$ connects the two copies of the $i$th node and makes
a triangle with $Q_i$ and $\Qt_i$, as in Fig.~\ref{fig:linear-quiver}.
Drawn in this form, it is clear that the $\CN = 2$ linear quiver is a
special case of the triangle quiver described in section~\ref{sec:BT},
except that the vertical arrow is missing between the flavor nodes at
the right end.  The corresponding brane tiling diagram is therefore
essentially the same, as shown in Fig.~\ref{fig:linear-quiver}.  Note
that the cubic superpotentials, generated around the triangles by
worldsheet instantons, are precisely what we need for the theory to
have $\CN = 2$ supersymmetry.

\begin{figure}
  \centering
  \begin{tikzpicture}[scale=0.8, baseline=(x.base)]
    \node (x) at (0,0) {\vphantom{x}};

    \draw (0,0) circle (1);

    \node[maxp, label={below:$w$}] (w) at (-0.7,0) {};
    \node[maxp, label={below:$z$}] (z) at (0.7,0) {};
    \node[minp, label={below:$\alpha$}] (min) at (0,0.7) {};
  \end{tikzpicture}
  \
  {\large $\leadsto$}
  \
  \begin{tikzpicture}[baseline=(x.base)]
    \node (x) at (1,1) {\vphantom{x}};

    \node[fnode] (w) at (0.35,0.35) {$w$};
    \node[fnode] (w') at (0.35,1.65) {$w$};
    \node[fnode] (z) at (1.5,0.35) {$z$};
    \node[fnode] (z') at (1.5,1.65) {$z$};

    \draw[q->] (z') -- node[above left=-4pt] {$\sqrt{t}/\alpha$} (w);
    \draw[q->] (w) -- node[below] {$\sqrt{t} \alpha$} (z);
  \end{tikzpicture}
  \
  {\large =}
  \begin{tikzpicture}[baseline=(x.base)]
    \node (x) at (0,1) {\vphantom{x}};

    \node (w) at (0.35,0.35) {$w$};
    \node (w') at (0.35,1.65) {$w$};
    \node (z) at (1.5,0.35) {$z$};
    \node (z') at (1.5,1.65) {$z$};

    \fill[dshaded] (0.7,0) rectangle (1,0.7);
    \fill[lshaded] (1,0.7) rectangle (2,1);
    \fill[dshaded] (0,1) -- (1,1) -- (1,2) -- (0.7,2) -- (0.7,1.3)
    -- (0,1.3) -- cycle;

    \draw[z->] (1,2) node[above] {$a$} -- (1,0);
    \draw[z->] (2,1) node[right] {$b$} -- (0,1);

    \draw[z->] (0.7,0) node[below] {$c$}
    -- (0.7,0.7) -- (2,0.7);
    \draw[z->] (0,1.3) node[left] {$c$}
    -- (0.7,1.3) -- (0.7,2);
  \end{tikzpicture}

  \caption{The building block of class-$\CS$ theories associated to a
    trinion.}
  \label{fig:trinion}
\end{figure}
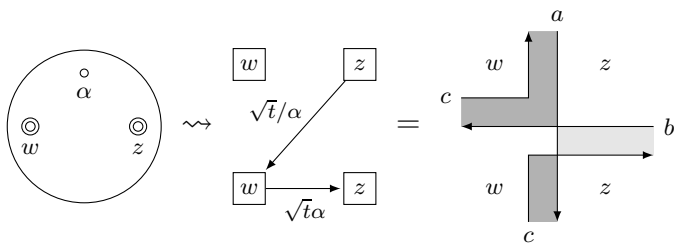

As we can split the $(m+1)$-punctured sphere into a collection of $m$
trinions, we can also break the brane tiling diagram into basic
pieces.  Each piece represents a single trinion and is made of three
zigzag paths; see Fig.~\ref{fig:trinion}.  Gluing two trinions
corresponds to concatenating two such diagrams side by side.  In the
course of this operation, we must interchange the positions of the
zigzag paths labeled $b$ and $c$ near the glued side of one of the
diagrams.  This results in an additional vertical arrow in the
combined quiver, which is the adjoint chiral multiplet in the
$\CN = 2$ vector multiplet used in the gauging.

Let us find the relationship between the convention we use for brane
tilings and that used above.  The R-charge $R$ in the brane tiling
model is given in terms of the charges of the $\CN = 2$ theory by
\begin{equation}
  R = R_0 + \frac12 \sum_i F_{\alpha_i}
  \,;
  \quad
  R_0 = -r + I_3
  \,.
\end{equation}
The flavor charges associated to the zigzag paths can be written as
\begin{equation}
  F_{a_i} = -F_{\alpha_i}
  \,,
  \quad 
  F_b = -\CF + \frac12 \sum_i F_{\alpha_i}
  \,,
  \quad 
  F_c = \CF + \frac12 \sum_i F_{\alpha_i}
  \,.
\end{equation}
Without loss of generality, we can set
\begin{equation}
  a_i = \frac{1}{\alpha_i}
  \,.
\end{equation}
Plugging these relations into the combination
$(pq)^{R/2} \prod_i a_i^{F_{a_i}} b^{F_b} c^{F_c}$ that enters the
indices of the bifundamental chiral multiplets, we deduce
\begin{equation}
  \label{eq:bc}
  b = \frac{1}{\sqrt{t}}
  \,,
  \quad
  c = \sqrt{\frac{t}{pq}}
  \,.
\end{equation}

Before proceeding, we should mention a peculiarity in the $A_1$ case.
When $N = 2$, the $\U(1)$ flavor symmetry of a bifundamental
hypermultiplet is enhanced to $\SU(2)$ due to the fact that the
fundamental representation of $\SU(2)$ is pseudoreal.  For this
reason there is no distinction between minimal and maximal punctures,
and each trinion can be regarded as a half-hypermultiplet in the
trifundamental representation of $\SU(2)^3$.  This is reflected in the
index of a trinion,
\begin{equation}
  \IB\bigl(w, z; \sqrt{t} \alpha\bigr) 
  \IB\bigl(z, w; \sqrt{t}/\alpha\bigr) 
  =
  \Gamma\bigl(\sqrt{t} \alpha^{\pm1} z^{\pm1} w^{\pm 1}\bigr)
  \,,
\end{equation}
which is manifestly symmetric under permutations of $\alpha$, $z$ and
$w$.

\subsection{\texorpdfstring{Surface defects in $A_1$ class-$\CS$
    theories}{Surface defects in A1 class-S theories}}

In~\cite{Gaiotto:2012xa}, it was explained how to construct a surface
defect labeled with a pair of integers~$(r,s)$, and how to determine its
action on the supersymmetric index.  Although the method applies to
general $\CN=2$ theories with $\SU(N)$ flavor symmetry, here we review
it in the language of class-$\CS$ theories.

Suppose we have a class-$\CS$ theory $\CT_\IR$ associated to a Riemann
surface that contains a maximal puncture, whose flavor group we call
$\SU(N)_z$.  To this surface we introduce an extra minimal puncture.
Concretely, we can do this as follows.  First, we rename the flavor
group $\SU(N)_z$ to $\SU(N)_{w'}$.  Then, we take a trinion
representing a hypermultiplet $(Q, \Qt)$ with flavor symmetry
$\SU(N)_{w''} \times \SU(N)_z \times \U(1)_\alpha$, and glue it to
$\CT_\IR$ by gauging the diagonal subgroup $\SU(N)_w$ of
$\SU(N)_{w'} \times \SU(N)_{w''}$.  The resulting theory $\CT_\UV$ has
one more flavor symmetry, $\U(1)_\alpha$, than $\CT_\IR$.
Correspondingly, the surface associated to $\CT_\UV$ has one more
minimal puncture than the original surface.

The theory $\CT_\UV$ is related to $\CT_\IR$ via the RG flow induced
by a diagonal constant vev given to the quark $Q$, or equivalently, to
the baryon $B = \det Q$.  (We may instead give a vev to the antibaryon
$\Bt = \det \Qt$, but this does not lead to anything different because
of the $\SU(2)_I$ symmetry.)  The vev higgses the gauge group
$\SU(N)_w$ and breaks $\SU(N)_w \times \SU(N)_z$ down to the diagonal
subgroup.  Moreover, it turns the cubic superpotential $\Qt \Phi Q$
into a quadratic one that makes $\Qt$ and $\Phi$ massive, where $\Phi$
is the adjoint chiral multiplet introduced in the gluing.  Up to
Nambu--Goldstone multiplets that survive the higgsing, in the
infrared the multiplets we added are gone and we recover $\CT_\IR$,
with $\SU(N)_w$ replaced with $\SU(N)_z$.  In effect, the minimal
puncture introduced by gluing the trinion is ``closed.''  The R-charge
$I_3$ is broken by the vev, but the combination $I_3 + F_\alpha/2$ is
preserved and identified with a Cartan generator of the infrared
$\SU(2)$ R-symmetry.

To create a surface defect in $\CT_\IR$, we instead give the baryon a
position-dependent vev $\vev{B} = \zeta_1^r \zeta_2^s$.  Here, as
before, $\zeta_1$ and $\zeta_2$ are complex coordinates of the two
orthogonal planes rotated by $j_p = j_1 + j_2$ and $j_q = j_1 - j_2$,
respectively.  Away from the origin, the effect of the
position-dependent vev is the same as that of the constant vev, so we
get $\CT_\IR$ in the infrared.  If $r \neq 0$, however, the infrared
theory is modified on the plane $\{\zeta_1 = 0\}$ since the vev
vanishes there.  By the same token, the theory is modified on the
plane $\{\zeta_2 = 0\}$ if $s \neq 0$.  Hence, in general we obtain
$\CT_\IR$ with the insertion of a surface defect labeled with the pair
of integers $(r,s)$, supported on the planes $\{\zeta_1 = 0\}$ and
$\{\zeta_2 = 0\}$.  This surface defect is to be identified with the
surface defect labeled with the pair
$(\underbrace{\yng(2) \dotsm \yng(1)}_r, \underbrace{\yng(2) \dotsm
  \yng(1)}_s$) of symmetric representations of $\SU(N)$ discussed in
the previous section~\cite{Gadde:2013ftv}.

The index of $\CT_\UV$ has a pole in the $\alpha$-plane at
$\alpha = \sqrt{t} p^{r/N} q^{s/N}$, and the residue there gives the
index of $\CT_\IR$ in the presence of the surface defect of type
$(r,s)$.  The reason is the following.  The position-dependent vev
$\vev{B} = \zeta_1^r \zeta_2^s$ breaks $\U(1)_p$, $\U(1)_q$, and
$\SU(2)_I$.  At this value of $\alpha$, however, the only combinations
of charges that enter the trace defining the index are those that are
preserved by the vev.  Thus, we can still define the index in this
background.  As explained above, $\CT_\UV$ flows to $\CT_\IR$ plus
Nambu--Goldstone multiplets in the infrared.  The latter contains
massless degrees of freedom, and they contribute to the index by a
diverging factor, in fact a simple pole in the $\alpha$-plane.
Therefore, the residue at this pole gives the index of $\CT_\IR$,
together with some factor associated with the Nambu--Goldstone
multiplets.

We wish to compute this residue and determine the action of the
surface defect on the index in the simplest nontrivial case, namely
when $N = 2$ and $(r,s) = (0,1)$.  But first, let us look at the
trivial case $(r,s) = (0,0)$ to gain a better understanding of the
computation.

In the construction of a surface defect described above, $\Qt$ and
$\Phi$ actually play no role.  The essential point is that the vev
given to the baryon built from $Q$ replaces $\SU(N)_w$ with $\SU(N)_z$
in the infrared.  So we couple $\CT_\IR$ just to $Q$ for the moment.
The index of the combined theory is given by
\begin{equation}
  \int_{\T} \frac{\rmd w}{2\pi iw}
  \IV(w)
  \IB(z, w; \rho) 
  \CI_{\CT_\IR}(w) 
  =
  \kappa
  \int_{\T} \frac{\rmd w}{2\pi iw}
  \frac{\Gamma(\rho z^{\pm1} w^{\pm1})}
          {\Gamma(w^{\pm2})}
  \CI_{\CT_\IR}(w)
  \,,
\end{equation}
where $\rho = \sqrt{t}/\alpha$ is the fugacity of $Q$ and
$\kappa = (p;p)_\infty (q;q)_\infty/2$.  In this integral, $|\rho| < 1$
is assumed, but we can analytically continue $\rho$ to a complex
parameter and study its pole structure.  At $\rho = 1$, a constant vev
can be turned on for $B$ without conflicting with the definition of
the index.  The integral should have a pole at this point in the
$\rho$-plane, and we want to calculate the residue there.

The integrand has two pairs of poles in the $w$-plane at
\begin{equation}
  w
  =
  \begin{cases}
    \rho z \,,
    \\
    \rho^{-1} z \,;
  \end{cases}
  \begin{cases}
    \rho z^{-1} \,,
    \\
    \rho^{-1} z^{-1} \,.
  \end{cases}
\end{equation}
As $\rho \to 1$, the first pair of poles collide and pinch the
integration contour, and the integral diverges.  Likewise, the second
pair also collide in this limit.  The pole of the integral in the
$\rho$-plane arises from the contributions from these poles in $w$.
Using formula~\eqref{eq:Res-Gamma}, we find that the contribution from
the pole at $w = \rho z$ is
\begin{equation}
  \frac12
  \frac{\Gamma(\rho^2 z^2) \Gamma(z^{-2})}
          {\Gamma(\rho^2 z^2) \Gamma(\rho^{-2} z^{-2})}
  \Gamma(\rho^2)
  \CI_{\CT_\IR}(\rho z)
  \,.
\end{equation}
The last factor $\Gamma(\rho^2)$ indeed has a pole at $\rho = 1$, with
residue $1/4\kappa$.  The pole at $w = \rho z^{-1}$ makes an equal
contribution, and we get
\begin{equation}
  \Res_{\rho = 1}
  \biggl[ 
  \int_{\T} \frac{\rmd w}{2\pi i w}
  \IV(w)
  \IB(z, w; \rho) 
  \CI_{\CT_\IR}(w)
  \biggr]
  =
  \frac{1}{2} \frac{1}{2\kappa} \CI_{\CT_\IR}(z)
  \,.
\end{equation}
As expected, the residue reproduces the index of $\CT_\IR$, multiplied
by some factors.  The factor of $1/2$ comes from the fact that $B$ has
fugacity $\rho^2$, and disappears if we add the equal contribution
from the pole at $\rho = -1$.  The factor $1/2\kappa$ is the
contribution from a decoupled free chiral multiplet contained in a
Nambu--Goldstone multiplet.  It is the inverse of the index of a free
vector multiplet since higgsing of a $\U(1)$ gauge theory with a
single chiral multiplet leads to a trivial theory whose index is $1$.

In order to express this result in a concise form, we introduce the
notation of ``striking out an arrow'' in a quiver diagram to indicate
that a constant vev is given to the baryonic operator built from the
bifundamental chiral multiplet represented by that arrow, and the
contributions from the accompanying Nambu--Goldstone multiplets are
discarded.  In this notation, what we just found is the identity
\begin{equation}
  \label{eq:giving-vev}
  \begin{tikzpicture}[baseline=(z.base)]
    \node[fnode] (z) at (0,0) {$z$};
    \node[gnode] (w) at (1,0) {$w$};

    \draw[q->, vev] (z) -- node[above] {$\rho$} (w);
  \end{tikzpicture}
  \ =
  4\kappa \Res_{\rho = 1}
  \Bigl[\,
  \begin{tikzpicture}[baseline=(z.base)]
    \node[fnode] (z) at (0,0) {$z$};
    \node[gnode] (w) at (1,0) {$w$};

    \draw[q->] (z) -- node[above] {$\rho$} (w);
  \end{tikzpicture}
  \,\Bigr]
  = \
  \begin{tikzpicture}[baseline=(z.base)]
    \node[fnode] (z) at (0,0) {$z$};
    \node[gnode] (w) at (1,0) {$w$};

    \draw[eq-] (z) -- (w);
  \end{tikzpicture}
  \ ,
\end{equation}
where the right-hand side is the delta function defined by the
relation~\eqref{eq:z=w-def}.  This identity holds when the index of
any theory with $\SU(2)$ flavor symmetry (or more generally, any
meromorphic function $f(w)$ such that $f(w) = f(1/w)$) is coupled to
the right node.

With the help of this identity, we can readily show that when a
constant vev is turned on for~$B$ (and the Nambu--Goldstone multiplets
are thrown away), the index of $\CT_\UV$ reduces to that of $\CT_\IR$.
All we have to do is to look at the part of $\CT_\UV$ describing the
coupling to the trinion, and compute the relevant residue:
\begin{equation}
  \begin{tikzpicture}[baseline=(x.base)]
    \node (x) at (0,0.5) {\vphantom{x}};

    \node[fnode] (z') at (1,1) {$z$};
    \node[gnode] (w) at (0,0) {$w$};
    \node[gnode] (w') at (0,1) {$w$};

    \draw[q->, vev] (z') -- (w)
    node[below right=-6pt, pos=0.4] {$\sqrt{t}/\alpha$};
    \draw[q->] (w) -- node[left] {$pq/t$} (w');
    \draw[q->] (w') -- node[above] {$\sqrt{t} \alpha$} (z');
  \end{tikzpicture}
  \ =
  \begin{tikzpicture}[baseline=(x.base)]
    \node (x) at (0,0.5) {\vphantom{x}};

    \node[fnode] (z') at (1,1) {$z$};
    \node[gnode] (w) at (0,0) {$w$};
    \node[gnode] (w') at (0,1) {$w$};

    \draw[eq-] (z') -- (w);
    \draw[q->] (w) -- node[left] {$pq/t$} (w');
    \draw[q->] (w') -- node[above] {$t$} (z');
  \end{tikzpicture}
  \ = \
  \begin{tikzpicture}[baseline=(z.base)]
    \node[gnode] (w) at (0,0) {$w$};
    \node[fnode] (z) at (1,0) {$z$};
    \draw[eq-] (w) -- (z);
  \end{tikzpicture}
  \ .
\end{equation}
In the first equality we used the identity~\eqref{eq:giving-vev} and
set $\rho = 1$, and in the second we canceled the pair of arrows
making a loop.  Thus, the vev transforms the trinion into the original
flavor node of $\CT_\IR$.

We can compute the index of $\CT_\IR$ in the presence of a surface
defect in a similar manner.  To indicate that the position-dependent
vev $\vev{B} = \zeta_1^r \zeta_2^s$ is turned on, we put the label
$(r,s)$ on the struck-out arrow:
\begin{equation}
  \label{eq:res-rs}
  \begin{tikzpicture}[baseline=(z.base)]
    \node[fnode] (z) at (0,0) {$z$};
    \node[gnode] (w) at (1,0) {$w$};

    \draw[q->, vev] (z) -- node[above] {$\rho$} node[below] {$(r,s)$} (w);
  \end{tikzpicture}
  \ =
  4\kappa
  \Res_{\rho = p^{-r/2} q^{-s/2}}
  \Bigl[\,
  \begin{tikzpicture}[baseline=(z.base)]
    \node[fnode] (z) at (0,0) {$z$};
    \node[gnode] (w) at (1,0) {$w$};

    \draw[q->] (z) -- node[above] {$\rho$} (w);
  \end{tikzpicture}
  \,\Bigr]
  \,.
\end{equation}
Then the action of the surface defect of type $(r,s)$ on the index is
encoded in the diagram
\begin{equation}
  \begin{tikzpicture}[baseline=(x.base)]
    \node (x) at (0,0.5) {\vphantom{x}};

    \node[fnode] (z') at (1,1) {$z$};
    \node[gnode] (w) at (0,0) {$w$};
    \node[gnode] (w') at (0,1) {$w$};

    \draw[q->, vev] (z') --
    node[below, sloped] {$(r,s)$} (w);
    \draw[q->] (w) -- (w');
    \draw[q->] (w') -- (z');
  \end{tikzpicture}
  \ .
\end{equation}

Let us calculate the residue~\eqref{eq:res-rs} for $(r,s) = (0,1)$.
At $\rho = q^{-1/2}$, the index
$\IB(z,w; \rho) = \Gamma(\rho z^{\pm 1} w^{\pm 1})$ of $Q$ has four
sets of colliding poles in the $w$-plane.  Two of them are
\begin{equation}
  w
  =
  \begin{cases}
    \rho qz \,,
    \\
    \rho^{-1} z \,;
  \end{cases}
  \begin{cases}
    \rho z^{-1} \,,
    \\
    \rho^{-1} q^{-1} z^{-1} \,,
  \end{cases}
\end{equation}
while the other two are
\begin{equation}
  w
  =
  \begin{cases}
    \rho q z^{-1} \,,
    \\
    \rho^{-1} z^{-1} \,;
  \end{cases}
  \begin{cases}
    \rho z,
    \\
    \rho^{-1} q^{-1} z \,.
  \end{cases}
\end{equation}
The contributions to the residue come from these poles.  A small
calculation shows that the first two sets of poles contribute in the
same way: they set $w = q^{1/2} z$ and give a factor of
$1/\theta(q^{-1}) \theta(z^2)$ in total.  Similarly, the contributions
from the last two set $w = q^{-1/2} z$ and give a factor of
$1/\theta(q^{-1}) \theta(z^{-2})$.  Altogether, we find that the
result can be expressed as
\begin{equation}
  \label{eq:vev-(0,1)}
  \begin{tikzpicture}[baseline=(z1.base)]
    \node[fnode] (z1) at (0,0) {$z$};
    \node[gnode] (y2) at (1,0) {$w$};

    \draw[q->, vev] (z1) -- node[above] {$\rho$} node[below] {$(0,1)$} (y2);
  \end{tikzpicture}
  \ =
  \frac{1}{\theta(q^{-1})}
  \sum_{s=\pm1}
  \frac{1}{\theta(z^{2s})}
  \Delta^{s/2}
  \
  \begin{tikzpicture}[baseline=(z1.base)]
    \node[fnode] (z1) at (0,0) {$z$};
    \node[gnode] (y2) at (1,0) {$w$};

    \draw[eq-] (z1) -- (y2);
  \end{tikzpicture}
  \ .
\end{equation}
We remind the reader that $\theta(z) = \theta(z; p)$ and $\Delta^{\pm
  1/2}$ act on functions of $z$ as $(\Delta^{\pm 1/2} f)(z) = f(q^{\pm
  1/2} z)$.

Unlike the case of the constant vev, this identity does not cause a
complete cancelation of the indices of $\Qt$ and $\Phi$.  Rather, for
$\rho = q^{-1/2}$ and $w = q^{\pm1/2} z$, we have
\begin{equation}
  \label{eq:wwz}
  \begin{tikzpicture}[baseline=(x.base)]
    \node (x) at (0,-0.5) {\vphantom{x}};

    \node[fnode] (y1) at (0,0) {$w$};
    \node[fnode] (y2) at (0,-1) {$w$};
    \node[fnode] (z1) at (1,0) {$z$};

    \draw[q->] (y2) --
    node[left] {$pq/t$} (y1);
    \draw[q->] (y1) --
    node[above] {$\sqrt{t} \alpha$} (z1);
  \end{tikzpicture}
  \ =
  \theta\biggl(\frac{t}{q} z^{\mp 2}\biggr)
  \theta(t)
  \,. 
\end{equation}
Therefore, the effect of introducing the surface defect of type $(0,1)$
on the index is realized by the difference operator
\begin{equation}
  \label{eq:S(0,1)}
  \mathfrak{S}_{(0,1)}
  =
  \frac{\theta(t)}{\theta(q^{-1})}
  \sum_{s=\pm1}
  \frac{1}{\theta(z^{2s})}
  \theta\biggl(\frac{t}{q} z^{-2s}\biggr)
  \Delta^{s/2}
  \,.
\end{equation}
The prefactor $\theta(t)/\theta(q^{-1})$ is equal to the index of a
free chiral field in two dimensions, and represents the center-of-mass
degree of freedom of the surface defect.

The difference operator $\mathfrak{S}_{(0,1)} $ acts on the fugacity
for the maximal puncture on which the surface defect was constructed.
This fact has a natural interpretation.  To construct the surface
defect, we first introduced an extra minimal puncture, and then took
the residue of a pole in the fugacity of the associated flavor
symmetry.  The latter step can be thought of as transforming the
minimal puncture to another kind of puncture which represents the
surface defect.  By construction, this puncture is located in the
neighborhood of a maximal puncture contained in a trinion.  We can
take the surface defect puncture and collide it to the maximal
puncture.  The collision produces a new puncture, and defines the
action of the surface defect on the maximal puncture.

\subsection{Comparison with the transfer matrix}

\begin{figure}
  \centering
  \begin{tikzpicture}
    \fill[dshaded] (0.7,0) rectangle (1,0.7);
    \fill[lshaded] (1,0.7) rectangle (2.5,1);
    \fill[dshaded] (0,1) -- (1,1) -- (1,2) -- (0.7,2) -- (0.7,1.3)
    -- (0,1.3) -- cycle;

    \draw[z->] (1,2) node[above] {$a$} -- (1,0);
    \draw[z->] (2.5,1) node[right] {$b$} -- (0,1);

    \draw[z->] (0.7,0) node[below] {$c$}
    -- (0.7,0.7) -- (2.5,0.7);
    \draw[z->] (0,1.3) node[left] {$c$}
    -- (0.7,1.3) -- (0.7,2);

    \draw[dz->] (1.5,2) node[above] {$d$} -- (1.5,0);

    \node (z) at (2,0.35) {$z$};
    \node (z') at (2,1.65) {$z$};
  \end{tikzpicture}

  \caption{The brane tiling representation of a surface defect acting
    on a maximal puncture.}
  \label{fig:trinion-defect}
\end{figure}
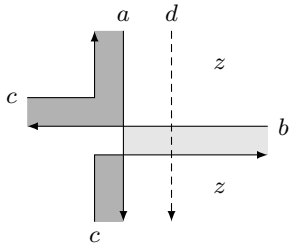

Let us compare the result with our proposal.  For clarity of
presentation, take a minimal puncture in $\CT_\IR$ and move it close
to the maximal puncture on which the surface defect acts.  Then the
neighborhood of these punctures looks like a trinion glued to another
maximal puncture, and is represented by zigzag paths as in
Fig.~\ref{fig:trinion-defect}.

According to our proposal, the surface defect creates a dashed line
with some spectral parameter $d$, also drawn in the picture.  It acts
on the lattice model as the transfer matrix
\begin{equation}
  \label{eq:TM-S}
  \Tr\Bigl(\Ldia\bigl(d, (c,b)\bigr)\Bigr)
  =
  \sum_{s=\pm1}
  \frac{1}{\theta(z^{2s})}
  \theta\biggl(\sqrt{\frac{p}{q}} \frac{bc}{d^2}\biggr) 
  \theta\biggl(\sqrt{\frac{p}{q}} \frac{c}{b} \frac{1}{z^{2s}}\biggr)
  \Delta^{s/2}
  \,.
\end{equation}
From the relation~\eqref{eq:bc}, we see that  if we set 
\begin{equation}
  \label{eq:d}
  d = \frac{1}{\sqrt{qt}}
  \,,
\end{equation}
the transfer matrix indeed reproduces the difference
operator~\eqref{eq:S(0,1)}, up to an overall factor which cannot be
fixed by the Yang--Baxter equations.

As noted in~\cite{Gaiotto:2012xa}, the above transfer matrix is
essentially the Hamiltonian of the elliptic Ruijsenaars--Schneider
model~\cite{Ruijsenaars:1986vq, Ruijsenaars:1986pp} of type~$A_1$.
This fact follows from a general result obtained in~\cite {MR1463830}.

Here we have considered only the surface defect of type $(0,1)$, but
the general story is similar.  The surface defect of type $(r,s)$ acts
on the index by a difference operator $\mathfrak{S}_{(r,s)}$.  This
operator is expected to coincide with the transfer matrix for an
appropriate L-operator.  If so, by the RLL relation~\eqref{eq:RLL-2},
the operators $\mathfrak{S}_{(r,s)}$ for all $(r,s)$ should commute
with one another.  This is indeed true~\cite{Gaiotto:2012xa}.  From
the class-$\CS$ point of view, the mutual commutativity is guaranteed
by the fact that the index is independent of the positions of
punctures representing surface defects.  Therefore, the order in which
they act on a maximal puncture is irrelevant.  Note that this argument
also exploits the existence of an extra dimension, which is the
M-theory circle that emerges as the type IIA brane configuration is
lifted to M-theory.

For the same reason, a surface defect puncture can be placed between
any two punctures, whether minimal or maximal, and still yield the
same result.  From the point of view of the type IIA system, this
property appears to be quite nontrivial and is known as the ``hopping
invariance'' of the index~\cite{Gadde:2013ftv}.  From the lattice
model viewpoint, this is guaranteed by the other RLL
relation~\eqref{eq:RLL-1}.

\subsection{\texorpdfstring{$\CN=1$ theories of class $\CS$}{N = 1
    theories of class S}}
\label{sec:class-S-N=1}

There are generalizations of class-$\CS$ theories that preserve only
$\CN = 1$ supersymmetry.  For these theories, we can compute the index
in the presence of a surface defect either by the residue method or
using the transfer matrix, and compare the results.

Suppose we have an $\CN = 2$ theory of class $\CS$, obtained by
compactification of M5-branes on a punctured Riemann surface $\CC$.
In the ordinary class-$\CS$ case, $\CC$ is embedded in the cotangent
bundle $T^*\CC$.  If we modify this setup in such a way that $\CC$
becomes a holomorphic curve in a generic Calabi--Yau threefold, the
$\CN = 2$ supersymmetry gets broken to $\CN = 1$.  A situation
commonly studied in the literature is when $\CC$ is the zero section
of the total space of the direct sum of two line bundles over $\CC$,
satisfying an appropriate topological condition~\cite{Bah:2012dg,
  Gadde:2013fma, Xie:2013gma, Bah:2013aha, Bonelli:2013pva,
  Agarwal:2013uga, Agarwal:2014rua}.

For our purpose, it is sufficient to consider $\CN = 1$ theories that
are realized by simple modifications of the type IIA brane
configuration~\eqref{eq:linear-quiver-IIA} for $\CN = 2$ linear quiver
theories.  In order to break supersymmetry by half, we rotate some of
the NS5-branes so that they span the $012389$ directions.  We refer to
these rotated NS5-branes as $\text{NS5}_-$, while calling the
unrotated ones $\text{NS5}_+$.  Lifted to M-theory, the two types of
NS5-branes, $\text{NS}_+$ and $\text{NS}_-$, both become M5-branes
supported at points on a cylinder.  Correspondingly, there are now two
types of minimal punctures labeled with a sign $\sigma = \pm 1$.  We
denote positive and negative minimal punctures by
\tikz[baseline={([yshift=-1pt]min.base)}]{\node[minp] (min) at (0,0)
  {$+$};} and \tikz[baseline={([yshift=-1pt]min.base)}]{\node[minp]
  (min) at (0,0) {$-$};}, respectively.

Recall that in the $A_1$ case, there is no distinction between minimal
and maximal punctures.  This fact suggests that maximal punctures also
come in two types, positive and negative, denoted by
\tikz[baseline={([yshift=-1pt]z.base)}]{\node[maxp] (z) at (0,0)
  {$+$};} and \tikz[baseline={([yshift=-1pt]z.base)}]{\node[maxp] (z)
  at (0,0) {$-$};}.  To incorporate maximal punctures of different
signs, we have to modify the brane setup slightly.  In the type IIA
picture, we terminate the D4-branes on a D6-brane on each side of the
brane system, rather than letting them continue to $X^6 = \pm\infty$.
Then, a D6-brane $\text{D6}_+$ extending along the $0123789$
directions represents a maximal puncture with $\sigma = +1$, and
$\text{D6}_-$ along the $0123457$ directions represents one with
$\sigma = -1$.  The total brane configuration is summarized as
follows:
\begin{equation}
  \begin{tabular}{|l|cccccccccc|}
    \hline
    & 0 & 1 & 2 & 3 & 4 & 5 & 6 & 7 & 8 & 9
    \\ \hline
    D4 & $\times$ & $\times$ & $\times$ & $\times$ & & & $\times$
    &&&
    \\
    $\text{NS5}_+$
    & $\times$ & $\times$ & $\times$ & $\times$ & $\times$ & $\times$
    &&&&
    \\
    $\text{NS5}_-$
    & $\times$ & $\times$ & $\times$ & $\times$ & & & & &
    $\times$ & $\times$
    \\
    $\text{D6}_+$
    & $\times$ & $\times$ & $\times$ & $\times$ & & & &
    $\times$ & $\times$ &  $\times$
    \\
    $\text{D6}_-$
    & $\times$ & $\times$ & $\times$ & $\times$ & $\times$ & $\times$
    & & $\times$ & &
    \\ \hline
  \end{tabular}
\end{equation}

D4-branes suspended between two NS5-branes of the same sign give rise
to an $\CN = 2$ vector multiplet as before.  From those suspended
between NS5-branes of different signs, we get an $\CN = 1$ vector
multiplet.  D4-branes suspended between an NS5-brane and a D6-brane of
different signs produce an extra chiral multiplet in the adjoint
representation of the $\SU(N)$ flavor symmetry of the maximal
puncture.

\begin{figure}
  \centering
  \begin{tikzpicture}[scale=0.8, baseline=(x.base)]
    \node (x) at (0,0) {\vphantom{x}};

    \draw (0,0) circle (1);

    \node[maxp, label={below:$w$}] (w) at (-0.7,0) {$-$};
    \node[maxp, label={below:$z$}] (z) at (0.7,0) {$-$};
    \node[minp, label={below:$\alpha$}] (min) at (0,0.7) {$-$};
  \end{tikzpicture}
  \
  {\large $\leadsto$}
  \
  \begin{tikzpicture}[baseline=(x.base)]
    \node (x) at (1,1) {\vphantom{x}};

    \node[fnode] (w) at (0.35,0.35) {$w$};
    \node[fnode] (w') at (0.35,1.65) {$w$};
    \node[fnode] (z) at (1.5,0.35) {$z$};
    \node[fnode] (z') at (1.5,1.65) {$z$};

    \draw[q->] (z) -- node[below left=-4pt] {$\sqrt{t'} \alpha$} (w');
    \draw[q->] (w') -- node[above] {$\sqrt{t'}/\alpha$} (z');
  \end{tikzpicture}
  \
  {\large =}
  \begin{tikzpicture}[baseline=(x.base)]
    \node (x) at (0,1) {\vphantom{x}};

    \node (w) at (0.35,0.35) {$w$};
    \node (w') at (0.35,1.65) {$w$};
    \node (z) at (1.5,0.35) {$z$};
    \node (z') at (1.5,1.65) {$z$};

    \fill[dshaded] (1,1) rectangle (2,1.3);
    \fill[lshaded] (0.7,1.3) rectangle (1,2);
    \fill[lshaded] (1,0) -- (1,1) -- (0,1) -- (0,0.7) -- (0.7,0.7)
    -- (0.7,0) -- cycle;

    \draw[z->] (1,0) node[below] {$a$} -- (1,2);
    \draw[z->] (2,1) node[right] {$b$} -- (0,1);

    \draw[z->] (0.7,2) node[above] {$c$}
    -- (0.7,1.3) -- (2,1.3);
    \draw[z->] (0,0.7) node[left] {$c$}
    -- (0.7,0.7) -- (0.7,0);
  \end{tikzpicture}

  \caption{A trinion with three negative punctures.  Here
    $t' = pq/t$.}
  \label{fig:trinion---}
\end{figure}
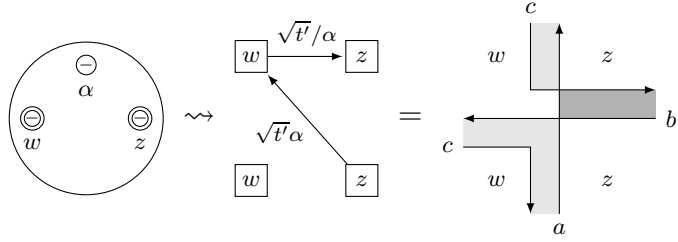

Flipping the sign of a puncture can be understood geometrically in
terms of an operation on zigzag paths.  The trinion in
Fig.~\ref{fig:trinion} has three punctures with $\sigma = +1$.
Changing the sign of a maximal puncture to $\sigma = -1$ amounts to
interchanging the positions of the corresponding pair of zigzag paths
so that the adjoint chiral multiplet arises from the crossing; see
Fig.~\ref{fig:trinion++-}.  Reversing the sign of a minimal puncture
entails flipping of the orientation of the corresponding vertical
zigzag path.  A trinion with all punctures having $\sigma = -1$ is
shown in Fig.~\ref{fig:trinion---}.
 
\begin{figure}
  \centering
  \begin{tikzpicture}[scale=0.8, baseline=(x.base)]
    \node (x) at (0,0) {\vphantom{x}};
 
    \draw (0,0) circle (1);

    \node[maxp, label={below:$w$}] (w) at (-0.7,0) {$+$};
    \node[maxp, label={below:$z$}] (z) at (0.7,0) {$-$};
    \node[minp, label={below:$\alpha$}] (min) at (0,0.7) {$+$};
  \end{tikzpicture}
  \
  {\large $\leadsto$}
  \
  \begin{tikzpicture}[baseline=(x.base)]
    \node (x) at (0,1) {\vphantom{x}};

    \node[fnode] (w) at (0.35,0.35) {$w$};
    \node[fnode] (w') at (0.35,1.65) {$w$};
    \node[fnode] (z) at (1.5,0.35) {$z$};
    \node[fnode] (z') at (1.5,1.65) {$z$};

    \draw[q->] (z') -- node[above left=-4pt] {$\sqrt{t}/\alpha$} (w);
    \draw[q->] (w) -- node[below] {$\sqrt{t} \alpha$} (z);
    \draw[q->] (z) -- node[right] {$pq/t$} (z');
  \end{tikzpicture}
  {\large =}
  \begin{tikzpicture}[baseline=(x.base)]
    \node (x) at (1,1) {\vphantom{x}};

    \node (w) at (0.35,0.35) {$w$};
    \node (w') at (0.35,1.65) {$w$};
    \node (z) at (1.5,0.35) {$z$};
    \node (z') at (1.5,1.65) {$z$};

    \fill[lshaded] (1,0.7) rectangle (1.3,1);
    \fill[dshaded] (0.7,0) rectangle (1,0.7);
    \fill[dshaded] (1.3,1) rectangle (2,1.3);
    \fill[dshaded] (0,1) -- (1,1) -- (1,2) -- (0.7,2) -- (0.7,1.3)
    -- (0,1.3) -- cycle;

    \draw[z->] (1,2) node[above] {$a$} -- (1,0);
    \draw[z->] (2,1) node[right] {$b$} -- (0,1);

    \draw[z->] (0.7,0) node[below] {$c$}
    -- (0.7,0.7) -- (1.3,0.7) -- (1.3,1.3) -- (2,1.3);
    \draw[z->] (0,1.3) node[left] {$c$}
    -- (0.7,1.3) -- (0.7,2);
  \end{tikzpicture}

  \caption{A trinion with one positive minimal, one positive maximal
    and one negative maximal puncture.}
  \label{fig:trinion++-}
\end{figure}
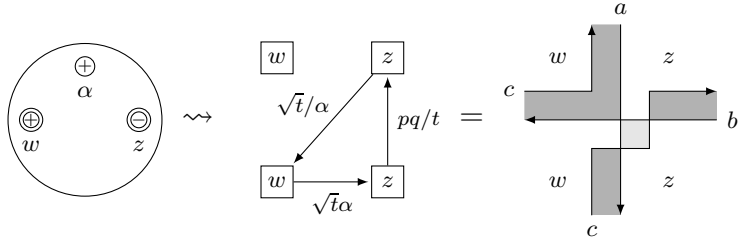

Let us compute the action of the surface defect of type $(0,1)$ on the
index of an $A_1$ theory that contains a negative maximal puncture.
To this end, we couple the trinion in Fig.~\ref{fig:trinion++-} to
the theory by connecting the positive maximal puncture of the former
and the negative maximal puncture of the latter, and take the residue
of the index of the resulting theory at the pole
$\alpha = \sqrt{t} q^{1/2}$.  The computation is the same as in the
case when the surface defect acts on a positive maximal puncture,
except that the factor~\eqref{eq:wwz} is replaced with
\begin{equation}
  \begin{tikzpicture}[baseline=(x.base)]
    \node (x) at (0,0.5) {\vphantom{x}};

    \node[fnode] (w) at (0,0) {$w$};
    \node[fnode] (z1) at (1,0) {$z$};
    \node[fnode] (z2) at (1,1) {$z$};

    \draw[q->] (z1) -- node[right] {$pq/t$} (z2);
    \draw[q->] (w) -- node[below] {$\sqrt{t} \alpha$} (z1);
  \end{tikzpicture}
  =
  \theta\bigl(t z^{\pm 2}\bigr)
  \theta(t)
  \,. 
\end{equation}
Thus, the action of the surface defect on a negative maximal puncture
is represented by the difference operator
\begin{equation}
  \frac{\theta(t)}{\theta(q^{-1})}
  \sum_{s=\pm1}
  \frac{1}{\theta(z^{2s})}
  \theta\bigl(t z^{2s}\bigr)
  \Delta^{s/2}
  \,.
\end{equation}

\begin{figure}
  \centering
  \begin{tikzpicture}
    \fill[dshaded] (0,0.7) rectangle (1.5,1);

    \draw[z->]  (0,1) node[left] {$c$} -- (1.5,1);
    \draw[z<-] (0,0.7) -- (1.5,0.7) node[right] {$b$};
    \draw[dz->] (0.5,1.7) node[above] {$d$} -- (0.5,0);

    \node at (1,0.35) {$z$};
    \node at (1,1.35) {$z$};
  \end{tikzpicture}

  \caption{The brane tiling representation of a surface defect acting
    on a negative maximal puncture.}
  \label{fig:trinion-defect-minus}
\end{figure}
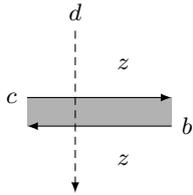

In the brane tiling picture, the surface defect is represented by a
dashed line traversing zigzag paths sandwiching an $(N,1)$ 5-brane
region, as illustrated in~Fig.~\ref{fig:trinion-defect-minus}.
Plugging the relations~\eqref{eq:bc} and~\eqref{eq:d} into
formula~\eqref{eq:TM-Ldiabar}, we see that the transfer matrix
reproduces the above difference operator.

\section{\texorpdfstring{Surface defects in $A_1$ theories of class
    $\CS_k$}{Surface defects in A1 theories of class Sk}}
\label{sec:class-Sk}

Lastly, we study surface defects in $A_1$ theories of class
$\CS_k$~\cite{Gaiotto:2015usa}, which are 4d $\CN = 1$ superconformal
theories obtained by compactification of the 6d $\CN = (1,0)$
superconformal theory of type $(A_1, \Z_k)$, or two M5-branes probing
a $\C^2/\Z_k$ orbifold singularity~\cite{Intriligator:1997kq,
  Blum:1997mm}.  After reviewing basic elements of class-$\CS_k$
theories with emphasis on their relation to brane tilings, we compute
their supersymmetric indices in the presence of simple surface
defects, extending calculations in~\cite{Gaiotto:2015usa}.  The
results agree with our proposal based on the lattice model approach:
these surface defects are represented by transfer matrices constructed
from $k$ copies of the relevant L-operator.

\subsection{\texorpdfstring{Class-$\CS_k$ theories}{Class-Sk
    theories}}

As in our discussion on class-$\CS$ theories in the previous section,
we first treat $A_{N-1}$ theories for general $N$.  We will later set
$N = 2$ when we actually carry out the index computation.

\begin{figure}
  \centering
  \begin{tikzpicture}[baseline=(x)]
    \node[fnode] (a0) at (0,0) {};
    \node[fnode] (b0) at (0,-1) {};
    \node[fnode] (c0) at (0,-2) {};
    \node[gnode] (a1) at (1,0) {};
    \node[gnode] (b1) at (1,-1) {};
    \node[gnode] (c1) at (1,-2) {};
    \node[gnode] (a2) at (2,0) {};
    \node[gnode] (b2) at (2,-1) {};
    \node[gnode] (c2) at (2,-2) {};

    \begin{scope}[shift={(0.5,0)}]
      \node[gnode] (a3) at (3,0) {};
      \node[gnode] (b3) at (3,-1) {};
      \node[gnode] (c3) at (3,-2) {};
      \node[fnode] (a4) at (4,0) {};
      \node[fnode] (b4) at (4,-1) {};
      \node[fnode] (c4) at (4,-2) {};
    \end{scope}

    \draw[q->] (a1) -- (b0);
    \draw[q->] (b1) -- (c0);

    \draw[q->] (a2) -- (b1);
    \draw[q->] (b2) -- (c1);

    \draw[q->] (a4) -- (b3);
    \draw[q->] (b4) -- (c3);

    \draw[q->] (a0) -- (a1);
    \draw[q->] (a1) -- (a2);
    \draw[q->] (a3) -- (a4);

    \draw[q->] (b0) -- (b1);
    \draw[q->] (b1) -- (b2);
    \draw[q->] (b3) -- (b4);

    \draw[q->] (c0) -- (c1);
    \draw[q->] (c1) -- (c2);
    \draw[q->] (c3) -- (c4);

    \draw[q->] (b1) -- (a1);
    \draw[q->] (c1) -- (b1);

    \draw[q->] (b2) -- (a2);
    \draw[q->] (c2) -- (b2);

    \draw[q->] (b3) -- (a3);
    \draw[q->] (c3) -- (b3);

    \draw[q->] ($(a0) + (0.5,0.5)$) -- (a0);
    \draw[q->] ($(a1) + (0.5,0.5)$) -- (a1);
    \draw[q->] ($(a2) + (0.5,0.5)$) -- (a2);
    \draw[q->] ($(a3) + (0.5,0.5)$) -- (a3);
    \draw[q->] ($(b2) + (0.5,0.5)$) -- (b2);
    \draw[q->] ($(c2) + (0.5,0.5)$) -- (c2);

    \draw[q-] (a3) -- ($(a3) + (-0.5,-0.5)$);
    \draw[q-] (b3) -- ($(b3) + (-0.5,-0.5)$);
    \draw[q-] (c1)  -- ($(c1) + (-0.5,-0.5)$);
    \draw[q-] (c2)  -- ($(c2) + (-0.5,-0.5)$);
    \draw[q-] (c3)  -- ($(c3) + (-0.5,-0.5)$);
    \draw[q-] (c4)  -- ($(c4) + (-0.5,-0.5)$);

    \draw[q-] (a1) -- ($(a1) + (0,0.5)$);
    \draw[q-] (a2) -- ($(a2) + (0,0.5)$);
    \draw[q-] (a3) -- ($(a3) + (0,0.5)$);

    \draw[q->] ($(c1) + (0,-0.5)$) -- (c1);
    \draw[q->] ($(c2) + (0,-0.5)$) -- (c2);
    \draw[q->] ($(c3) + (0,-0.5)$) -- (c3);

    \draw[q-] (a2) -- ($(a2) + (0.5,0)$);
    \draw[q-] (b2) -- ($(b2) + (0.5,0)$);
    \draw[q-] (c2) -- ($(c2) + (0.5,0)$);

    \draw[q->] ($(a3) + (-0.5,0)$) -- (a3);
    \draw[q->] ($(b3) + (-0.5,0)$) -- (b3);
    \draw[q->] ($(c3) + (-0.5,0)$) -- (c3);

    \node (x) at (2.8,-1) {$\dots$};
  \end{tikzpicture}

  \bigskip

  \begin{tikzpicture}
    \node[maxp] at (0.2,0) {};
    \node[minp] (min1) at (0.5,0.3) {};
    \node[minp] (min2) at (1.5,0.3) {};
    \node[minp] (min5) at (4,0.3) {};
    \node[maxp] (maxt) at (4.3,0) {};

    \node at (2.8,0) {$\dots$};

    \clip (0,-0.5) rectangle (2.4,0.5) (3.1,-0.5) rectangle (4.5,0.5);
    \draw (0.5,0) circle (0.45);
    \draw (1.5,0) circle (0.45);
    \draw (2.5,0) circle (0.45);
    \draw (3,0) circle (0.45);
    \draw (4,0) circle (0.45);
    \fill[white] (0.8,-0.1) to[bend left] (1.2, -0.1) -- (1.2,0.1) 
    to[bend left]  (0.8,0.1) -- cycle;
    \draw (0.8,-0.1) to[bend left] (1.2, -0.1) (1.2,0.1)  to[bend left]  (0.8,0.1);
    \fill[white] (1.8,-0.1) to[bend left] (2.2, -0.1) -- (2.2,0.1) 
    to[bend left]  (1.8,0.1) -- cycle;
    \draw (1.8,-0.1) to[bend left] (2.2, -0.1) (2.2,0.1)  to[bend left]  (1.8,0.1);
    \fill[white] (3.3,-0.1) to[bend left] (3.7, -0.1) -- (3.7,0.1) 
    to[bend left]  (3.3,0.1) -- cycle;
    \draw (3.3,-0.1) to[bend left] (3.7, -0.1) (3.7,0.1)  to[bend left]  (3.3,0.1);
  \end{tikzpicture}
  \caption{A brane tiling model as a class-$\CS_k$ theory associated
    to a punctured sphere.}
  \label{fig:class-Sk}
\end{figure}
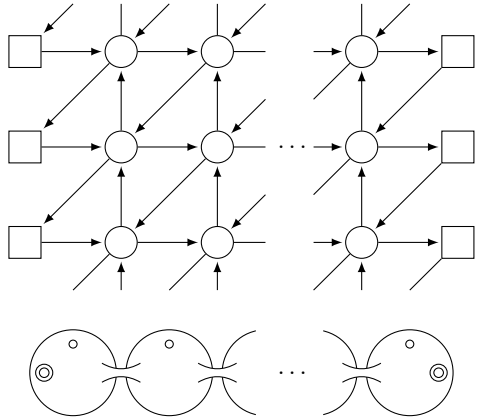

Let us consider a brane tiling model with $\SU(N)$ gauge groups
described by the quiver shown in Fig.~\ref{fig:class-Sk}.  The
quiver consists of $m+1$ columns, each containing $k$ nodes.  The
vertical direction is periodic, whereas the horizontal direction is a
finite interval.  When $k=1$, the theory reduces to the $\CN = 2$
linear quiver theory considered in the previous section.  Like that
case, we could make the horizontal direction also periodic by gluing
the leftmost and rightmost columns in a consistent manner, but we will
leave the quiver as it is in the following discussion.

If we apply T-duality to the vertical direction (which we take to be
the $X^4$-direction), we arrive at the type IIA brane
configuration~\eqref{eq:linear-quiver-IIA} for an $\CN = 2$ linear
quiver theory, superposed on a $\C^2/\Z_k$ orbifold singularity, with
$\Z_k$ acting on $v = \exp(i(X^4 + iX^5))$ and $w = \exp(i(X^8 + iX^9))$
by
\begin{equation}
  v \to \exp(2\pi i/k) v
  \,,
  \quad  
  w \to \exp(-2\pi i/k) w
  \,.
\end{equation}
This is a situation studied in~\cite{Lykken:1997gy, Uranga:1998vf}.
Lifting this type IIA system to M-theory, we obtain $N$ M5-branes
placed on the orbifold and wrapped on a sphere with $2$ maximal and
$m$ minimal punctures.  The low-energy dynamics of the $N$ M5-branes
probing a $\C^2/\Z_k$ orbifold singularity is governed by the 6d
$\CN = (1,0)$ theory of type $(A_{N-1}, \Z_k)$.  The quiver gauge
theory under consideration is therefore an example of a class-$\CS_k$
theory.

Various symmetries of the 4d theory arise from six dimensions as
follows.  The global symmetry of the 6d theory is
$\SU(2)_R \times \U(1)_t \times \SU(k)_\beta \times \SU(k)_\gamma$.
The theory is topologically twisted along the punctured sphere by a
subgroup $\U(1)_R$ of the R-symmetry $\SU(2)_R$ (i.e., the
structure group $\U(1)_C$ of the sphere is replaced with the diagonal
subgroup of $\U(1)_C \times \U(1)_R$).  Due to the twisting, only the
$\U(1)_R$ part of $\SU(2)_R$ commutes with the rotation group, and it
descends to an R-symmetry of the 4d theory.%
\footnote{In general, its generator differs from the R-charge that
  appears in the infrared superconformal algebra by a linear
  combination of other $\U(1)$ charges.  The superconformal R-charge
  can be determined by $a$-maximization~\cite{Intriligator:2003jj}.}
Also, we turn on Wilson lines for $\SU(k)_\beta \times \SU(k)_\gamma$
so that the 4d theory has a nice Lagrangian description.  The Wilson
lines break the flavor symmetry to its abelian part
$\U(1)_t \times \mathrm{S}[\prod_{i=1}^k \U(1)_{\beta_i}] \times
\mathrm{S}[\prod_{i=1}^k \U(1)_{\gamma_i}]$.
Additionally, the 4d theory inherits flavor symmetries
$\U(1)_{\alpha_j}$, $j = 1$, $\dotsc$, $m$ from the minimal punctures.
The symmetries associated to the zigzag paths come from these $\U(1)$
flavor symmetries.  Finally, each maximal puncture gives rise to a set
of $k$ $\SU(N)$ flavor symmetries, represented in the quiver by a
column of $k$ flavor nodes.

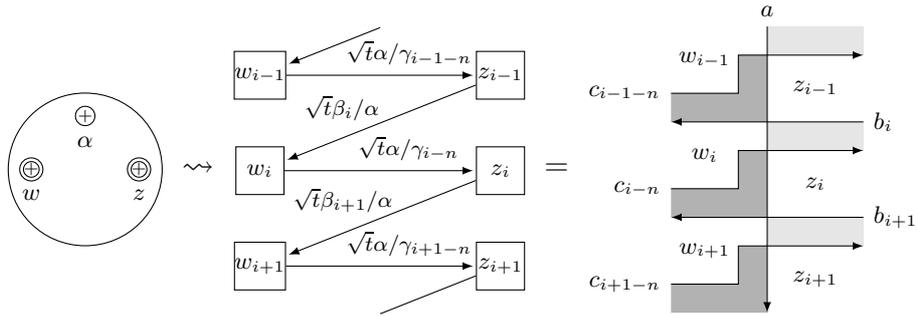
\begin{figure}
  \centering
  \begin{tikzpicture}[scale=0.8, baseline=(x.base)]
    \node (x) at (0,0) {\vphantom{x}};

    \draw (0,0) circle (1);

    \node[maxp, label={below:$w$}] (w) at (-0.7,0) {$+$};
    \node[maxp, label={below:$z$}] (z) at (0.7,0) {$+$};
    \node[minp, label={below:$\alpha$}] (min) at (0,0.7) {$+$};
  \end{tikzpicture}
  \
  {\large $\leadsto$}
  \
  \begin{tikzpicture}[xscale=2.5, baseline=(wi.base)]
    \node[fnode, minimum size=18pt] (wi-1) at (0,1) {$w_{i-1}$};
    \node[fnode, minimum size=18pt] (wi) at (0,0) {$w_i$};
    \node[fnode, minimum size=18pt] (wi+1) at (0,-1) {$w_{i+1}$};
    \node[fnode, minimum size=18pt] (zi-1) at (1,1) {$z_{i-1}$};
    \node[fnode, minimum size=18pt] (zi) at (1,0) {$z_i$};
    \node[fnode, minimum size=18pt] (zi+1) at (1,-1) {$z_{i+1}$};

    \draw[q->] (0.5,1.5) -- (wi-1);
    \draw[q->] (zi-1) -- (wi)
    node[above=3pt, pos=0.7]{$\sqrt{t} \beta_i/\alpha$};
    \draw[q->] (zi) -- (wi+1)
    node[above=4pt, pos=0.7] {$\sqrt{t} \beta_{i+1}/\alpha$};
    \draw[q-] (zi+1) -- (0.5,-1.5);

    \draw[q->] (wi-1) -- (zi-1)
    node[above, pos=0.65] {$\sqrt{t} \alpha/\gamma_{i-1-n}$};
    \draw[q->] (wi) -- (zi)
    node[above, pos=0.65] {$\sqrt{t} \alpha/\gamma_{i-n}$};
    \draw[q->] (wi+1) -- (zi+1)
    node[above, pos=0.65] {$\sqrt{t} \alpha/\gamma_{i+1-n}$};
  \end{tikzpicture}
  \
  {\large =}
  \begin{tikzpicture}[baseline=(x.base)]
    \node (x) at (0,1.5) {\vphantom{x}};

    \begin{scope}
      \fill[lshaded] (1,0.7) rectangle (2,1);
      \fill[dshaded] (0,0) -- (1,0) -- (1,0.7) -- (0.7,0.7)
      -- (0.7,0.3) -- (0,0.3) -- cycle;

      \draw[z->] (0,0.3) node[left] {$c_{i+1-n}$}
      -- (0.7,0.3) -- (0.7,0.7) -- (2,0.7);

      \node (wi+1) at (0.35,0.65) {$w_{i+1}$};
      \node (zi+1) at (1.5,0.35) {$z_{i+1}$};
    \end{scope}

    \begin{scope}[shift={(0,1)}]
      \fill[lshaded] (1,0.7) rectangle (2,1);
      \fill[dshaded] (0,0) -- (1,0) -- (1,0.7) -- (0.7,0.7)
      -- (0.7,0.3) -- (0,0.3) -- cycle;

      \draw[z->] (0,0.3) node[left] {$c_{i-n}$}
      -- (0.7,0.3) -- (0.7,0.7) -- (2,0.7);

      \node (wi+1) at (0.35,0.65) {$w_i$};
      \node (zi+1) at (1.5,0.35) {$z_i$};
    \end{scope}

    \begin{scope}[shift={(0,2)}]
      \fill[lshaded] (1,0.7) rectangle (2,1);
      \fill[dshaded] (0,0) -- (1,0) -- (1,0.7) -- (0.7,0.7)
      -- (0.7,0.3) -- (0,0.3) -- cycle;

      \draw[z->] (0,0.3) node[left] {$c_{i-1-n}$}
      -- (0.7,0.3) -- (0.7,0.7) -- (2,0.7);

      \node (wi+1) at (0.35,0.65) {$w_{i-1}$};
      \node (zi+1) at (1.5,0.35) {$z_{i-1}$};
    \end{scope}

    \draw[z->] (1,3) node[above] {$a$} -- (1,0);
    \draw[z->] (2,1) node[right] {$b_{i+1}$} -- (0,1);
    \draw[z->] (2,2) node[right] {$b_i$} -- (0,2);
  \end{tikzpicture}

  \caption{The building block of class-$\CS_k$ theories associated to
    a trinion.  The left maximal puncture has $(\pc, \po) = (n+1, +1)$
    and the right one has $(\pc, \po) = (n, -1)$.}
  \label{fig:trinion+++Sk}
\end{figure}

A building block of quivers of this kind is a strip of bifundamental
chiral multiplets depicted in Fig.~\ref{fig:trinion+++Sk}.  It is
associated to a trinion with one minimal and two maximal punctures.
(Anticipating introduction of punctures of different types, we have
drawn the minimal and maximal punctures with a plus sign as
\tikz[baseline={([yshift=-1pt]min.base)}]{\node[minp] (min) at (0,0)
  {$+$};} and \tikz[baseline={([yshift=-1pt]z.base)}]{\node[maxp] (z)
  at (0,0) {$+$};}\,, respectively.)  A puncture is labeled with its
flavor symmetry, $\U(1)$ or $\SU(N)^k$.  In addition, a maximal
puncture carries labels called ``color'' $\pc \in \Z_k$ and
``orientation'' $\po \in \{\pm 1\}$.  The orientation simply
distinguishes the two maximal punctures in the trinion.  In our
pictures, we will always place the maximal puncture with positive
orientation on the left and the one with negative orientation on the
right.  The color is defined by the relation between fugacities:
arrows with fugacities $\sqrt{t} \beta_i/\alpha$ and
$\sqrt{t} \alpha/\gamma_{i-\pc+\po}$ start from or end at the same
node in the column of nodes corresponding to a maximal puncture
labeled $(\pc, \po)$.  The color of the positively oriented puncture
is greater than that of the negatively oriented puncture by $1$.

The brane tiling diagram for a trinion is also shown in
Fig.~\ref{fig:trinion+++Sk}.  The minimal puncture corresponds to
the vertical zigzag path in the middle, while the maximal punctures
correspond to the unshaded regions on the sides of the diagram.  The
two sets of fugacities $(\alpha, \beta_i, \gamma_i)$ and
$(a, b_i, c_i)$ are related by
\begin{equation}
  \label{eq:abc-rel-Sk}
  a = \frac{1}{\alpha}
  \,,
  \quad
  b_i = \frac{1}{\sqrt{t} \beta_i}
  \,,
  \quad
  c_i = \sqrt{\frac{t}{pq}} \frac{1}{\gamma_i}
  \,.
\end{equation}
From the brane tiling point of view, the color of a maximal puncture
can be defined by the rule that the $(N,\po)$ 5-brane regions are
sandwiched by pairs of zigzag paths with fugacities
$(b_i, c_{i-\pc})$.

To reconstruct the quiver with $m+1$ columns that we started with, we
glue together $m$ copies of trinions to get the $(m+2)$-punctured
sphere.  Gluing can be done only between two maximal punctures with
opposite orientation and the same color.  This operation gauges the
diagonal combination of the $\SU(N)^k$ flavor symmetries of the
maximal punctures, and at the same time, adds in bifundamental chiral
multiplets corresponding to arrows going upward between the gauged
nodes.  The restriction on the color and orientation ensures that the
mixed anomalies for $\U(1)_{\beta_i}$ and $\U(1)_{\gamma_i}$ cancel.

The rule for gluing is transparent in the brane tiling picture.  When
we concatenate two brane tiling diagrams, we must connect the zigzag
paths in a way consistent with their labels, or the associated flavor
symmetries would be lost.  Therefore, the colors of the maximal
punctures glued together are required to match.  Furthermore, each
pair of horizontal paths near the glued sides is forced to cross once,
resulting in the additional vertical arrows in the combined quiver.

From the brane tiling perspective, it is also clear that the color of
the positively oriented maximal puncture increases by $1$ as we glue a
trinion to it, since the zigzag paths with fugacities $c_i$ shift
upward when they cross a vertical path.  In particular, the color
comes back to the original value after $k$ trinions are glued.

\subsection{Turning on flux}

By definition, class-$\CS_k$ theories arise from compactification of
the 6d theory on punctured Riemann surfaces.  In order to completely
specify a class-$\CS_k$ theory, however, we need more data than just a
punctured Riemann surface.  When we compactify the 6d theory, we can
turn on flux for the abelian part of its flavor symmetry, i.e., we
have a choice of the associated line bundles.  Consequently, there are
different theories associated to the same punctured Riemann surface,
corresponding to different flux backgrounds for $\U(1)_t$,
$\U(1)_{\beta_i}$, and $\U(1)_{\gamma_i}$.  In fact, for $k = 1$ we
have already analyzed the case with $\U(1)_{\CF}$ flux in
section~\ref{sec:class-S}.  This is the class-$\CS$ counterpart of the
case with $\U(1)_t$ flux, which we will treat in
section~\ref{sec:class-Sk-flux}.  Here we discuss flux for
$\U(1)_{\beta_i}$ and $\U(1)_{\gamma_i}$.

A procedure for turning on flux in class-$\CS_k$ theories was proposed
in~\cite{Gaiotto:2015usa}.  Suppose we want to create flux for
$\U(1)_{\beta_*}$ in a class-$\CS_k$ theory, where $* \in \Z_k$ is a
fixed index.  To do that, we glue a trinion (of the sort depicted in
Fig.~\ref{fig:trinion+++Sk}) to a maximal puncture of the associated
Riemann surface.  The new surface thus obtained has one more minimal
puncture than the original surface does, hence one more flavor
symmetry $\U(1)_\alpha$.  Then we ``close'' this puncture: we give a
constant vev to the baryon made of the bifundamental chiral multiplet
with fugacity $\sqrt{t} \beta_*/\alpha$, and ``flip'' the other
baryons in the same column, whose fugacities are
$(\sqrt{t} \beta_i/\alpha)^N$ with $i \neq *$.  By ``flipping a chiral
operator $X$,'' we mean coupling $X$ to an external chiral multiplet
$\phi_X$ through a superpotential $\phi_X X$.  After closing the
puncture, we obtain a theory associated to the same surface as the
original one, but with the color of the maximal puncture shifted by
$1$.  The result is interpreted as a theory with one unit of
$\U(1)_{\beta_*}$ flux turned on.

Similarly, we can turn on (minus) one unit of $\U(1)_{\gamma_*}$ flux
by giving a vev to the antibaryon with fugacity
$(\sqrt{t} \alpha/\gamma_*)^N$.  More generally, we can repeat the
above procedure to add any amount of flux for $\U(1)_{\beta_i}$ and
$\U(1)_{\gamma_i}$.  If we turn on flux for more than one flavor
symmetries, there are different orders of doing this.  However, they
all lead to the same result due to the S-duality permuting minimal
punctures.

An important point is that adding one unit of $\U(1)_{\beta_i}$ flux
for every $i$ (or one unit of $\U(1)_{\gamma_i}$ flux for every $i$)
is equivalent to doing nothing.  This is because the $\U(1)_{\beta_i}$
symmetries come from the $\SU(N)_\beta$ symmetry of the 6d theory,
hence the sum of their charges is zero.  We can see this property more
explicitly as follows.

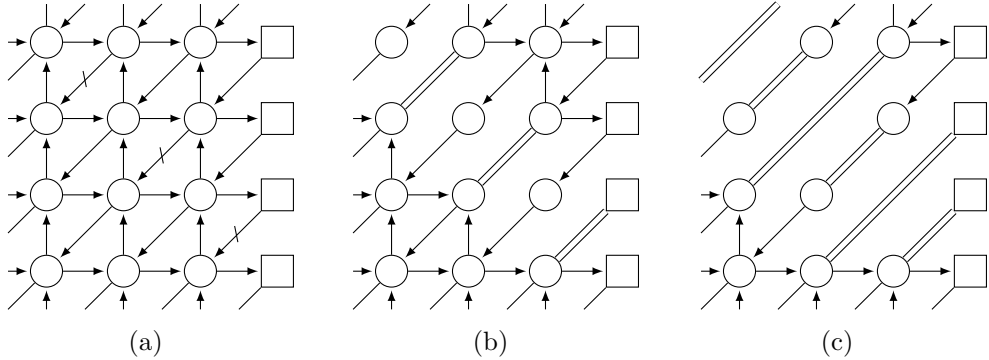
\begin{figure}
\centering
\subcaptionbox{\label{fig:closing-all-a}}{
  \begin{tikzpicture}[scale=0.8]
    \node[gnode] (a1) at (1,0) {};
    \node[gnode] (b1) at (1,-1) {};
    \node[gnode] (c1) at (1,-2) {};
    \node[gnode] (d1) at (1,-3) {};
    \node[gnode] (a2) at (2,0) {};
    \node[gnode] (b2) at (2,-1) {};
    \node[gnode] (c2) at (2,-2) {};
    \node[gnode] (d2) at (2,-3) {};
    \node[gnode] (a3) at (3,0) {};
    \node[gnode] (b3) at (3,-1) {};
    \node[gnode] (c3) at (3,-2) {};
    \node[gnode] (d3) at (3,-3) {};
    \node[fnode] (a4) at (4,0) {};
    \node[fnode] (b4) at (4,-1) {};
    \node[fnode] (c4) at (4,-2) {};
    \node[fnode] (d4) at (4,-3) {};

    \draw[q->,vev] (a2) -- (b1);
    \draw[q->] (b2) -- (c1);
    \draw[q->] (c2) -- (d1);

    \draw[q->] (a3) -- (b2);
    \draw[q->,vev] (b3) -- (c2);
    \draw[q->] (c3) -- (d2);

    \draw[q->] (a4) -- (b3);
    \draw[q->] (b4) -- (c3);
    \draw[q->,vev] (c4) -- (d3);

    \draw[q->] (a1) -- (a2);
    \draw[q->] (a2) -- (a3);
    \draw[q->] (a3) -- (a4);

    \draw[q->] (b1) -- (b2);
    \draw[q->] (b2) -- (b3);
    \draw[q->] (b3) -- (b4);

    \draw[q->] (c1) -- (c2);
    \draw[q->] (c2) -- (c3);
    \draw[q->] (c3) -- (c4);

    \draw[q->] (d1) -- (d2);
    \draw[q->] (d2) -- (d3);
    \draw[q->] (d3) -- (d4);

    \draw[q->] (b1) -- (a1);
    \draw[q->] (c1) -- (b1);
    \draw[q->] (d1) -- (c1);

    \draw[q->] (b2) -- (a2);
    \draw[q->] (c2) -- (b2);
    \draw[q->] (d2) -- (c2);

    \draw[q->] (b3) -- (a3);
    \draw[q->] (c3) -- (b3);
    \draw[q->] (d3) -- (c3);

    \draw[q->] ($(a1) + (0.5,0.5)$) -- (a1);
    \draw[q->] ($(a2) + (0.5,0.5)$) -- (a2);
    \draw[q->] ($(a3) + (0.5,0.5)$) -- (a3);

    \draw[q-] (a1) -- ($(a1) + (0,0.5)$);
    \draw[q-] (a2) -- ($(a2) + (0,0.5)$);
    \draw[q-] (a3) -- ($(a3) + (0,0.5)$);

    \draw[q->] ($(d1) + (0,-0.5)$) -- (d1);
    \draw[q->] ($(d2) + (0,-0.5)$) -- (d2);
    \draw[q->] ($(d3) + (0,-0.5)$) -- (d3);

    \draw[q->] ($(a1) + (-0.5,0)$) -- (a1);
    \draw[q->] ($(b1) + (-0.5,0)$) -- (b1);
    \draw[q->] ($(c1) + (-0.5,0)$) -- (c1);
    \draw[q->] ($(d1) + (-0.5,0)$) -- (d1);

    \draw[q-] (a1) -- ($(a1) + (-0.5,-0.5)$);
    \draw[q-] (b1) -- ($(b1) + (-0.5,-0.5)$);
    \draw[q-] (c1) -- ($(c1) + (-0.5,-0.5)$);
    \draw[q-] (d1) -- ($(d1) + (-0.5,-0.5)$);

    \draw[q-] (d2) -- ($(d2) + (-0.5,-0.5)$);
    \draw[q-] (d3) -- ($(d3) + (-0.5,-0.5)$);
    \draw[q-] (d4) -- ($(d4) + (-0.5,-0.5)$);
  \end{tikzpicture}
}
\quad
\subcaptionbox{\label{fig:closing-all-b}}{
  \begin{tikzpicture}[scale=0.8]
    \node[gnode] (a1) at (1,0) {};
    \node[gnode] (b1) at (1,-1) {};
    \node[gnode] (c1) at (1,-2) {};
    \node[gnode] (d1) at (1,-3) {};
    \node[gnode] (a2) at (2,0) {};
    \node[gnode] (b2) at (2,-1) {};
    \node[gnode] (c2) at (2,-2) {};
    \node[gnode] (d2) at (2,-3) {};
    \node[gnode] (a3) at (3,0) {};
    \node[gnode] (b3) at (3,-1) {};
    \node[gnode] (c3) at (3,-2) {};
    \node[gnode] (d3) at (3,-3) {};
    \node[fnode] (a4) at (4,0) {};
    \node[fnode] (b4) at (4,-1) {};
    \node[fnode] (c4) at (4,-2) {};
    \node[fnode] (d4) at (4,-3) {};

    \draw[eq-] (a2) -- (b1);
    \draw[q->] (b2) -- (c1);
    \draw[q->] (c2) -- (d1);

    \draw[q->] (a3) -- (b2);
    \draw[eq-] (b3) -- (c2);
    \draw[q->] (c3) -- (d2);

    \draw[q->] (a4) -- (b3);
    \draw[q->] (b4) -- (c3);
    \draw[eq-] (c4) -- (d3);

    \draw[q->] (a2) -- (a3);
    \draw[q->] (a3) -- (a4);

    \draw[q->] (b3) -- (b4);

    \draw[q->] (c1) -- (c2);

    \draw[q->] (d1) -- (d2);
    \draw[q->] (d2) -- (d3);
    \draw[q->] (d3) -- (d4);

    \draw[q->] (c1) -- (b1);
    \draw[q->] (d1) -- (c1);

    \draw[q->] (d2) -- (c2);

    \draw[q->] (b3) -- (a3);

    \draw[q->] ($(a1) + (0.5,0.5)$) -- (a1);
    \draw[q->] ($(a2) + (0.5,0.5)$) -- (a2);
    \draw[q->] ($(a3) + (0.5,0.5)$) -- (a3);

    \draw[q-] (a2) -- ($(a2) + (0,0.5)$);
    \draw[q-] (a3) -- ($(a3) + (0,0.5)$);

    \draw[q->] ($(d1) + (0,-0.5)$) -- (d1);
    \draw[q->] ($(d2) + (0,-0.5)$) -- (d2);
    \draw[q->] ($(d3) + (0,-0.5)$) -- (d3);

    \draw[q->] ($(b1) + (-0.5,0)$) -- (b1);
    \draw[q->] ($(c1) + (-0.5,0)$) -- (c1);
    \draw[q->] ($(d1) + (-0.5,0)$) -- (d1);

    \draw[q-] (a1) -- ($(a1) + (-0.5,-0.5)$);
    \draw[q-] (b1) -- ($(b1) + (-0.5,-0.5)$);
    \draw[q-] (c1) -- ($(c1) + (-0.5,-0.5)$);
    \draw[q-] (d1) -- ($(d1) + (-0.5,-0.5)$);

    \draw[q-] (d2) -- ($(d2) + (-0.5,-0.5)$);
    \draw[q-] (d3) -- ($(d3) + (-0.5,-0.5)$);
    \draw[q-] (d4) -- ($(d4) + (-0.5,-0.5)$);
  \end{tikzpicture}
}
\quad
\subcaptionbox{\label{fig:closing-all-c}}{
  \begin{tikzpicture}[scale=0.8]
    \node (a1) at (1,0) {};
    \node[gnode] (b1) at (1,-1) {};
    \node[gnode] (c1) at (1,-2) {};
    \node[gnode] (d1) at (1,-3) {};
    \node[gnode] (a2) at (2,0) {};
    \node[gnode] (c2) at (2,-2) {};
    \node[gnode] (d2) at (2,-3) {};
    \node[gnode] (a3) at (3,0) {};
    \node[gnode] (b3) at (3,-1) {};
    \node[gnode] (d3) at (3,-3) {};
    \node[fnode] (a4) at (4,0) {};
    \node[fnode] (b4) at (4,-1) {};
    \node[fnode] (c4) at (4,-2) {};
    \node[fnode] (d4) at (4,-3) {};

    \draw[eq-] (a3) -- (c1);
    \draw[eq-] (b4) -- (d2);

    \draw[eq-] (a2) -- (b1);
    \draw[q->] (c2) -- (d1);

    \draw[eq-] (b3) -- (c2);

    \draw[q->] (a4) -- (b3);
    \draw[eq-] (c4) -- (d3);

    \draw[q->] (a3) -- (a4);

    \draw[q->] (d1) -- (d2);
    \draw[q->] (d2) -- (d3);
    \draw[q->] (d3) -- (d4);

    \draw[q->] (d1) -- (c1);

    \draw[eq-] ($(a1) + (0.5,0.5)$) -- ($(a1) + (-0.5,-0.5)$);

    \draw[q->] ($(a2) + (0.5,0.5)$) -- (a2);
    \draw[q->] ($(a3) + (0.5,0.5)$) -- (a3);

    \draw[q-] (a3) -- ($(a3) + (0,0.5)$);

    \draw[q->] ($(d1) + (0,-0.5)$) -- (d1);
    \draw[q->] ($(d2) + (0,-0.5)$) -- (d2);
    \draw[q->] ($(d3) + (0,-0.5)$) -- (d3);

    \draw[q->] ($(c1) + (-0.5,0)$) -- (c1);
    \draw[q->] ($(d1) + (-0.5,0)$) -- (d1);

    \draw[q-] (b1) -- ($(b1) + (-0.5,-0.5)$);
    \draw[q-] (c1) -- ($(c1) + (-0.5,-0.5)$);
    \draw[q-] (d1) -- ($(d1) + (-0.5,-0.5)$);

    \draw[q-] (d2) -- ($(d2) + (-0.5,-0.5)$);
    \draw[q-] (d3) -- ($(d3) + (-0.5,-0.5)$);
    \draw[q-] (d4) -- ($(d4) + (-0.5,-0.5)$);
  \end{tikzpicture}
}
\caption{Closing $k$ minimal punctures.  In the first step, we give
  vevs to the baryons at the arrows on the diagonal.  In the
  subsequent steps we apply relation~\eqref{eq:I-inv} repeatedly.}
  \label{fig:closing-all}
\end{figure}

To create one unit of flux for each and every $\U(1)_{\beta_i}$ in a
given theory, we attach to the theory a sphere with $2$ maximal and
$k$ minimal punctures, and give vevs to $k$ baryons charged under
distinct $\U(1)_{\beta_i}$.  Let us suppose that the maximal puncture
to which we attach the sphere has $\po = -1$.  We align the minimal
punctures horizontally and number them $1$, $\dotsc$, $k$ from left to
right.  Denoting by $\alpha_i$ the fugacity for the $\U(1)$ flavor
symmetry associated to the $i$th puncture, we choose to give vevs to
the baryons with fugacities $(\sqrt{t} \beta_i/\alpha_i)^N$, $i = 1$,
$\dotsc$, $k$.  Fig.~\ref{fig:closing-all-a} illustrates the
situation.

The vevs identify pairs of nodes as we explained before, and also turn
the cubic superpotentials involving the stuck-out arrows into
quadratic ones that give masses to the other arrows participating in
these superpotentials.  After the massive arrows are integrated out,
we have the situation in Fig.~\ref{fig:closing-all-b}.  The quiver now
contains a number of gauge nodes with two arrows attached.  These
nodes exhibit confinement and chiral symmetry breaking, and as a
result, further pairs of nodes are identified, as in
Fig.~\ref{fig:closing-all-c}.  The identification of nodes results in
more gauge nodes with two arrows attached, which again equate pairs of
nodes.  This process continues until all flavor nodes are identified
with the gauge nodes coming from the maximal puncture to which the
sphere was attached.  There are baryons left over from the
confinement, but they couple to the scalars introduced in the flip
operation and together become massive.  In the end, all minimal
punctures are gone, and we recover the original theory.

\subsection{\texorpdfstring{Surface defects in $A_1$ theories}{Surface
    defects in A1 theories}}
\label{subsec:surfaceSk}

Surface defects in class-$\CS_k$ theories can be realized via RG flows
in much the same way as in their counterparts in class-$\CS$ theories.
Given a theory $\CT_\IR$, we construct another theory~$\CT_\UV$ by
gluing a trinion to it.  Then, we give a position-dependent vev
$\vev{B} = \zeta_1^r \zeta_2^s$ to a baryon $B$ charged under the
flavor symmetry of the new minimal puncture, while flipping the other
baryons in the same column.  The vev triggers $\CT_\UV$ to flow to the
original theory~$\CT_\IR$, but in the presence of a surface defect
labeled with the pair of integers $(r,s)$, or the pair
$(\underbrace{\yng(2) \dotsm \yng(1)}_r, \underbrace{\yng(2) \dotsm
  \yng(1)}_s$) of symmetric representations of $\SU(N)$.

A novelty in class-$\CS_k$ theories is that maximal punctures have
colors.  For the color of the maximal puncture of $\CT_\IR$ to remain
unchanged by the above operation, the trinion we attach to it must
have two maximal punctures of the same color.  This is not the case
for the trinion in Fig.~\ref{fig:trinion+++Sk}, as it has maximal
punctures differing in their colors by $1$.

Luckily, we know how to make the colors of the maximal punctures
match: we glue to it $k-1$ more trinions of the same type, and close
all minimal punctures introduced in the process.  More precisely, we
prepare a sphere with $2$ maximal and $k$ minimal punctures, pick
$* \in \Z_k$, and for all $i \neq *$, give constant vevs to the
baryons with fugacities $(\sqrt{t} \beta_i/\alpha_i)^N$ and flip those
with fugacities $(\sqrt{t} \beta_j/\alpha_i)^N$, $j \neq i$.  Then,
the $(k+2)$-punctured sphere flows to a trinion with one minimal
puncture and two maximal punctures of the same color, with minus one
unit of flux turned on for $\U(1)_{\beta_*}$.

In summary, a surface defect in $\CT_\IR$ is realized as follows.
First, we construct $\CT_\UV$ by coupling to $\CT_\IR$ a sphere with
$2$ maximal and $k$ minimal punctures.  Next, we give constant vevs to
the baryons with fugacities $(\sqrt{t} \beta_i/\alpha_i)^N$,
$i \neq *$.  Finally, we give the position-dependent vev to the baryon
with fugacity $(\sqrt{t} \beta_*/\alpha_*)^N$, and flip all other
baryons.  There are $k$ choices for the index $*$, and each choice
leads to a different surface defect.  Of course, we could follow the
same procedure with the roles of baryons and antibaryons exchanged, so
in total there are $2k$ inequivalent surface defects that may be
constructed in this way for a given pair of integers~$(r,s)$.

What we have to do for the computation of the supersymmetric index
with a surface defect is clear now.  We couple the quiver shown in
Fig.~\ref{fig:closing-all-a} to $\CT_\IR$ by gauging the diagonal
combination of the $\SU(N)^k$ flavor symmetry in the leftmost column
of the quiver and the $\SU(N)^k$ flavor symmetry of a maximal puncture
of $\CT_\IR$.  Then we replace one of the
$\tikz[baseline={([yshift=-2pt]x.base)}]{\node (x) at (0,0) {};
  \draw[q->, vev] (0,0) -- (1,0);}$
arrows with
$\tikz[baseline={([yshift=-2pt]x.base)}]{\node (x) at (0,0) {};
  \draw[q->, vev] (0,0) -- node[below] {$(r,s)$}
  (1,0);}$,
and compute the residues accordingly.  The result is the index of
$\CT_\IR$ in the presence of a surface defect of type $(r,s)$.

Let us carry out this computation in the simplest case of $A_1$
theories and $(r,s) = (0,1)$, when the surface defect is labeled with
the fundamental representation of $\SU(2)$.

Without loss of generality, we assume that the maximal puncture of
$\CT_\IR$ to which we glue the $(k+2)$-punctured sphere has color
$\pc = 0$.  The baryon $B$ that is given the position-dependent vev
$\vev{B} = \zeta_2$ is made of the bifundamental chiral multiplet with
fugacity $\rho_*$, where $\rho_i = \sqrt{t} \beta_i/\alpha_i$.  Since
the index is independent of the position of punctures, we can
rearrange the minimal punctures so that the $*$th puncture comes to
the rightmost position and the rest follow in descending order in
their labels.  After the rearrangement, the neighborhood of the baryon
looks as in Fig.~\ref{fig:surface-defect-a}, where we have introduced
the symbols $\sharp = * - 1$ and $\flat = *-2$ for convenience.

The identity~\eqref{eq:giving-vev} obtained in the previous section
states that the constant vevs given to $k-1$ arrows identify pairs of
nodes and make the neighboring arrows massive, producing the quiver
shown in Fig.~\ref{fig:surface-defect-b}.

\begin{figure}
\centering
\subcaptionbox{\label{fig:surface-defect-a}}{
  \begin{tikzpicture}
    \node[gnode] (a2) at (2,0) {$x_\flat$};
    \node[gnode] (b2) at (2,-1) {$x_\sharp$};
    \node[gnode] (c2) at (2,-2) {$x_*$};
    \node[gnode] (a3) at (3,0) {$w_\flat$};
    \node[gnode] (b3) at (3,-1) {$w_\sharp$};
    \node[gnode] (c3) at (3,-2) {$w_*$};
    \node[fnode] (a4) at (4,0) {$z_\flat$};
    \node[fnode] (b4) at (4,-1) {$z_\sharp$};
    \node[fnode] (c4) at (4,-2) {$z_*$};

    \draw[q->,vev] (a3) -- (b2);
    \draw[q->] (b3) -- (c2);

    \draw[q->] (a4) -- (b3);
    \draw[q->,vev] (b4) -- node[below, sloped] {$(0,1)$} (c3);

    \draw[q->] (a2) -- (a3);
    \draw[q->] (a3) -- (a4);

    \draw[q->] (b2) -- (b3);
    \draw[q->] (b3) -- (b4);

    \draw[q->] (c2) -- (c3);
    \draw[q->] (c3) -- (c4);

    \draw[q->] (b2) -- (a2);
    \draw[q->] (c2) -- (b2);

    \draw[q->] (b3) -- (a3);
    \draw[q->] (c3) -- (b3);

    \draw[q->] ($(a2) + (0.5,0.5)$) -- (a2);
    \draw[q->] ($(a3) + (0.5,0.5)$) -- (a3);

    \draw[q-] (a2) -- ($(a2) + (0,0.5)$);
    \draw[q-] (a3) -- ($(a3) + (0,0.5)$);

    \draw[q->] ($(c2) + (0,-0.5)$) -- (c2);
    \draw[q->] ($(c3) + (0,-0.5)$) -- (c3);

    \draw[q->] ($(a2) + (-0.5,0)$) -- (a2);
    \draw[q->] ($(b2) + (-0.5,0)$) -- (b2);
    \draw[q->] ($(c2) + (-0.5,0)$) -- (c2);

    \draw[q-] (a2) -- ($(a2) + (-0.5,-0.5)$);
    \draw[q-] (b2) -- ($(b2) + (-0.5,-0.5)$);
    \draw[q-] (c2) -- ($(c2) + (-0.5,-0.5)$);
    \draw[q-] (c3) -- ($(c3) + (-0.5,-0.5)$);
    \draw[q-] (c4) -- ($(c4) + (-0.5,-0.5)$);
  \end{tikzpicture}
}
\qquad
\subcaptionbox{\label{fig:surface-defect-b}}{
  \begin{tikzpicture}
    \node[gnode] (a2) at (2,0) {$x_\flat$};
    \node[gnode] (b2) at (2,-1) {$x_\sharp$};
    \node[gnode] (c2) at (2,-2) {$x_*$};
    \node[gnode] (a3) at (3,0) {$w_\flat$};
    \node[gnode] (b3) at (3,-1) {$w_\sharp$};
    \node[gnode] (c3) at (3,-2) {$w_*$};
    \node[fnode] (a4) at (4,0) {$z_\flat$};
    \node[fnode] (b4) at (4,-1) {$z_\sharp$};
    \node[fnode] (c4) at (4,-2) {$z_*$};

    \draw[eq-] (a3) -- (b2);
    \draw[q->] (b3) -- (c2);

    \draw[q->] (a4) -- (b3);
    \draw[q->,vev] (b4) -- node[below, sloped] {$(0,1)$} (c3);

    \draw[q->] (a3) -- (a4);

    \draw[q->] (b3) -- (b4);

    \draw[q->] (c2) -- (c3);
    \draw[q->] (c3) -- (c4);

    \draw[q->] (c2) -- (b2);

    \draw[q->] (c3) -- (b3);

    \draw[q->] ($(a2) + (0.5,0.5)$) -- (a2);
    \draw[q->] ($(a3) + (0.5,0.5)$) -- (a3);

    \draw[q-] (a3) -- ($(a3) + (0,0.5)$);

    \draw[q->] ($(c2) + (0,-0.5)$) -- (c2);
    \draw[q->] ($(c3) + (0,-0.5)$) -- (c3);

    \draw[q->] ($(b2) + (-0.5,0)$) -- (b2);
    \draw[q->] ($(c2) + (-0.5,0)$) -- (c2);

    \draw[q-] (a2) -- ($(a2) + (-0.5,-0.5)$);
    \draw[q-] (b2) -- ($(b2) + (-0.5,-0.5)$);
    \draw[q-] (c2) -- ($(c2) + (-0.5,-0.5)$);
    \draw[q-] (c3) -- ($(c3) + (-0.5,-0.5)$);
    \draw[q-] (c4) -- ($(c4) + (-0.5,-0.5)$);
  \end{tikzpicture}
}
\caption{(a) A quiver engineering a surface defect of type
  $(r,s) = (0,1)$.  (b) The same quiver after the
  relation~\eqref{eq:giving-vev} is used.}
\label{fig:surface-defect}
\end{figure}

For the arrow with the position-dependent vev, we use
relation~\eqref{eq:vev-(0,1)}.  This relation says that $w_*$ is set
to $q^{1/2} z_\sharp$ or $q^{-1/2} z_\sharp$ after taking the residue.
To be specific, let us consider the former case.  In this case, at
$\rho_* = q^{-1/2}$ the $w_\sharp$-integral involves 12 factors of
elliptic gamma functions of the form
$\prod_{a=1}^6 \Gamma(t_a w_\sharp^{\pm1})$:
\begin{equation}
  \label{eq:GGGG}
  \Gamma\biggl(q^{1/2} t \frac{\beta_*}{\gamma_\sharp}
  w_\sharp^{\pm1} z_\sharp^{-1}\biggr) 
  \Gamma\biggl(\frac{pq^{3/2}}{t} \frac{\gamma_\sharp}{\beta_*}
  z_\sharp w_\sharp^{\pm1}\biggr)
  \Gamma\biggl(q^{-1/2} \frac{\beta_\sharp}{\beta_*}
  z_\flat^{\pm 1} w_\sharp^{\pm 1}\biggr)
  \Gamma\biggl(\frac{\beta_*}{\beta_\sharp}
  w_\sharp^{\pm 1} x_*^{\pm 1}\biggr)
  \,.
\end{equation}
Using identity~\eqref{eq:elliptic-beta}, we see that the
$w_\sharp$-integral evaluates to a product of elliptic gamma functions
which contains the factors
\begin{equation}
 \Gamma\bigl(pq^2\bigr)
 \Gamma\bigl(q^{-1/2} z_\flat^{\pm1} x_*^{\pm1}\bigr)
 \,.
\end{equation}
The first factor is zero.  However, the second factor has
singularities in the $x_*$-plane, namely double poles at four values
$x_* = q^{\pm 1/2} z_\flat^{\pm 1}$, of which two lie inside the
$x_*$-contour.  Therefore, the $w_\sharp$-integral vanishes for a
generic value of $x_*$, but may receive contributions from these two
double poles.

To evaluate the $w_\sharp$- and $x_*$-integrals properly, we can
multiply the fugacities of the first and last factors in the product
\eqref{eq:GGGG} by $\epsilon^2$ and $\epsilon^{-1}$, respectively, and
later take the limit $\epsilon \to 1$.  If we apply
identity~\eqref{eq:elliptic-beta} after this shift of fugacities, we
find that the double poles $x_* = q^{1/2} z_\flat^{\pm 1}$ are
resolved into four simple poles located at
$x_* = \epsilon^{\pm1} q^{1/2} z_\flat^{\pm 1}$.  It is
straightforward to compute the residues.  The end result of the
calculation is that from the $w_\sharp$- and $x_*$-integration, we get
\begin{equation}
  \sum_{s_\flat=\pm1}
  \frac{1}{\theta(z_\flat^{2s_\flat})}
  \theta\biggl(t \frac{\beta_*^2}{\beta_\sharp \gamma_\sharp}
  \frac{z_\flat^{s_\flat}}{z_\sharp}\biggr) 
  \theta\biggl(\frac{t}{q} \frac{\beta_\sharp}{\gamma_\sharp} \frac{1}{z_\flat^{s_\flat}
    z_\sharp}\biggr) 
  \Delta_\flat^{s_\flat/2}
  \
  \begin{tikzpicture}[baseline=(zk.base)]
    \node[fnode] (zk) at (0,0) {$z_\flat$};
    \node[gnode] (x2) at (1,0) {$x_*$};

    \draw[eq-] (zk) -- (x2);
  \end{tikzpicture}
  \ .
\end{equation}
The result for the case when $w_*$ is set to $q^{-1/2} z_\sharp$ is
obtained by replacing $z_\sharp$ with $z_\sharp^{-1}$.

Proceeding to the $w_\flat$-integral, we encounter the same
calculation after the $x_\flat$-integral is performed, and the index
receives contributions analogous to those found above for the
$w_\sharp$-integral.  The same is true for every other $w_i$-integral.
Apart from these contributions, the structure of the calculation is
essentially identical to the constant vev case illustrated in
Fig.~\ref{fig:closing-all}.  All in all, the effect of the surface
defect of type $(0,1)$ is represented in the index by the action of
the difference operator
\begin{equation}
  \mathfrak{S}_{(0,1)}^{(\beta_*)}
  =
  \sum_{(s_i) \in \{\pm 1\}^k}
  \frac{1}{\theta(q^{-1})}  
  \prod_{i \in \Z_k}
  \frac{1}{\theta(z_i^{2s_i})}
  \theta\biggl(t \frac{\beta_*^2}{\beta_i \gamma_i}
  \frac{z_{i-1}^{s_{i-1}}}{z_i^{s_i}}\biggr) 
  \theta\biggl(\frac{t}{q} \frac{\beta_i}{\gamma_i}
  \frac{1}{z_{i-1}^{s_{i-1}} z_i^{s_i}}\biggr) 
  \prod_{j \in \Z_k}
  \Delta_j^{s_j/2}
  \,,
\end{equation}
where $\{z_i\}$ are the fugacities for the $\SU(2)^k$ flavor symmetry
of the maximal puncture on which we constructed the surface defect.

We can perform a similar computation for the case when the antibaryon
charged under $\U(1)_{\gamma_*}$ is given the position-dependent vev.
The corresponding difference operator is
\begin{equation}
  \mathfrak{S}_{(0,1)}^{(\gamma_*)}
  =
  \sum_{(s_i) \in \{\pm 1\}^k}
  \frac{1}{\theta(q^{-1})}
  \prod_{i \in \Z_k}
  \frac{1}{\theta(z_i^{2s_i})}
  \theta\biggl(t \frac {\beta_i \gamma_i}{\gamma_*^2}
  \frac{z_i^{s_i}}{z_{i-1}^{s_{i-1}}}\biggr) 
  \theta\biggl(\frac{t}{q} \frac{\beta_i}{\gamma_i}
  \frac{1}{z_{i-1}^{s_{i-1}} z_i^{s_i}}\biggr) 
  \prod_{j \in \Z_k}
  \Delta_j^{s_j/2}
  \,.
\end{equation}
Note that in either case, the difference operator depends on the
choice of the index $*$ only through the value of the fugacity
$\beta_*$ or $\gamma_*$.

\subsection{Comparison with transfer matrices}

In the brane tiling picture, the above surface defects should be
represented by a dashed line that crosses $k$ pairs of horizontal
zigzag paths as in Fig.~\ref{fig:trinion-defectSk}.  We see that it
inserts the transfer matrix constructed from the product of
$\Ldia_i(d, (c_i,b_{i+1}))$.  Using formula~\eqref{eq:TM-Ldia} and
relation~\eqref{eq:abc-rel-Sk}, we can check
\begin{align}
  \mathfrak{S}_{(0,1)}^{(\beta_*)}
  &=
  \Tr_\W\Bigl(
  \Ldia_k\bigl(d_{\beta_*}, (c_k,b_{k+1})\bigr)
  \circ_\W \dotsb \circ_\W 
  \Ldia_1\bigl(d_{\beta_*}, (c_1,b_2)\bigr)
  \Bigr),
  \\
  \mathfrak{S}_{(0,1)}^{(\gamma_*)}
  &=
  \Tr_\W\Bigl(
  \Ldia_k\bigl(d_{\gamma_*}, (c_k,b_{k+1})\bigr)
  \circ_\W \dotsb \circ_\W 
  \Ldia_1\bigl(d_{\gamma_*}, (c_1,b_2)\bigr)
  \Bigr)
    \,,
\end{align}
up to an overall factor independent of the spectral parameters, with
\begin{equation}
  \label{eq:d-beta-gamma}
  d_{\beta_*} = \frac{1}{\sqrt{qt} \beta_*}
  \,,
  \quad
  d_{\gamma_*} = \sqrt{\frac{t}{pq}} \frac{1}{\gamma_*}
  \,.
\end{equation}
As expected, the transfer matrix reproduces the effects of these
surface defects on the index.

\begin{figure}
  \centering
  \begin{tikzpicture}
    \fill[lshaded] (0,0.7) rectangle (1.5,1);
    \fill[lshaded] (0,1.7) rectangle (1.5,2);

    \draw[z->] (1.5,1) node[right] {$b_{i+1}$} -- (0,1);
    \draw[z->] (1.5,2) node[right] {$b_i$} -- (0,2);

    \draw[z->] (0,0.7) node[left] {$c_{i+1}$} -- (1.5,0.7);
    \draw[z->] (0,1.7) node[left] {$c_i$} -- (1.5,1.7);

    \draw[dz->] (0.5,2.7) node[above] {$d$} -- (0.5,0);

    \node at (1,0.35) {$z_{i+1}$};
    \node at (1,1.35) {$z_i$};
    \node at (1,2.35) {$z_{i-1}$};
  \end{tikzpicture}

  \caption{The brane tiling representation of a surface defect acting
    on a maximal puncture with $(\pc,\po) = (0,-1)$.}
  \label{fig:trinion-defectSk}
\end{figure}

Interestingly, the transfer matrix unifies the $2k$ difference
operators $\mathfrak{S}_{(0,1)}^{(\beta_i)}$,
$\mathfrak{S}_{(0,1)}^{(\gamma_i)}$, corresponding to the $2k$ choices
of the position-dependent vev given to the trinion attached to
$\CT_\IR$, into a single one-parameter family of difference operators.
These operators differ simply in the value of the spectral parameter
for the dashed line.

\subsection{\texorpdfstring{The case with $\U(1)_t$ flux}{The case
    with U(1)t flux}}
\label{sec:class-Sk-flux}

Finally, we consider surface defects in theories with $\U(1)_t$
flux~\cite{Gaiotto:2015usa, Franco:2015jna, Hanany:2015pfa}.  This is
the $\C^2/\Z_k$-orbifold version of the $\CN = 1$ class-$\CS$ theories
discussed in section~\ref{sec:class-S-N=1}.  The type IIA brane
construction for theories associated to a cylinder or torus is the
same as in the class-$\CS$ case, except that the branes are placed at
the orbifold singularity.  Therefore, each puncture has a sign
$\sigma \in \{\pm 1\}$ specifying which type of NS5- or D6-brane it
comes from.

\begin{figure}
  \centering
  \begin{tikzpicture}[scale=0.8, baseline=(x.base)]
    \node (x) at (0,0) {\vphantom{x}};

    \draw (0,0) circle (1);

    \node[maxp, label=below:{$w$}] (w) at (-0.7,0) {$-$};
    \node[maxp, label=below:{$z$}] (z) at (0.7,0) {$-$};
    \node[minp, label=below:{$\alpha$}] (min) at (0,0.7) {$-$};
  \end{tikzpicture}
  \
  {\large $\leadsto$}
  \
  \begin{tikzpicture}[xscale=2.5, baseline=(wi.base)]
    \node[fnode, minimum size=18pt] (wi-1) at (0,1) {$w_{i-1}$};
    \node[fnode, minimum size=18pt] (wi) at (0,0) {$w_i$};
    \node[fnode, minimum size=18pt] (wi+1) at (0,-1) {$w_{i+1}$};
    \node[fnode, minimum size=18pt] (zi-1) at (1,1) {$z_{i-1}$};
    \node[fnode, minimum size=18pt] (zi) at (1,0) {$z_i$};
    \node[fnode, minimum size=18pt] (zi+1) at (1,-1) {$z_{i+1}$};

    \draw[q-] (zi-1) -- (0.5,1.5);
    \draw[q->] (zi) -- (wi-1)
    node[above=3pt, pos=0.2]
    {$\sqrt{t'} \alpha/\beta_i$};
    \draw[q->] (zi+1) -- (wi)
    node[above=4pt, pos=0.2] {$\sqrt{t'} \alpha/\beta_{i+1}$};
    \draw[q->] (0.5,-1.5) -- (wi+1);

    \draw[q->] (wi-1) -- (zi-1)
    node[above=-2pt, pos=0.4]{$\sqrt{t'} \gamma_{i-1-n}/\alpha$};
    \draw[q->] (wi) -- (zi)
    node[above=-2pt, pos=0.4] {$\sqrt{t'} \gamma_{i-n}/\alpha$};
    \draw[q->] (wi+1)-- (zi+1)
    node[above=-2pt, pos=0.4] {$\sqrt{t'} \gamma_{i+1-n}/\alpha$};
  \end{tikzpicture}
  \
  {\large =}
  \begin{tikzpicture}[baseline=(x.base)]
    \node (x) at (0,1.5) {\vphantom{x}};

    \begin{scope}
      \fill[dshaded] (1,0) rectangle (2,0.3);
      \fill[lshaded] (0,1) -- (1,1) -- (1,0.3) -- (0.7,0.3)
      -- (0.7,0.7) -- (0,0.7) -- cycle;

      \draw[z->] (0,0.7) node[left] {$c_{i+1-n}$}
      -- (0.7,0.7) -- (0.7,0.3) -- (2,0.3);

      \node (wi+1) at (0.35,0.35) {$w_{i+1}$};
      \node (zi+1) at (1.5,0.65) {$z_{i+1}$};
    \end{scope}

    \begin{scope}[shift={(0,1)}]
      \fill[dshaded] (1,0) rectangle (2,0.3);
      \fill[lshaded] (0,1) -- (1,1) -- (1,0.3) -- (0.7,0.3)
      -- (0.7,0.7) -- (0,0.7) -- cycle;

      \draw[z->] (0,0.7) node[left] {$c_{i-n}$}
      -- (0.7,0.7) -- (0.7,0.3) -- (2,0.3);

      \node (wi) at (0.35,0.35) {$w_i$};
      \node (zi) at (1.5,0.65) {$z_i$};
    \end{scope}

    \begin{scope}[shift={(0,2)}]
      \fill[dshaded] (1,0) rectangle (2,0.3);
      \fill[lshaded] (0,1) -- (1,1) -- (1,0.3) -- (0.7,0.3)
      -- (0.7,0.7) -- (0,0.7) -- cycle;

      \draw[z->] (0,0.7) node[left] {$c_{i-1-n}$}
      -- (0.7,0.7) -- (0.7,0.3) -- (2,0.3);

      \node (wi-1) at (0.35,0.35) {$w_{i-1}$};
      \node (zi-1) at (1.5,0.65) {$z_{i-1}$};
    \end{scope}

    \draw[z->] (1,0) node[below] {$a$} -- (1,3);
    \draw[z->] (2,1) node[right] {$b_{i+1}$} -- (0,1);
    \draw[z->] (2,2) node[right] {$b_i$} -- (0,2);
  \end{tikzpicture}

  \caption{A trinion with three negative punctures.  The colors of the
    maximal punctures on the left and right are $\pc = n$ and $n+1$,
    respectively, and $t' = pq/t$.}
  \label{fig:trinion---Sk}
\end{figure}
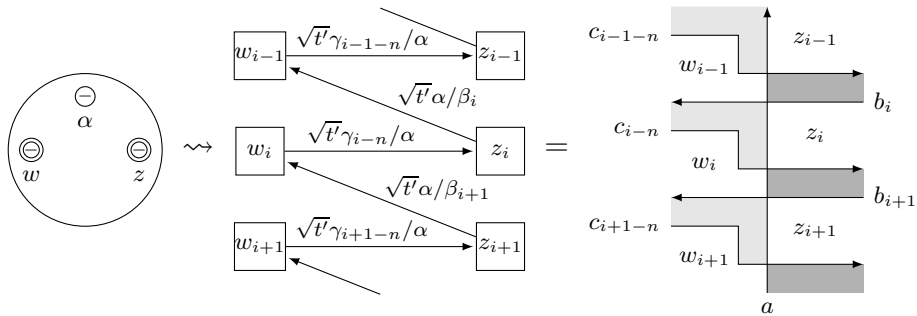

The trinion in Fig.~\ref{fig:trinion+++Sk} has $\sigma = +1$ for all
punctures.  If we flip the sign of a maximal puncture, there arise
additional bifundamental chiral multiplets between the flavor nodes
coming from that puncture; see Fig.~\ref{fig:trinion++-Sk}.  On the
other hand, flipping the signs of all punctures gives the quiver in
Fig.~\ref{fig:trinion---Sk}.

\begin{figure}
  \centering
  \begin{tikzpicture}[scale=0.8, baseline=(x.base)]
    \node (x) at (0,0) {\vphantom{x}};

    \draw (0,0) circle (1);

    \node[maxp, label={below:$w$}] (w) at (-0.7,0) {$+$};
    \node[maxp, label={below:$z$}] (z) at (0.7,0) {$-$};
    \node[minp, label={below:$\alpha$}] (min) at (0,0.7) {$+$};
  \end{tikzpicture}
  \
  {\large $\leadsto$}
  \
  \begin{tikzpicture}[xscale=2.5, baseline=(wi.base)]
    \node[fnode, minimum size=18pt] (wi-1) at (0,1) {$w_{i-1}$};
    \node[fnode, minimum size=18pt] (wi) at (0,0) {$w_i$};
    \node[fnode, minimum size=18pt] (wi+1) at (0,-1) {$w_{i+1}$};
    \node[fnode, minimum size=18pt] (zi-1) at (1,1) {$z_{i-1}$};
    \node[fnode, minimum size=18pt] (zi) at (1,0) {$z_i$};
    \node[fnode, minimum size=18pt] (zi+1) at (1,-1) {$z_{i+1}$};

    \draw[q->] (0.5,1.5) -- (wi-1);
    \draw[q->] (zi-1) -- (wi)
    node[above=3pt, pos=0.7]{$\sqrt{t} \beta_i/\alpha$};
    \draw[q->] (zi) -- (wi+1)
    node[above=4pt, pos=0.7] {$\sqrt{t} \beta_{i+1}/\alpha$};
    \draw[q-] (zi+1) -- (0.5,-1.5);

    \draw[q->] (wi-1) -- (zi-1)
    node[above, pos=0.65] {$\sqrt{t} \alpha/\gamma_{i-1-n}$};
    \draw[q->] (wi) -- (zi)
    node[above, pos=0.65] {$\sqrt{t} \alpha/\gamma_{i-n}$};
    \draw[q->] (wi+1) -- (zi+1)
    node[above, pos=0.65] {$\sqrt{t} \alpha/\gamma_{i+1-n}$};

    \draw[q->] (1,-1.5) -- (zi+1);
    \draw[q->] (zi+1) -- (zi);
    \draw[q->] (zi) -- (zi-1);
    \draw[q-] (zi-1) -- (1,1.5);
  \end{tikzpicture}
  \
  {\large =}  
  \begin{tikzpicture}[baseline=(x.base)]
    \node (x) at (0,1.5) {\vphantom{x}};

    \begin{scope}
      \fill[lshaded] (1,0.7) rectangle (1.3,1);
      \fill[dshaded] (1.3,0) rectangle (2,0.3);
      \fill[dshaded] (0,0) -- (1,0) -- (1,0.7) -- (0.7,0.7)
      -- (0.7,0.3) -- (0,0.3) -- cycle;

      \draw[z-] (0,0.3) node[left] {$c_{i+1-n}$}
      -- (0.7,0.3) -- (0.7,0.7) -- (1.3,0.7) -- (1.3,1);

      \draw[z->] (1.3,0) -- (1.3,0.3) -- (2,0.3);

      \node (wi+1) at (0.35,0.65) {$w_{i+1}$};
      \node (zi+1) at (1.65,0.65) {$z_{i+1}$};
    \end{scope}

    \begin{scope}[shift={(0,1)}]
      \fill[lshaded] (1,0.7) rectangle (1.3,1);
      \fill[dshaded] (1.3,0) rectangle (2,0.3);
      \fill[dshaded] (0,0) -- (1,0) -- (1,0.7) -- (0.7,0.7)
      -- (0.7,0.3) -- (0,0.3) -- cycle;

      \draw[z-] (0,0.3) node[left] {$c_{i-n}$}
      -- (0.7,0.3) -- (0.7,0.7) -- (1.3,0.7) -- (1.3,1);

      \draw[z->] (1.3,0) -- (1.3,0.3) -- (2,0.3);

      \node (wi) at (0.35,0.65) {$w_i$};
      \node (zi) at (1.65,0.65) {$z_i$};
    \end{scope}

    \begin{scope}[shift={(0,2)}]
      \fill[lshaded] (1,0.7) rectangle (1.3,1);
      \fill[dshaded] (1.3,0) rectangle (2,0.3);
      \fill[dshaded] (0,0) -- (1,0) -- (1,0.7) -- (0.7,0.7)
      -- (0.7,0.3) -- (0,0.3) -- cycle;

      \draw[z-] (0,0.3) node[left] {$c_{i-1-n}$}
      -- (0.7,0.3) -- (0.7,0.7) -- (1.3,0.7) -- (1.3,1);

      \draw[z->] (1.3,0) -- (1.3,0.3) -- (2,0.3);

      \node (wi-1) at (0.35,0.65) {$w_{i-1}$};
      \node (zi-1) at (1.65,0.65) {$z_{i-1}$};
    \end{scope}

    \draw[z->] (1,3) node[above] {$a$} -- (1,0);
    \draw[z->] (2,1) node[right] {$b_{i+1}$} -- (0,1);
    \draw[z->] (2,2) node[right] {$b_i$} -- (0,2);
  \end{tikzpicture}

  \caption{A trinion with one positive minimal, one positive maximal
    and one negative maximal puncture.}
  \label{fig:trinion++-Sk}
\end{figure}
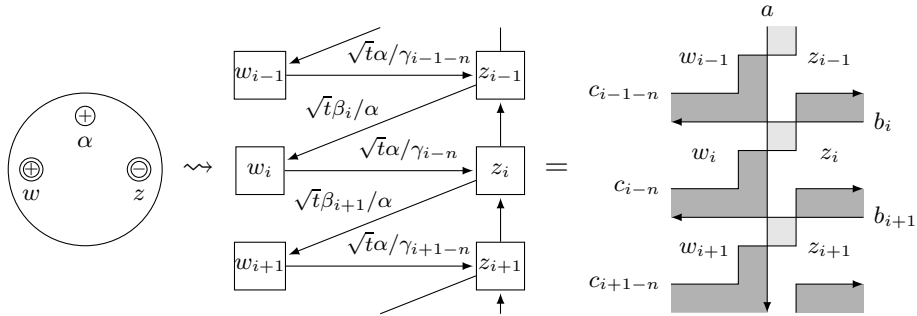

Looking at the brane tiling diagrams for these quivers, we notice that
the sign of a puncture is correlated with the 5-brane charge
distribution in the relevant area: a minimal puncture with sign
$\sigma$ corresponds to a column of $k$ $(N, \sigma)$ 5-brane regions
in the middle, while a maximal puncture of orientation $\po$ and sign
$\sigma$ corresponds to a column of $k$ $(N,\po\sigma)$ 5-brane
regions on a side. (This point of view makes it clear that each
$\sigma$ is really a $k$-tuple of signs, as pointed out
in~\cite{Hanany:2015pfa}.)  The change in the color between the left
and right maximal punctures is equal to $\sigma$ of the minimal
puncture.  When we glue two trinions together, the maximal punctures
connected by a tube must have opposite signs in order for the 5-brane
charges to be conserved.  Otherwise, we have to flip the sign of one
of the punctures, as we did in the case of gluing two trinions with
positive punctures.

Let us determine how surface defects act on a negative maximal
puncture.  We take a trinion with three positive punctures and glue it
to the trinion in Fig.~\ref{fig:trinion---Sk} from the left.  Then,
we close the minimal puncture contained in the latter trinion by
giving a vev to its baryon charged under $\U(1)_{\beta_*}$ for some
$* \in \Z_k$, to obtain a trinion with one minimal and two maximal
punctures having the same color and different signs.  This trinion has
minus one unit of flux for $\U(1)_{\beta_*}$.  We attach it to a
negative maximal puncture (which we assume to have color $\chi = 0$)
in some theory from the right, and give the position-dependent vev
$\vev{B} = \zeta_2$ to the baryon charged under $\U(1)_{\beta_*}$ in
the trinion.  This gives a surface defect of type $(0,1)$ acting on
the negative maximal puncture.

A calculation similar to the one we performed before shows that the
surface defect acts on the index as the difference operator
\begin{equation}
  \mathfrak{S}_{(0,1)}^{(\beta_*,-)}
  =
  \sum_{(s_i) \in \{\pm 1\}^k}
  \frac{1}{\theta(q^{-1})}  
  \prod_{i \in \Z_k}
  \frac{1}{\theta(z_i^{2s_i})}
  \theta\biggl(t \frac{\beta_*^2}{\beta_i \gamma_i}
  \frac{z_{i-1}^{s_{i-1}}}{z_i^{s_i}}\biggr) 
  \theta\biggl(t \frac{\beta_i}{\gamma_i}
  z_{i-1}^{s_{i-1}} z_i^{s_i}\biggr)
  \prod_{j \in \Z_k}
  \Delta_j^{s_j/2}
  \,.
\end{equation}
We can also make a surface defect by giving the position-dependent vev
to the antibaryon charged under $\U(1)_{\gamma_*}$.  This surface
defect is represented by
\begin{equation}
  \mathfrak{S}_{(0,1)}^{(\gamma_*,-)}
  =
  \sum_{(s_i) \in \{\pm 1\}^k}
  \frac{1}{\theta(q^{-1})}
  \prod_{i \in \Z_k}
  \frac{1}{\theta(z_i^{2s_i})}
  \theta\biggl(t \frac {\beta_i \gamma_i}{\gamma_*^2}
  \frac{z_i^{s_i}}{z_{i-1}^{s_{i-1}}}\biggr) 
  \theta\biggl(t \frac{\beta_i}{\gamma_i}
  z_{i-1}^{s_{i-1}} z_i^{s_i}\biggr) 
  \prod_{j \in \Z_k}
  \Delta_j^{s_j/2}
  \,.
\end{equation}

In the brane tiling diagram, the surface defect creates a dashed line
as in Fig.~\ref{fig:trinion---defect}.  From
formula~\eqref{eq:TM-Ldiabar} and relations~\eqref{eq:abc-rel-Sk}
and~\eqref{eq:d-beta-gamma}, we can check that the above difference
operators are reproduced from the transfer matrix.

\begin{figure}
  \centering
  \begin{tikzpicture}
    \fill[dshaded] (0,0.7) rectangle (1.5,1);
    \fill[dshaded] (0,1.7) rectangle (1.5,2);

    \draw[z->] (0,1) node[left] {$c_{i+1}$} -- (1.5,1);
    \draw[z->] (0,2) node[left] {$c_i$} -- (1.5,2);

    \draw[z<-] (0,0.7) -- (1.5,0.7) node[right] {$b_{i+1}$};
    \draw[z<-] (0,1.7) -- (1.5,1.7) node[right] {$b_i$};

    \draw[dz->] (0.5,2.7) node[above] {$d$} -- (0.5,0);

    \node at (1,0.35) {$z_{i+1}$};
    \node at (1,1.35) {$z_i$};
    \node at (1,2.35) {$z_{i-1}$};
  \end{tikzpicture}

  \caption{The brane tiling representation of a surface defect acting
    on a negative maximal puncture with $(\pc,\po) = (0,-1)$.}
  \label{fig:trinion---defect}
\end{figure}

\section{Outlook}
\label{sec:con}

The key element underlying various connections between 4d $\CN = 1$
supersymmetric field theories and integrable lattice models is the
emergence of the structure of a 2d TQFT equipped with line operators
that are localized in extra dimensions~\cite{Costello:2013zra,
  Costello:2013sla, Yagi:2015lha}.  Branes in string theory, combined
with protected quantities such as supersymmetric indices, provide a
natural framework in which such structures may be found.

In this paper we have utilized this framework to elucidate the
integrability aspect of surface defects in 4d $\CN = 1$ theories.  As
we have seen, under the correspondence between brane tilings and
integrable lattice models, a class of half-BPS surface defects labeled
with representations of $\SU(N)$ are mapped to transfer matrices
constructed from L-operators.  In the case of the fundamental
representation of $\SU(2)$, the relevant L-operator is that of
Sklyanin, which satisfies an RLL relation with Baxter's R-operator for
the eight-vertex model.  We have shown that the corresponding transfer
matrix unifies the $2k$ difference operators obtained by the residue
method for $A_1$ theories of class $\CS_k$.

Our analysis is far from complete, however.  Obviously lacking is the
answer to the following question: What is the L-operator for a general
representation of $\SU(N)$?

We may approach this important question either from the lattice model
side or from the field theory side.  The strategy on the lattice model side
would be to search for an L-operator that solves the appropriate
Yang--Baxter equations, as we have done for the fundamental
representation of $\SU(2)$.

From the field theory side, the strategy is to somehow compute the
indices of brane tiling models in the presence of general surface
defects, and read off the L-operator from them.  For example, we may
combine the residue method for class-$\CS_k$ theories, which can
handle the symmetric representations, and analysis of the algebra
generated by the resulting difference operators.  This program had
some success in the class-$\CS$ case~\cite{Alday:2013kda,
  Bullimore:2014nla}.  A different method is to realize a surface
defect as a 2d $\CN = (0,2)$ theory, and compute the index of the
coupled 4d--2d system by localization of the path integral.  This was
done in~\cite{Gadde:2013ftv} for $k = 1$ and symmetric
representations.  A related computation appeared
in~\cite{Chen:2014rca}.

Either strategy is not without shortcomings.  The Yang--Baxter
equations do not uniquely determine the L-operator.  The
supersymmetric indices, on the other hand, encode transfer matrices
but not the L-operator directly.  We would therefore need to combine
approaches from both sides.

Another direction we have left unexplored is the study of the 2d TQFT
itself, which in our discussion just served as an intermediate step
whereby brane tilings and integrable lattice models were connected.
String theory predicts that this TQFT is related to the 2d TQFT
arising from the indices of class-$\CS_k$ theories through a duality
exchanging line operators in the former and local operators in the
latter.  It may be possible to determine these TQFTs by localization
computations along the lines of~\cite{Kawano:2012up, Fukuda:2012jr,
  Yagi:2013fda, Lee:2013ida, Cordova:2013cea, Kawano:2015ssa}.

There are many more interesting questions to be asked in relation to
4d $\CN = 1$ supersymmetric field theories and integrable lattice
models.  We wish to answer some of them in future work.

\section*{Acknowledgements}

We are grateful to Giulio Bonelli and Alessandro Tanzini for their
invitation to the workshop ``V Workshop on Geometric Correspondences
of Gauge Theories'' at SISSA, during which this project was initiated.
We also thank Hironori Mori, Jaewon Song, and Shigeki Sugimoto for
useful discussions, and Michio Jimbo, Saburo Kekei, Satoshi Nawata,
Shlomo Razamat, Vyacheslav Spiridonov, Piotr Su\l kowski, and Yuji
Yamada for helpful comments.  K.M. would like to thank Piotr Su\l
kowski for his hospitality at the University of Warsaw, where part of
this work was carried out.  The work of K.M. is supported by the EPSRC
Programme Grant EP/K034456/1 ``New Geometric Structures from String
Theory.''  The work of J.Y. is supported by the ERC Starting Grant
no.~335739 ``Quantum fields and knot homologies'' funded by the
European Research Council under the European Union's Seventh Framework
Programme.

\appendix
\section{Definitions and useful formulas}

In this appendix we collect definitions and useful formulas concerning
special functions we encounter in this paper.

\subsection{Theta functions}

The Jacobi theta functions are defined by
\begin{align}
  \theta_1(\zeta|\tau)
  &=
  -\sum_{n \in \Z}
  \exp(\pi i(n + 1/2)^2 \tau)
  \exp(2\pi i(n + 1/2)(\zeta + 1/2))
    \,,
  \\
  \theta_2(\zeta|\tau)
  &=
  \theta_1(\zeta + 1/2|\tau)
    \,,
  \\
  \theta_3(\zeta|\tau)
  &=
  \exp(\pi i\tau/4 + \pi i\zeta) \theta_2(\zeta + \tau/2|\tau)
    \,,
  \\
  \theta_4(\zeta|\tau)
  &=
  \theta_3(\zeta + 1/2|\tau)
    \,,
\end{align}
where $\zeta$ is a complex variable and $\tau$ is a complex parameter
in the upper half-plane.  The first of these, $\theta_1$, is an odd
function of $\zeta$ and satisfies
\begin{align}
  \theta_1(\zeta+1|\tau) 
  &=
  -\theta_1(\zeta|\tau)
    \,,
  \\
  \theta_1(\zeta+\tau|\tau) 
  &=
  -\exp(2\pi i\zeta - \pi i\tau) \theta_1(\zeta|\tau)
    \,.
\end{align}
The other three are even functions.  We have
\begin{align}
  2\theta_1(\zeta+\zeta') \theta_1(\zeta-\zeta')
  &=
  \thetab_4(\zeta) \thetab_3(\zeta') - \thetab_4(\zeta')
    \thetab_3(\zeta)
    \,,
  \\
  2\theta_2(\zeta+\zeta') \theta_2(\zeta-\zeta')
  &=
  \thetab_3(\zeta) \thetab_3(\zeta') - \thetab_4(\zeta')
    \thetab_4(\zeta)
    \,,
  \\
  2\theta_3(\zeta+\zeta') \theta_3(\zeta-\zeta')
  &=
  \thetab_3(\zeta) \thetab_3(\zeta') + \thetab_4(\zeta')
    \thetab_3(\zeta)
    \,,
  \\
  2\theta_4(\zeta+\zeta') \theta_4(\zeta-\zeta')
  &=
  \thetab_4(\zeta) \thetab_3(\zeta') + \thetab_4(\zeta')
    \thetab_3(\zeta)
    \,,
\end{align}
with $\theta_a(\zeta) = \theta_a(\zeta|\tau)$ and
$\thetab_a(\zeta) = \theta_a(\zeta|\tau/2)$.

Closely related to the Jacobi theta functions is the function
\begin{equation}
  \theta(z;p) = (z;p)_\infty (p/z;p)_\infty
  \,;
  \quad
  (z;p)_\infty = \prod_{k=0}^\infty (1 - p^k z)
  \,,
  \
  |p| < 1
  \,.
\end{equation}
It satisfies
\begin{equation}
  \theta(z;p) = \theta(p/z;p)
  \,.
\end{equation}
In multiplicative notation,
\begin{equation}
  \theta_a(z; p) = \theta_a(\zeta|\tau)
  \,;
  \quad
  z = \exp(2\pi i\zeta)
  \,,
  \
  p = \exp(2\pi i\tau)
  \,,
\end{equation}
the Jacobi theta functions can be written in terms of $\theta$ as
\begin{align}
  \theta_1(z;p)
  &=
  i p^{1/8} (p;p)_\infty z^{-1/2} \theta(z;p)
    \,,
  \\
  \theta_2(z;p)
  &=
    p^{1/8} (p;p)_\infty z^{-1/2} \theta(-z;p)
    \,,
  \\
  \theta_3(z;p)
  &= (p;p)_\infty \theta(-\sqrt{p} z; p)
    \,,
  \\
  \theta_4(z;p)
  &= (p;p)_\infty \theta(\sqrt{p} z; p)
    \,.
\end{align}

\subsection{Elliptic gamma function}

The elliptic gamma function depends on two complex parameters $p$ and
$q$:
\begin{equation}
  \Gamma(z; p,q)
  =
  \prod_{j,k=0}^\infty
  \frac{1 - z^{-1} p^{j+1} q^{k+1}}{1 - zp^j q^k}
  \,;
  \quad
  |p|,\,|q| < 1
  \,.
\end{equation}
It satisfies the identities
\begin{equation}
  \Gamma(z; p,q)  \Gamma(pq/z; p,q) = 1
\end{equation}
and
\begin{align}
  \Gamma(pz; p,q)
  &= \theta(z; q) \Gamma(z; p,q)
    \,,
  \\
  \Gamma(qz; p,q)
  &= \theta(z; p) \Gamma(z; p,q)
    \,.
\end{align}

The function $\Gamma(z; p,q) $ has a pole at $z = p^{-j} q^{-k}$,
where $j$, $k$ are nonnegative integers.  The residue at this pole is
given by
\begin{equation}
  \label{eq:Res-Gamma}
  \Res_{z = p^{-j} q^{-k}} \bigl[\Gamma(z; p,q)\bigr]
  =
  \frac{(-1)^{jk + j + k} p^{(k+1)j(j+1)/2} q^{(j+1)k(k+1)/2}}
          {(p;p)_\infty (q;q)_\infty
            \theta(p, \dotsc, p^j; q) \theta(q, \dotsc, q^k; p)}
          \,,
\end{equation}
where we have introduced the notation
$\theta(z_1, \dotsc, z_n; q) = \theta(z_1; q) \dotsm \theta(z_n; q)$.

Let $t_j$, $j = 1$, $\dotsc$, $6$ be six complex parameters such that
$|t_j| < 1$ and $\prod_{j=1}^6 t_j = pq$.  Then, we have the following
identity proved in~\cite{MR1846786}:
\begin{equation}
  \label{eq:elliptic-beta}
  \frac{(p;p)_\infty (q;q)_\infty}{2}
  \int_\T \frac{\rmd z}{2\pi iz}
  \frac{\prod_{j=1}^6 \Gamma(t_j z^{\pm 1}; p,q)}
          {\Gamma(z^{\pm 2}; p,q)}
  =
  \prod_{1 \leq j < k \leq 6} \Gamma(t_j t_k; p,q)
  \,.
\end{equation}
Here $\T$ is the unit circle with counterclockwise orientation and
\begin{equation}
  \Gamma(z^{\pm n}; p,q) = \Gamma(z^n; p,q) \Gamma(z^{-n}; p,q)
  \,.
\end{equation}
The left-hand side of the above formula is known as the elliptic beta
integral.

\end{document}